\documentclass[english]{article}
\usepackage[latin9]{inputenc}
\usepackage{geometry}
\usepackage{mathtools}
\usepackage{amsmath}
\usepackage{graphicx}
\usepackage{esint}
\usepackage[authoryear]{natbib}

\makeatletter

\DeclareFontEncoding{LGR}{}{}

\ProvideTextCommand{\~}{LGR}[1]{\char126#1}

\usepackage{tikz}
\usetikzlibrary{calc}

\newcommand{\lyxaddress}[1]{
	\par {\raggedright #1
	\vspace{1.4em}
	\noindent\par}
}
\newcommand{\Real}{\mbox{Re}} 
\newcommand{\Imag}{\mbox{Im}} 

\newcommand{\Ma}{\text{Ma}}
\newcommand{\Bo}{\text{Bo}}
\@addtoreset{equation}{section}
\numberwithin{equation}{section}
\usepackage{babel}

\makeatother

\usepackage{babel}
\begin{document}
\title{Surfactant and gravity dependent instability of two-layer channel
flows: Linear theory covering all wave lengths }
\author{Alexander L. Frenkel, David Halpern and Adam J. Schweiger}
\maketitle

\lyxaddress{Department of Mathematics, University of Alabama, Tuscaloosa AL 35487,
USA}

\section*{Abstract}

A linear stability analysis of a two-layer plane Couette flow of two
immiscible fluid layers with different densities, viscosities and
thicknesses, bounded by two infinite parallel plates moving at a constant
relative velocity to each other, with an insoluble surfactant monolayer
along the interface and in the presence of gravity is carried out.
The normal modes approach is applied to the equations governing flow
disturbances in the two layers. These equations, together with boundary
conditions at the plates and the interface, yield a linear eigenvalue
problem. When inertia is neglected the velocity amplitudes are the
linear combinations of certain hyperbolic functions, and a quadratic
dispersion equation for the increment, that is the complex growth
rate, is obtained where coefficients depend on the aspect ratio, the
viscosity ratio, the basic velocity shear, the Marangoni number $\Ma$
that measures the effects of surfactant, and the Bond number $\Bo$
that measures the influence of gravity. An extensive investigation
is carried out that examines the stabilizing or destabilizing influences
of these parameters. Since the dispersion equation is quadratic in
the growth rate, there are two continuous branches of the normal modes:
a robust branch that exists even with no surfactant, and a surfactant
branch that, to the contrary, vanishes when $\Ma\downarrow0$. Due
to the availability of explicit forms for the growth rates, in many
instances the numerical results are corroborated with analytical asymptotics.
For the less unstable branch, a mid-wave interval of unstable wavenumbers
(\citet{Halpern2003}) sometimes co-exists with a long-wave one. We
study the instability landscape, determined by the threshold curve
of the long-wave instability and the critical curve of the mid-wave
instability in the ($\textrm{Ma, Bo}$)-plane. The changes of the
extremal points of the critical curves with the variation of the other
parameters, such as the viscosity ratio, and the extrema bifurcation
points are investigated.

\section{Introduction}

\label{sec:Intro} Surfactants are surface active compounds that reduce
the surface tension between two fluids, or between a fluid and a solid.
\citet{Frenkel2002} (hereafter referred to as FH) and \citet{Halpern2003}
(from now on referred to as HF) uncovered that certain stable surfactant-free
Stokes flows become unstable if an interfacial surfactant is introduced.
For this, the interfacial shear of velocity must be nonzero; in particular,
this instability disappears if the basic flow is stopped. In contrast
to the well-known instability of two viscous fluids (\citet{Yih1967})
which needs inertia effects for its existence, this instability may
exist in the absence of fluid inertia. With regard to multi-fluid
horizontal channel flows, this instability has been further studied
in a number of papers, such as \citet{Blyth2004b}, \citet{Pozrikidis2004a},
\citet{Blyth2004a}, \citet{Frenkel2005}, \citet{Wei2005c}, \citet{Frenkel2006},
\citet{Halpern2008}, \citet{Bassom2010}, \citet{JIE2010}, \citet{kalogirou2016},
\citet{picardo2016}, and \citet{Frenkel2017}. In the latter paper,
we have added gravity to the long-wave considerations of FH. Since
in the absence of surfactants gravity can be either stabilizing or
destabilizing depending on the flow parameters, the interaction of
the Rayleigh-Taylor instability with the surfactant instability leads
to interesting phenomena.

In the present work, we expand the linear stability analysis of \citet{Frenkel2017},
which was confined to long waves, by including disturbances of arbitrary
wavenumbers. The current paper can also be regarded as an extension
of HF, who considered arbitrary wavenumbers, by incorporating the
effects of gravity. As was indicated in \citet{Frenkel2017}, one
can expect a rich landscape of stability properties, especially since
there are two active normal modes of infinitesimal disturbances corresponding
to the presence of two interfacial functions: the interface displacement
function and the interfacial surfactant concentration (FH, HF). Since
the growth rates of the normal modes satisfy a (complex) quadratic
equation, and thus are relatively simple, in many instances numerical
results may enjoy analytic (asymptotic) corroboration. The stability
properties of two-layer Couette flows with both the interfacial surfactant
and gravity effects for arbitrary wavenumbers were the subject of
the dissertation \citet{schweiger2013gravity}. These studies are
further developed and expanded in the present paper. In section \ref{sec:GovEqnsStabForm},
the stability problem is formulated. In section \ref{sec:FiniteCaseDerivationDispRelation},
the dispersion equation is obtained. The long-wave stability properties
are considered in section \ref{sec:FiniteCaseLongwaveApprox}, while
in section \ref{sec:FiniteCaseMidWave} we consider normal modes of
arbitrary wavelengths and study the so-called mid-wave instability
(uncovered in HF but significantly modified by gravity effects). In
section \ref{subsec:MA-Bo plane stability}, we consider the instability
landscape in the (Marangoni number, Bond number)-plane that is determined
by the threshold curve of the long-wave instability and the critical
curve of the mid-wave instability, and study how it changes with the
other parameters. Finally, section \ref{sec:Conclusions} contains
discussion and concluding remarks. Some of the more technical information
appears in Appendices.

\section{Stability problem formulation}

\label{sec:GovEqnsStabForm} 
\begin{figure}
\centerline{\includegraphics[width=0.85\textwidth]{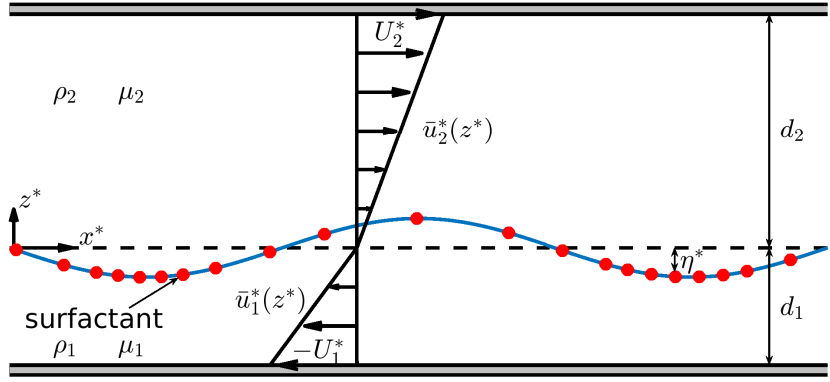}} \caption{Sketch of a disturbed two-layer Couette flow of two horizontal liquid
layers with different thicknesses, viscosities, and mass densities.
The insoluble surfactant monolayer is located at the interface and
is indicated by the dots. The (spanwise) uniform gravity field with
a constant acceleration $g$ is not shown. \label{fig:FigDefinitionSketch} }
\end{figure}

The general framework and governing equations
of the problem were given before (see \citet{schweiger2013gravity},
\citet{Frenkel2016}, \citet{Frenkel2017}) and are as follows. Two
immiscible Newtonian fluid layers with different densities, viscosities
and thicknesses are bounded by two infinite horizontal plates, a distance
$d=d_{1}+d_{2}$ apart, with the top plate moving at a constant relative
velocity, $U^{*}$,as shown in figure \ref{fig:FigDefinitionSketch}.
The $z^{\ast}$-axis is the spanwise, vertical, coordinate perpendicular
to the moving plates, with the upper plate located at $z^{\ast}=d_{2}$
and the lower plate located at $z^{\ast}=-d_{1}$, and with $z^{\ast}=0$
determining the location of the unperturbed liquid-liquid interface.
(The symbol $^{\ast}$ indicates a dimensional quantity.) The direction
of the horizontal $x^{\ast}$-axis is parallel to the plates. At the
interface, the surface tension, $\sigma^{\ast}$, depends on the concentration
of the insoluble surfactant monolayer, $\Gamma^{\ast}$. The frame
of reference is fixed at the liquid-liquid interface so that the velocity
of the lower plate is $-U_{1}^{*}$, and that of the upper plate is
$U_{2}^{*}$, where $U_{1}^{*}+U_{2}^{*}=U^{*}$, the velocity of
the top plate relative to the bottom plate. In the base state, the
horizontal velocity profiles are linear in $z^{\ast}$, the interface
is flat, and the surfactant concentration is uniform. Once disturbed,
the surfactant concentration is no longer uniform and the deflection
of the interface is represented by the function $\eta^{\ast}(x^{\ast},t^{\ast})$
where $t^{\ast}$ represents the time. The infinitesimal disturbances
may grow under the action of the Marangoni and/or gravity forces (\citet{Frenkel2017}).
The governing equations for this problem are given, for example, in
\citet{Frenkel2016}, in both dimensional and dimensionless forms.
(Also, the dimensionless form of these equations can be found in \citet{Frenkel2017}.)
We use the following notations (with $j=1$ for the bottom liquid
layer and $j=2$ for the top liquid layer): $\rho_{j}$ is the density;
$\boldsymbol{v}_{j}^{\ast}=(u_{j}^{\ast},w_{j}^{\ast})$ is the fluid
velocity vector with horizontal component $u_{j}^{\ast}$ and vertical
component $w_{j}^{\ast}$; $p_{j}^{\ast}$ is the pressure; $\mu_{j}$
is the viscosity; and $g$ is the gravity acceleration.

We assume the dependence of surface tension $\sigma^{\ast}$ on the
surfactant concentration $\Gamma^{\ast}$ to be given by the Langmuir
isotherm relation (e.g., \citet{Edwards1991}). For the small disturbances,
\begin{equation}
\sigma^{\ast}=\sigma_{0}-E(\Gamma^{\ast}-\Gamma_{0}),\label{eq:nineSigmaStar}
\end{equation}
where $\sigma_{0}$ is the base surface tension corresponding to the
base surfactant concentration $\Gamma_{0}$ and the known constant
$E\coloneqq\left.-(\partial\sigma^{\ast}/\partial\Gamma^{\ast})\right|_{\Gamma^{*}=\Gamma_{0}}$
is the elasticity parameter.

We use the following dimensionless variables: 
\[
(x,z,\eta)=\frac{(x^{\ast},z^{\ast},\eta^{\ast})}{d_{1}}\text{, }t=\frac{t^{\ast}}{d_{1}\mu_{1}/\sigma_{0}}\text{, }\boldsymbol{v}_{j}=(u_{j},w_{j})=\frac{(u_{j}^{\ast},w_{j}^{\ast})}{\sigma_{0}/\mu_{1}}\text{,}
\]
\begin{equation}
\text{ }p_{j}=\frac{p_{j}^{\ast}}{\sigma_{0}/d_{1}}\,\text{, }\Gamma=\frac{\Gamma^{\ast}}{\Gamma_{0}}\text{, }\sigma=\frac{\sigma^{\ast}}{\sigma_{0}}\text{.}\label{eq:scales}
\end{equation}
As in \citet{Frenkel2016,Frenkel2017}, the dimensionless velocity
field of the basic Couette flow, with a flat interface, $\eta=0$,
uniform surface tension, $\bar{\sigma}=1$, and corresponding surfactant
concentration, $\bar{\Gamma}=1$ (where the over-bar indicates a base
quantity), is 
\begin{equation}
\bar{u}_{1}(z)=sz,\;\bar{w}_{1}=0\text{, }\text{and }\bar{p}_{1}=-\text{Bo}_{1}z\text{ \ \ for \ }-1\leq z\leq0\text{,}\label{eq:u1w1p1BSprofiles}
\end{equation}
\begin{equation}
\bar{u}_{2}(z)=\frac{s}{m}z,\;\bar{w}_{2}=0\text{, and }\bar{p}_{2}=-\text{Bo}_{2}z\text{ \ \ for \ }0\leq z\leq n,\label{eq:u2w2p2BSprofiles}
\end{equation}
where $\text{Bo}_{j}:=\rho_{j}gd_{1}^{2}/\sigma_{0}$ is the Bond
number of the layer $j$, $m=\mu_{2}/\mu_{1}$ is the ratio of the
viscosities, and $n=d_{2}/d_{1}$ is the ratio of the thicknesses.
The constant $s$ represents the base interfacial shear rate of the
bottom layer, $s=D\bar{u}_{1}(0)$, where $D=d/dz$, and is used to
characterize the flow instead of the relative velocity of the plates.
It is straightforward to establish that $U=\mu_{1}U^{*}/\sigma_{0}=s(1+n/m)$.
The disturbed state with small deviations (indicated by the tilde,
$^{\sim}$) from the base flow is given by 
\begin{equation}
\eta=\tilde{\eta}\text{,}\ u_{j}=\bar{u}_{j}+{\tilde{u}}_{j}\text{, }w_{j}=\tilde{w}_{j}\text{, }p_{j}=\bar{p}_{j}+\tilde{p}_{j}\text{, }\Gamma=\bar{\Gamma}+\tilde{\Gamma}\text{.}\label{eq:transformations}
\end{equation}
The normal modes are disturbances of the form 
\begin{equation}
(\tilde{\eta}\text{, }\tilde{u}_{j}\text{, }\tilde{w}_{j}\text{, }\tilde{p}_{j}\text{, }\tilde{\Gamma})=[h\text{, }\hat{u}_{j}(z)\text{, }\hat{w}_{j}(z)\text{, }\hat{f}_{j}(z)\text{, }G]e^{i\alpha x+\gamma t}\text{,}\label{eq:ModesNormal}
\end{equation}
where $\hat{u}_{j}(z)$, $\hat{w}_{j}(z)$, and $\hat{f}_{j}(z)$
are the complex amplitudes that depend on the depth, $\alpha$ is
the wavenumber of the disturbance, $G$ is the constant amplitude
of $\tilde{\Gamma}$ ($G=\hat{\Gamma}$), $h$ is the constant amplitude
of $\tilde{\eta}$ ($h=\hat{\eta}$), and (complex) $\gamma$ is the
increment, $\gamma=\gamma_{R}+i\gamma_{I}$. The stability of the
flow depends on the sign of the growth rate $\gamma_{R}$: if $\gamma_{R}>0$
for some normal modes then the system is unstable; and if $\gamma_{R}<0$
for all normal modes then the system is stable. The linearized governing
equations for the disturbances translate into the following system
for the normal mode amplitudes (See \citet{Frenkel2016,Frenkel2017}
for the omitted details). The continuity equation becomes 
\begin{equation}
\hat{u}_{j}=\frac{i}{\alpha}D\hat{w}_{j}.\label{eq:ConEqnDisturb}
\end{equation}
Eliminating the pressure disturbances from the horizontal and vertical
components of the momentum equations with neglected inertia yields
the well-known Orr-Sommerfeld equations, here for the vertical velocity
disturbances, 
\begin{equation}
m_{j}(D^{2}-\alpha^{2})^{2}\hat{w}_{j}=0\text{, }\label{eq:OrrSommerfeldStokesEqn}
\end{equation}
where $m_{j}\coloneqq\mu_{j}/\mu_{1}$ (so that $m_{1}=1$ and $m_{2}=m$).
The disturbances of the velocities are subject to the boundary conditions
at the plates and at the interface. At the plates, the boundary conditions
are 
\begin{equation}
D\hat{w}_{1}(-1)=0\text{, }\hat{w}_{1}(-1)=0\text{, }D\hat{w}_{2}(n)=0\text{, }\hat{w}_{2}(n)=0\text{.}\label{eq:PlateDisturbBCs}
\end{equation}
The kinematic boundary condition and surfactant transport equation
yield, respectively, 
\begin{equation}
\gamma h-\hat{w}_{1}=0\text{ (}z=0\text{),}\label{eq:KineDisturbBC}
\end{equation}
\begin{equation}
\gamma G-D\hat{w}_{1}+si\alpha h=0\text{ (}z=0\text{).}\label{eq:SurfDisturbEqn}
\end{equation}
(Note that equation (\ref{eq:SurfDisturbEqn}) is the normal form
of equation (2.9) in \citet{Frenkel2017} which was derived in HF,
and was mentioned there to be consistent with the more general equation
of \citet{Wong1996}. The last term in (\ref{eq:SurfDisturbEqn})
comes from the Taylor expansion of the base state fluid velocities
at $z=\eta(x,t)$.) Continuity of velocity at the interface yields
\begin{equation}
\hat{w}_{1}-\hat{w}_{2}=0\text{ (}z=0\text{)}\label{eq:continVatIntfDisturb}
\end{equation}
and 
\begin{equation}
\text{ }D\hat{w}_{2}-D\hat{w}_{1}-i\alpha sh\left(\frac{1-m}{m}\right)=0\text{ (}z=0\text{).}\label{eq:continVatIntfDisturb2}
\end{equation}
To obtain the linearized homogeneous normal stress condition, the
pressure amplitude, $\hat{f}_{j}$, is first written in terms of $\hat{w}_{j}$.
From the horizontal momentum equation it is given by 
\begin{equation}
\alpha^{2}\hat{f}_{j}=m_{j}(D^{2}-\alpha^{2})D\hat{w}_{j}\text{.}\label{eq:xMomEqnDisturbPressure}
\end{equation}
The interfacial tangential stress condition is 
\begin{equation}
mD^{2}\hat{w}_{2}-D^{2}\hat{w}_{1}+\alpha^{2}(m\hat{w}_{2}-\hat{w}_{1})-\alpha^{2}G\text{Ma}=0\text{ (}z=0\text{),}\label{eq:TanDisturbBC}
\end{equation}
where 
\[
\text{Ma}:=E\Gamma_{0}/\sigma_{0}
\]
is the Marangoni number, and the normal stress condition is 
\begin{equation}
mD^{3}\hat{w}_{2}-3m\alpha^{2}D\hat{w}_{2}-D^{3}\hat{w}_{1}+\text{Bo}\alpha^{2}h+3\alpha^{2}D\hat{w}_{1}+\alpha^{4}h=0\text{ (}z=0\text{),}\label{eq:NormDisturbStokesBC}
\end{equation}
where $\Bo$ is the effective Bond number 
\begin{equation}
\text{Bo}=\text{Bo}_{1}-\text{Bo}_{2}=\frac{(\rho_{1}-\rho_{2})gd_{1}^{2}}{\sigma_{0}}.\label{eq:effectiveBoNumber}
\end{equation}
Note that $\text{Bo}$ can be negative, unlike the parameters $n$,
$m$, $s$ and $\text{Ma}$. Equations (\ref{eq:OrrSommerfeldStokesEqn})-(\ref{eq:continVatIntfDisturb2}),
(\ref{eq:TanDisturbBC}) and (\ref{eq:NormDisturbStokesBC}) form
the eigenvalue boundary value problem for the disturbances, which
determines the growth rate as a function of the wavenumber $\alpha$
and the parameters $s$, $m$, $n$, $\Ma$, and $\Bo$. The eigenvalue,
the increment $\gamma$, satisfies a quadratic equation which is obtained
in the next section.

\section{Dispersion relation; special points of dispersion curves}

\label{sec:FiniteCaseDerivationDispRelation} For finite aspect ratio,
$n$, the general solutions of (\ref{eq:OrrSommerfeldStokesEqn})
are given by 
\begin{equation}
\hat{w}_{j}(z)=a_{j}\cosh(\alpha z)+b_{j}\sinh(\alpha z)+c_{j}z\cosh(\alpha z)+d_{j}z\sinh(\alpha z)\text{,}\label{eq:whatGSfinite}
\end{equation}
where the coefficients $a_{j}$, $b_{j}$, $c_{j}$, and $d_{j}$
are determined by the boundary conditions up to a common normalization
factor. Equation (\ref{eq:continVatIntfDisturb}) yields $a_{2}=a_{1}$,
which is used to eliminate $a_{2}$ from the equations.

Applying the plate velocity conditions, equation (\ref{eq:PlateDisturbBCs}),
the coefficients $c_{1}$ and $d_{1}$ are expressed in terms of $a_{1}$
and $b_{1}$, and the coefficients $c_{2}$ and $d_{2}$ are expressed
in terms of $a_{1}$ and $b_{2}$: 
\begin{align}
\hat{w}_{1}(z) & =a_{1}\cosh(\alpha z)+b_{1}\sinh(\alpha z)+\frac{1}{\alpha}\left[-s_{\alpha}^{2}b_{1}+\left(s_{\alpha}c_{\alpha}+\alpha\right)a_{1}\right]z\cosh(\alpha z)\nonumber \\
 & +\frac{1}{\alpha}\left[-\left(s_{\alpha}c_{\alpha}-\alpha\right)b_{1}+c_{\alpha}^{2}a_{1}\right]z\sinh(\alpha z)\label{eq:w1hat}
\end{align}
and 
\begin{align}
\hat{w}_{2}(z) & =a_{1}\cosh(\alpha z)+b_{2}\sinh(\alpha z)-\frac{1}{\alpha n^{2}}\left[s_{\alpha n}^{2}b_{2}+\left(s_{\alpha n}c_{\alpha n}+\alpha n\right)a_{1}\right]z\cosh(\alpha z)\nonumber \\
 & +\frac{1}{\alpha n^{2}}\left[\left(s_{\alpha n}c_{\alpha n}-\alpha n\right)b_{2}+c_{\alpha n}^{2}a_{1}\right]z\sinh(\alpha z),\label{eq:w2hat}
\end{align}
where 
\begin{equation}
c_{\alpha}=\cosh(\alpha)\text{, }s_{\alpha}=\sinh(\alpha)\,\text{, }c_{\alpha n}=\cosh(\alpha n)\text{, }s_{\alpha n}=\sinh(\alpha n)\,\text{.}\label{eq:coshsinh}
\end{equation}
We substitute these velocity expressions into the interfacial conditions
(\ref{eq:continVatIntfDisturb2}), (\ref{eq:TanDisturbBC}), and (\ref{eq:NormDisturbStokesBC})
to obtain a linear nonhomogeneous system for $a_{1}$, $b_{1}$, and
$b_{2}$. Solving this system yields $a_{1}$, $b_{1}$, and $b_{2}$
in terms of $h$ and $G$. Hence, we have the velocities $\hat{w}_{j}(z)$
in terms of $h$ and $G$. Then the kinematic boundary condition (\ref{eq:KineDisturbBC})
and surfactant transport equation (\ref{eq:SurfDisturbEqn}) yield
a linear homogeneous system for $h$ and $G$, written in matrix form
as 
\begin{equation}
\begin{bmatrix}(\gamma+A_{11}) & A_{12}\\
A_{21} & (\gamma+A_{22})
\end{bmatrix}\begin{bmatrix}h\\
G
\end{bmatrix}=\begin{bmatrix}0\\
0
\end{bmatrix}\text{,}\label{eq:dispEqnSystem}
\end{equation}
where $A_{11}$, $A_{12}$, $A_{21}$, and $A_{22}$ are known functions
of the wavenumber $\alpha$ and the system parameters (see Appendix
\ref{sec:Coefficients-of-equations}). The condition for the existence
of nontrivial solutions is $\det(A)=(\gamma+A_{11})(\gamma+A_{22})-A_{12}A_{21}=0$;
this yields a quadratic equation for the mode increment $\gamma$.
We write this 'dispersion equation' in the form 
\begin{equation}
F_{2}\gamma^{2}+F_{1}\gamma+F_{0}=0,\label{eq:DispersionEquation}
\end{equation}
and its two solutions in the forms 
\begin{equation}
\gamma=\frac{1}{2F_{2}}\left(-F_{1}+\left[F_{1}^{2}-4F_{2}F_{0}\right]^{1/2}\right)\label{eq:QuadEqnGamma}
\end{equation}
or 
\begin{equation}
\gamma=-\frac{F_{1}}{2F_{2}}+\left[\left(\frac{F_{1}}{2F_{2}}\right)^{2}-\frac{F_{0}}{F_{2}}\right]^{1/2},\label{eq:QuadEqnGammaa}
\end{equation}
where $F_{2}$, $F_{1}$, and $F_{0}$ are as follows: 
\begin{align}
\operatorname{Re}(F_{2}) & =\frac{1}{\alpha^{4}}\left\{ \left(c_{\alpha n}^{2}+\alpha^{2}n^{2}\right)\left(s_{\alpha}^{2}-\alpha^{2}\right)m^{2}+2\left(s_{\alpha}c_{\alpha}s_{\alpha n}c_{\alpha n}-\alpha^{2}n+\alpha^{4}n^{2}\right)m\right.\nonumber \\
 & +\left.\left(s_{\alpha n}^{2}-\alpha^{2}n^{2}\right)\left(c_{\alpha}^{2}+\alpha^{2}\right)\right\} \text{,}\label{eq:F2Re}\\
\operatorname{Im}(F_{2}) & =0\text{,}\label{F2Im}\\
\operatorname{Re}(F_{1}) & =\frac{1}{2\alpha^{3}}\left\{ m\text{Ma}(s_{\alpha n}c_{\alpha n}+\alpha n)\left(s_{\alpha}^{2}-\alpha^{2}\right)+\text{Ma}(s_{\alpha n}^{2}-\alpha^{2}n^{2})\left(s_{\alpha}c_{\alpha}+\alpha\right)\right.\nonumber \\
 & +\frac{1}{\alpha^{2}}m(s_{\alpha n}c_{\alpha n}-\alpha n)\left(s_{\alpha}^{2}-\alpha^{2}\right)\left(\text{Bo}+\alpha^{2}\right)\nonumber \\
 & +\left.\frac{1}{\alpha^{2}}(s_{\alpha n}^{2}-\alpha^{2}n^{2})\left(s_{\alpha}c_{\alpha}-\alpha\right)\left(\text{Bo}+\alpha^{2}\right)\right\} \text{,}\label{eq:F1Re}\\
\operatorname{Im}(F_{1}) & =\frac{s}{\alpha^{2}}(1-m)(s_{\alpha n}c_{\alpha n}-\alpha n+n^{2}s_{\alpha}c_{\alpha}-\alpha n^{2})\text{,}\label{eq:F1Im}\\
\operatorname{Re}(F_{0}) & =\frac{\text{Ma}}{4\alpha^{4}}(s_{\alpha n}^{2}-\alpha^{2}n^{2})(s_{\alpha}^{2}-\alpha^{2})\left(\text{Bo}+\alpha^{2}\right)\text{,}\label{eq:F0Re}\\
\operatorname{Im}(F_{0}) & =-\frac{\text{Ma}}{2\alpha}s(s_{\alpha n}^{2}-s_{\alpha}^{2}n^{2})\text{.}\label{eq:F0Im}
\end{align}
Because the coefficients of the quadratic equation (\ref{eq:DispersionEquation})
are complex numbers, it is clear that in general the imaginary parts
of the solutions $\gamma_{1}$ and $\gamma_{2}$ are non zero which
signifies an oscillatory instability. One can see that the growth
rate $\gamma_{R}$ (as well as the increment $\gamma$ ) has the function
symmetry property 
\begin{equation}
\gamma_{R}(-n\alpha;\;ns,\;m^{-1},\;n^{-1},\;\text{Ma},\;n^{2}\text{Bo})=nm\gamma_{R}(\alpha;\;s,\;m,\;n,\;\text{Ma},\;\text{Bo}).\label{eq:gammasymmetrycondition}
\end{equation}
In view of this symmetry, it is sufficient to consider stability for
$n\ge1$. (See Frenkel and Halpern (2016) for comprehensive details.)
We also note the following facts. All the coefficients of the quadratic
equation (\ref{eq:DispersionEquation}) are continuous at each point
$(\alpha;s,m,n,\text{Ma},\text{Bo})$ for the physical values of $\alpha$
and the parameters. All parenthetical expressions in equations (\ref{eq:F2Re})
through (\ref{eq:F0Im}) containing hyperbolic functions are positive.
Therefore, $F_{2}>0$, and $\textrm{Re}(F_{1})$ and $\textrm{Re}(F_{0})$
are positive for $\text{Bo}\ge0$. For $\textrm{Bo}<0$, the functions
$\textrm{Re}(F_{1})$ and $\textrm{Re}(F_{0})$ are positive provided
$\alpha^{2}>-\Bo$. Also, $\textrm{Im}(F_{1})>0(<0)$ for $m<1(>1)$.
Furthermore, $\operatorname{Im}(F_{0})=0$ for $n=1$, and negative
for $n>1$. The zero gravity limit studied in FH and HF is recovered
when $\Bo=0$. We want to investigate the dependence of the growth
rates $\gamma_{R}=\operatorname{Re}(\gamma)$ on the wavenumber $\alpha$
and the parameters $n$, $m$, $s$, $\Ma$ and $\Bo$ in the ranges
$0<\alpha<\infty$, $1\leq n<\infty$, $0<m<\infty$, $0\le s<\infty$,
$0\le\text{Ma}<\infty$ and $-\infty<\text{Bo}<\infty$.

It is an elementary fact of complex analysis that there are two analytic,
and therefore continuous, branches of the complex square root function
in every simply connected domain not containing the origin (see e.g.
\citet{Bak2010} pages 114-115). Then, as the discriminant 
\begin{equation}
\zeta=F_{1}^{2}-4F_{0}F_{2}\label{eq:DiscZ}
\end{equation}
is clearly a smooth function of $\alpha$ and the parameters, there
are two continuous branches of the increment $\gamma$ (\ref{eq:QuadEqnGamma})
as functions of $\alpha$ and the parameters, and correspondingly
two continuous branches of the growth rate $\gamma_{R}$. If $\Ma\downarrow0$
then $\gamma_{1}\gamma_{2}=F_{0}/F_{2}\downarrow0$ and $\gamma_{1}+\gamma_{2}=-F_{2}/F_{1}\not\rightarrow0$
and so either $\gamma_{1}\downarrow0$ or $\gamma_{2}\downarrow0$.
We call the increment branch that is non-zero at $\Ma=0$ the ``robust
branch,'' and the other one, that vanishes as $\Ma\downarrow0$,
is named the ``surfactant branch''. Correspondingly, these are the
continuous robust and surfactant branches of the growth rate. In certain
cases, such as the one considered in section 4.3.1 with $m=1$, it
can be shown that the discriminant $\zeta$ never takes the zero value
and the range of the function $\zeta$$(\alpha;s,m,n,\Ma,\Bo)$ is
a simply connected domain in the complex $\zeta$-plane. Then, there
are two branches of the growth rate which are continuous functions
of $(\alpha;s,m,n,\Ma,\Bo)$.

However, as will be seen below, the discriminant (\ref{eq:DiscZ})
may become zero for some parameter values. This happens when $\textrm{Re}(\zeta)=0$
and $\textrm{Im(}\zeta)=0$. These two equations define a manifold
of co-dimension two in the $(\alpha;s,m,n,\Ma,\Bo)$ space that is
analogous to a branch point in the complex plane; and if we draw the
line of increasing $\alpha$ from each point of this manifold, that
is a ray parallel to the $\alpha$-axis, with all the parameter values
fixed, we obtain the ``branch cut'' hypersurface. The growth rates
are not defined on this branch cut, and there is a jump in the growth
rate when crossing from one side of the branch cut to the other. Still,
each of the two growth-rate branches is defined and continuous almost
everywhere in the $\alpha$-parameter space (with the branch cut hypersurface
excluded from it), and the growth-rate branches defined this way are
smooth in $\alpha$. The surfactant branch of the growth rate is again
defined as the one which vanishes as $\text{Ma}\downarrow0$. These
considerations are given in more detail in appendix \ref{sec:On-the-Continuous-Branches}.
It will be seen below, as for example in figure 9, that the discriminant
equal to zero corresponds to the reconnection point of the two growth
rate branches, when the crossing dispersion curves of the two branches
become non-crossing at a certain value of a changing parameter. There
is a jump discontinuity of the growth rate in the changing parameter
at its reconnection-point value, for all $\alpha$ exceeding the reconnection-point
value of $\alpha$. Except for such reconnection situations, all the
dispersion curves are smooth at all $\alpha$.

\begin{figure}
\centering{}\includegraphics[clip,width=0.6\textwidth]{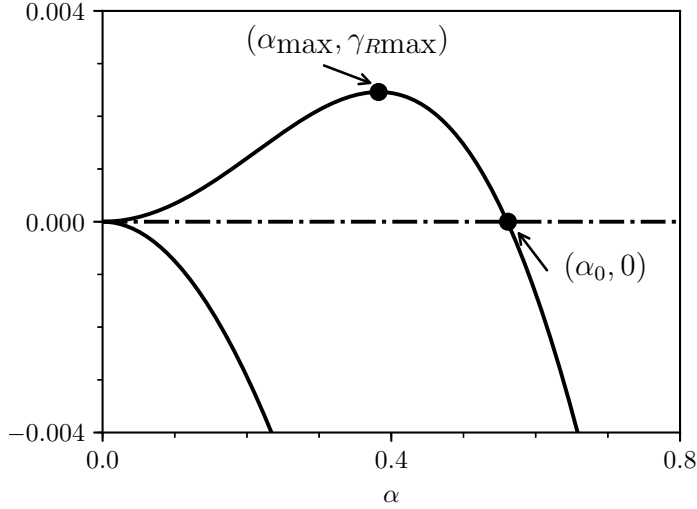}
\protect\caption{Typical dispersion curves of the two normal modes: (1) the unstable
mode, which has a maximum growth rate $\gamma_{R}=\gamma_{R\max}$
at a wavenumber $\alpha=\alpha_{\max}$ and then decays, eventually
becoming stable for $\alpha>\alpha_{0}$, and (2) the stable mode,
which has negative growth rates for all wavenumbers.}
\label{fig:FigTypicalDispCurve} 
\end{figure}

Typical dispersion curves of stable and unstable cases look like those
in figure \ref{fig:FigTypicalDispCurve}. The unstable branch starts
at $\alpha=0$ and $\gamma_{R}=0$, grows with $\alpha$, attains
a maximum value $\gamma_{R\max}$ at some $\alpha=\alpha_{\max}$,
then decreases and crosses the $\alpha$-axis so that $\gamma_{R}=0$
at some non-zero wavenumber, $\alpha_{0}$, called the marginal wavenumber.
The other, stable, branch also starts at $\alpha=0$ and $\gamma_{R}=0$
but then decreases with $\alpha$. The values of $\alpha_{0}$, $\gamma_{R\max}$,
and $\alpha_{\max}$ depend on the parameters $n$, $m$, $s$, $\Ma$,
and $\Bo$.

Each solution ($\gamma$;$h,G$) of the system (\ref{eq:dispEqnSystem})
determines the normal-mode amplitudes (and thus the complete structure
of the normal mode), since $h$ and $G$ determine the coefficients
$a_{1}$, $b_{1}$, and $b_{2}$, and thus the vertical velocities
$\hat{w}_{j}$ via equations (\ref{eq:w1hat}) and (\ref{eq:w2hat}),
then the horizontal velocities $\hat{u}_{j}$ via equations (\ref{eq:ConEqnDisturb})
and the pressures $\hat{f}_{j}$ via equations (\ref{eq:xMomEqnDisturbPressure}).

It is pointed out in FH (i.e., for the case $\Bo=0$) that at least
one of the modes for each given $\alpha$ is stable. This result holds
for $\Bo\geq0$ as well, which is seen as follows. (However, we will
see that for $\Bo<0$ both modes are unstable sometimes.) Let the
two solutions of (\ref{eq:QuadEqnGamma}) be $\gamma_{1}=\gamma_{1R}+i\gamma_{1I}$
and $\gamma_{2}=\gamma_{2R}+i\gamma_{2I}$. Then the real parts of
the solutions satisfy $\gamma_{1R}+\gamma_{2R}=-\operatorname{Re}(F_{1})/F_{2}<0$.
The latter inequality holds because, as was discussed before, $\textrm{Re}(F_{1})>0$
when $\Bo\ge0$. So, if one of the quantities $\gamma_{jR}$ is positive
(corresponding to an unstable mode), then the other must be negative,
thus giving a stable mode.

In order to compute the maximum growth rate, $\gamma_{R\max}$, the
wavenumber corresponding to the maximum growth rate, $\alpha_{\max}$,
and the marginal wavenumber, $\alpha_{0}$, it is convenient to split
the dispersion equation (\ref{eq:DispersionEquation}) into its real
and imaginary parts, 
\begin{equation}
F_{2}\gamma_{R}^{2}-F_{2}\gamma_{I}^{2}+\operatorname{Re}(F_{1})\gamma_{R}-\operatorname{Im}(F_{1})\gamma_{I}+\operatorname{Re}(F_{0})=0\text{, }\label{eq:ReDispEqn}
\end{equation}
\begin{equation}
2F_{2}\gamma_{R}\gamma_{I}+\operatorname{Re}(F_{1})\gamma_{I}+\operatorname{Im}(F_{1})\gamma_{R}+\operatorname{Im}(F_{0})=0\text{.}\label{eq:ImDispEqn}
\end{equation}
The imaginary part of the growth rate $\gamma_{I}$ is expressed in
terms of $\gamma_{R}$ using equation (\ref{eq:ImDispEqn}) (assuming
$\Real(F_{1})\ne0$) and then substituted it into (\ref{eq:ReDispEqn})
to obtain the following quartic equation for $\gamma_{R}$, 
\begin{gather}
4\,F_{2}{}^{3}\gamma_{R}^{4}+8F_{2}^{2}\operatorname{Re}(F_{1})\gamma_{R}^{3}+F_{2}\left[4\,F_{2}\operatorname{Re}(F_{0})+\operatorname{Im}(F_{1})^{2}+5\,\operatorname{Re}(F_{1})^{2}\right]\gamma_{R}^{2}\nonumber \\
+\operatorname{Re}(F_{1})\left[\operatorname{Re}(F_{1})^{2}+4F_{2}\operatorname{Re}(F_{0})+\operatorname{Im}(F_{1})^{2}\right]\gamma_{R}-F_{2}\operatorname{Im}(F_{0})^{2}\nonumber \\
+\operatorname{Re}(F_{1})^{2}\operatorname{Re}(F_{0\,})+\operatorname{Re}(F_{1})\operatorname{Im}(F_{1})\operatorname{Im}(F_{0})=0\text{.}\label{eq:MAXeqn}
\end{gather}
Since $\gamma_{R}=0$ at the marginal wavenumber, $\alpha_{0}$, equation
(\ref{eq:MAXeqn}) becomes 
\begin{equation}
-F_{2}\operatorname{Im}(F_{0})^{2}+\operatorname{Re}(F_{1})\operatorname{Im}(F_{1})\operatorname{Im}(F_{0})+\operatorname{Re}(F_{1})^{2}\operatorname{Re}(F_{0})=0\text{, }\label{eq:MaEquation}
\end{equation}
the marginal wavenumber equation. This equation (\ref{eq:MaEquation})
is a polynomial in $\Ma$ and $\Bo$ 
\begin{equation}
k_{20}\text{Ma}^{2}+k_{11}\text{Ma}B+k_{31}\text{Ma}^{3}B+k_{22}\text{Ma}^{2}B^{2}+k_{13}\text{Ma}B^{3}=0\text{ }\label{eq:MaEquation2}
\end{equation}
where $B:=\text{Bo}+\alpha^{2}$ and the coefficients $k_{ij}$ are
given in appendix \ref{sec:Coefficients-of-equations}. For $\text{Ma}=0$,
it transpires that these marginal wavenumber equations are not valid.
However, then the coefficient $F_{0}$ of the quadratic equation (\ref{eq:DispersionEquation})
vanishes, and there remains just one mode corresponding to the Rayleigh-Taylor
instability whose increment $\gamma=-F_{1}/F_{2}$. For the marginal
wavenumber, it follows that $\Real(F_{1})=0$, which implies that
$\alpha_{0}=(-\text{Bo})^{1/2}$. This corresponds to capillary forces
balancing the destabilizing gravitational forces provided $\text{Bo}<0$.

The wavenumber $\alpha_{\max}$ corresponding to the maximum growth
rate $\gamma_{R\max}$ is obtained by simultaneously solving (\ref{eq:MAXeqn})
and the equation obtained by differentiating (\ref{eq:MAXeqn}) with
respect to $\alpha$, taking into account that $d\gamma_{R}/d\alpha=0$
at the maximum. The latter equation is written as 
\begin{equation}
\gamma_{R}^{4}\frac{d}{d\alpha}C_{4}(\alpha)+\gamma_{R}^{3}\frac{d}{d\alpha}C_{3}(\alpha)+\gamma_{R}^{2}\frac{d}{d\alpha}C_{2}(\alpha)+\gamma_{R}\frac{d}{d\alpha}C_{1}(\alpha)+\frac{d}{d\alpha}C_{0}(\alpha)=0,\text{ }\label{eq:alphaeq:MAXeqn}
\end{equation}
where $C_{j}$ denotes the coefficient of the $\gamma_{R}^{j}$ term
that appears in equation (\ref{eq:MAXeqn}). (For example, $C_{4}=4F_{2}^{3}$.)

\section{Long-wave approximation}

\label{sec:FiniteCaseLongwaveApprox} 
\begin{figure}
\centering{}\includegraphics[clip,width=0.6\textwidth]{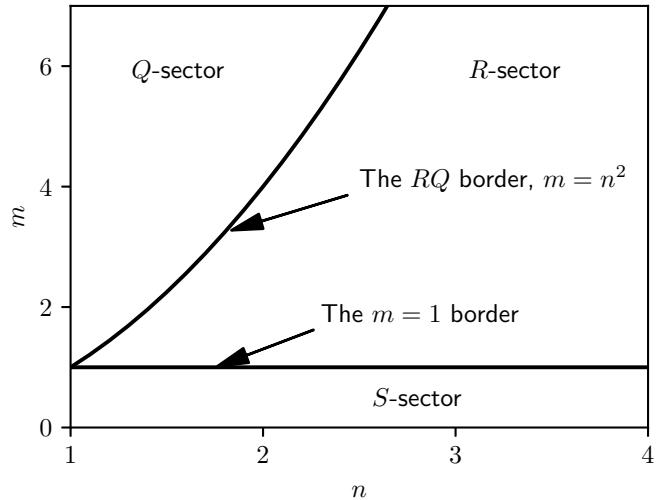}
\protect\caption{Partition of the $(n,m)$-plane of the system, $n\ge1$ and $m>0$,
into three sectors ($Q$, $R$, and $S$) and their borders corresponding
to differences in stability properties of the flow. \label{fig:FigRegions}}
\end{figure}

As was mentioned earlier, from
the long-wave approximation by FH ($\Bo=0$), three sectors in the
$(n,m)$-plane were identified that characterize the stability of
the flow for $n\ge1$. Based on the long-wave results of FH17, the
same three sectors are found to be relevant in the presence of gravity
effects: the $Q$ sector ($m>n^{2}$), the $R$ sector ($1<m<n^{2}$),
and the $S$ sector ($0<m<1$). Figure \ref{fig:FigRegions} shows
the three sectors and their borders. Stability properties of the robust
and surfactant branches can change significantly from sector to sector,
and can be special on borders as well. 

\subsection{General asymptotics for the three sectors}

\label{sec:FiniteCaseLongwaveApproxGeneralGammaExpressions}

\subsubsection{Increments and growth rates}

While it is straightforward to use equation (\ref{eq:QuadEqnGamma})
to evaluate and graph growth rates, the limit of long waves yields
some simpler asymptotic expressions. The general growth rate (and
the increment) expressions in the three sectors are given in this
subsection, but additional results in each sector will be discussed
in later sections. First, the coefficients $F_{2}$, $F_{1}$, and
$F_{0}$ (\ref{eq:F2Re})-(\ref{eq:F0Im}) in the dispersion equation
(\ref{eq:DispersionEquation}) are expanded in a Taylor series about
$\alpha=0$. The leading order terms are given in Appendix \ref{sec:Longwave-formulas-for-F}.
Unless $s=0$ and $\text{Bo}\ne0$, we have $\left\vert F_{1}^{2}\right\vert \gg\left\vert F_{2}F_{0}\right\vert $,
provided $\alpha\ll s$, since if $s\ne0$, then $\left\vert F_{1}^{2}\right\vert \approx\operatorname{Re}\left(F_{1}^{2}\right)\sim\alpha^{2}$
and $\left\vert F_{2}F_{0}\right\vert \approx\operatorname{Im}(F_{2}F_{0})\sim\alpha^{3}$;
and if $s=0$ and $\text{Bo}=0$ then $|F_{1}^{2}|\sim\alpha^{4}$
and $|F_{2}F_{0}|\sim\alpha^{6}$ (see Appendix \ref{sec:Longwave-formulas-for-F}).
Therefore, keeping the four leading members in the series for the
second term of equation (\ref{eq:QuadEqnGamma}), the two increments
are 
\begin{equation}
\gamma\approx\frac{1}{2F_{2}}\left(-F_{1}\pm F_{1}\left[1+\frac{1}{2}\left(-\frac{4F_{2}F_{0}}{F_{1}^{2}}\right)-\frac{1}{8}\left(-\frac{4F_{2}F_{0}}{F_{1}^{2}}\right)^{2}+\frac{1}{16}\left(-\frac{4F_{2}F_{0}}{F_{1}^{2}}\right)^{3}\right]\right)\text{,}\label{eq:gamGeneralEqnRSQsectors}
\end{equation}
or, keeping the terms necessary to obtain the growth rate $\gamma_{R}$
to the leading order, 
\begin{equation}
\gamma\approx-\frac{F_{1}}{F_{2}}+\frac{F_{0}}{F_{1}}+\frac{F_{0}^{2}F_{2}}{F_{1}^{3}}\label{eq:gamCeqn}
\end{equation}
and 
\begin{equation}
\gamma\approx-\frac{F_{0}}{F_{1}}-\frac{F_{2}F_{0}^{2}}{F_{1}^{3}}-2\frac{F_{0}^{3}F_{2}^{2}}{F_{1}^{5}}\text{.}\label{eq:gamSeqn}
\end{equation}
For $s\ne0$ the growth rates for the robust (\ref{eq:gamCeqn}) and
surfactant (\ref{eq:gamSeqn}) branches are found to be, respectively,
\begin{equation}
\gamma_{R}\approx\left({\frac{\,\varphi\left(m-{n}^{2}\right)}{4\left(1-m\right)\psi}}\text{Ma}-\frac{n^{3}(n+m)}{3\psi}\text{Bo}\right)\alpha^{2}\text{}\label{eq:gamCsmallAlphaApprox}
\end{equation}
and 
\begin{equation}
\gamma_{R}\approx\frac{(n-1)\text{Ma}}{4(1-m)}\alpha^{2}+k_{S}\alpha^{4},\label{eq:gamSsmallAlphaApprox}
\end{equation}
where 
\begin{equation}
\varphi={n}^{3}+3\,{n}^{2}+3\,mn+m\text{}\label{eq:phi}
\end{equation}
and 
\begin{equation}
\psi={n}^{4}+4\,m{n}^{3}+6\,m{n}^{2}+4\,mn+{m}^{2}\text{.}\label{eq:psi}
\end{equation}
We include the term with $k_{S}$ in equation (\ref{eq:gamSsmallAlphaApprox})
because the coefficient of the $\alpha^{2}$ term vanishes when $n=1$.
The expression for $k_{S}$ is given in appendix B, see equation (\ref{eq:ks}).
For the case $s=0$ and $\text{Bo}=0$, the growth rates for the robust
(\ref{eq:gamCeqn}) and surfactant (\ref{eq:gamSeqn}) branches are
found to be 
\[
\gamma_{R}\approx-\frac{n^{3}}{12(m+n^{3})}\alpha^{4}
\]
and 
\[
\gamma_{R}\approx-\frac{n(m+n^{3})\text{Ma}}{\psi}\alpha^{2}
\]
which is in agreement with FH.

Finally, for the case $s=0$ and $\text{Bo}\ne0$, we find that $|F_{1}^{2}|\sim\alpha^{4}\sim|F_{2}F_{0}|$.
So, the expansion (\ref{eq:gamGeneralEqnRSQsectors}) is no longer
valid. However, both modes are stable if $\text{Bo}>0$, but there
is instability if $\text{Bo}<0$. Indeed, if $\text{Bo}<0$ then $F_{0}\approx\frac{1}{36}n^{4}\alpha^{4}\text{Ma}\text{Bo}<0$
(see equation (\ref{eq:F0reApprox})). Therefore, the discriminant
$F_{1}^{2}-4F_{0}F_{2}>F_{1}^{2}$. Then equation (\ref{eq:QuadEqnGamma})
yields one of the two growth rates to be positive, so we have instability.
On the other hand, if $\text{Bo}>0$, then $\Real(F_{1})>0$ but the
discriminant can be either positive or negative. If it is negative,
then the square roots in equation (\ref{eq:QuadEqnGamma}) are purely
imaginary and therefore both values of $\gamma_{R}$ are negative.
If the discriminant is positive, then $|\sqrt{F_{1}^{2}-4F_{0}F_{2}}|<F_{1}$,
so that both values of $\gamma$ given by equation (\ref{eq:QuadEqnGamma})
are negative again. These leading-order results were obtained in a
different way and discussed in more detail in \citet{Frenkel2016}
and FH17.

\subsubsection{Marginal wavenumbers and their small $s$ asymptotics\label{subsec:Marginal-wavenumbers}}

When the marginal wavenumber determined by equation (\ref{eq:MaEquation})
happens to be small (typically, due to the smallness of some of the
three parameters s, Bo, and Ma), it is approximated by substituting
the long-wave expressions for the coefficients (\ref{eq:F2reApprox})-(\ref{eq:F0imApprox})
into (\ref{eq:MaEquation}) provided $\text{Ma}\ne0$. If $s\ne0$
is fixed, then by keeping only the two leading terms in $\alpha^{2}$,
we arrive at 
\begin{gather}
\zeta_{0}+\zeta_{2}\alpha^{2}=0\label{eq:MAappEqn}
\end{gather}
where $\zeta_{0}$ and $\zeta_{2}$ are polynomials in $\Ma$ and
$\Bo$ given by equations (\ref{eq:zeta0}) and (\ref{eq:zeta2}).
Therefore, at leading order, 
\[
\alpha_{0}=\sqrt{-(\zeta_{0}/\zeta_{2})}.
\]
Clearly, for this result to be consistent, $\zeta_{0}/\zeta_{2}$
must be negative and small, which is the case for appropriate parameter
values, such as, for example, those used in figures 4, 5, and 6. It
is interesting to investigate the transition from instability to stability
of the case $s=0$ by considering the limit $s\downarrow0$. In this
we should distinguish two cases: $\text{Bo}=0$ and $\text{Bo}\ne0$.
For $\text{Bo}\ne0$, the marginal wavenumber is given by 
\begin{equation}
\widetilde{\zeta_{0}}s^{2}+\zeta_{20}\alpha^{2}=0\label{eq:alpha0smalls1}
\end{equation}
instead of equation (\ref{eq:MAappEqn}), where, by definition the
coefficients $\widetilde{\zeta_{0}}=\zeta_{0}/s^{2}$ and $\zeta_{20}=\zeta_{2}(s=0)$
(see equations (\ref{eq:zeta0}) and (\ref{eq:zeta2})). These coefficients
are independent of $s$ and $\alpha$, and so, asymptotically $\alpha_{0}$
is proportional to $s$, with the coefficient of proportionality $\sqrt{-\widetilde{\zeta_{0}}/\zeta_{20}}$.

However, for $\text{Bo}=0$, the coefficient of the $\alpha^{2}$
term in equation (\ref{eq:alpha0smalls1}) vanishes, and, instead
the leading order equation for the marginal wavenumber is found to
be 
\[
\widetilde{\zeta_{0}}s^{2}+\zeta_{40}\alpha^{4}=0,
\]
where $\zeta_{40}=\frac{1}{324}n^{2}(m+n^{3})^{2}\text{Ma}^{2}$.
Then the marginal wavenumber is asymptotically $\alpha_{0}=(-\widetilde{\zeta_{0}}/\zeta_{40})^{1/4}s^{1/2}$.
\begin{figure}
\includegraphics[bb=0bp 0bp 576bp 383.76bp,clip,width=0.9\textwidth]{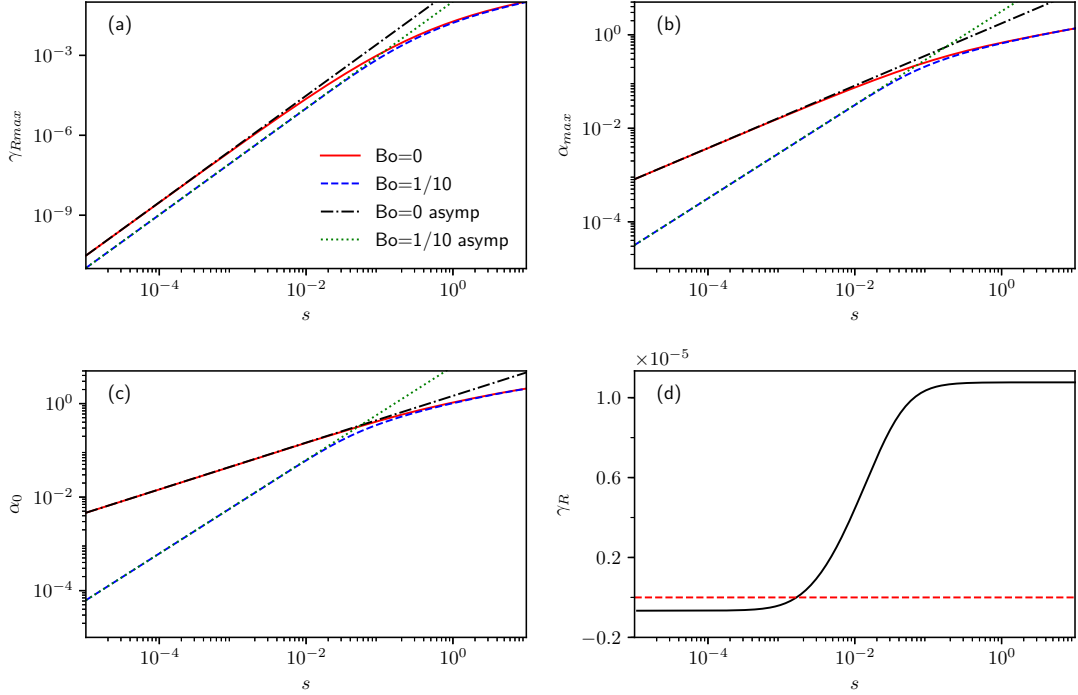}
\caption{(a) $\gamma_{R\text{max}}$, (b) $\alpha_{\text{max}}$, (c) $\alpha_{0}$,
for $\textrm{Bo}=0$ and $\textrm{Bo}=0.1$, along with their small-$s$
asymptotics, and (d) $\gamma_{R}$ at $\alpha=0.01$ and $\textrm{Bo}=0.1$,
vs $s$ for $n=2$ and $m=2$ (which is in the $R$ sector), with
$\text{Ma}=1$. \label{fig:fig4abcd}}
\end{figure}
Panel (c) of figure \ref{fig:fig4abcd} shows these asymptotes along
with the marginal wavenumbers obtained by solving equation (\ref{eq:MaEquation2})
for $\text{Bo=0}$ and some positive values of $\text{Bo}$ in the
$R$ sector. Panel (d) shows, for a fixed wavenumber, $\alpha=0.01$,
how the instability at the larger $s$ corresponding to the (positive)
growth rate (\ref{eq:gamCsmallAlphaApprox}), changes to stability
with the growth rate corresponding, in the leading order, to the case
of $s=0$ and nonzero $\textrm{Bo}$. The growth rate that crosses
the zero value at the $s$ for which $\alpha=0.01$ is the marginal
wavenumber.

\begin{figure}
\includegraphics[bb=0bp 0bp 576bp 383.76bp,clip,width=0.9\textwidth]{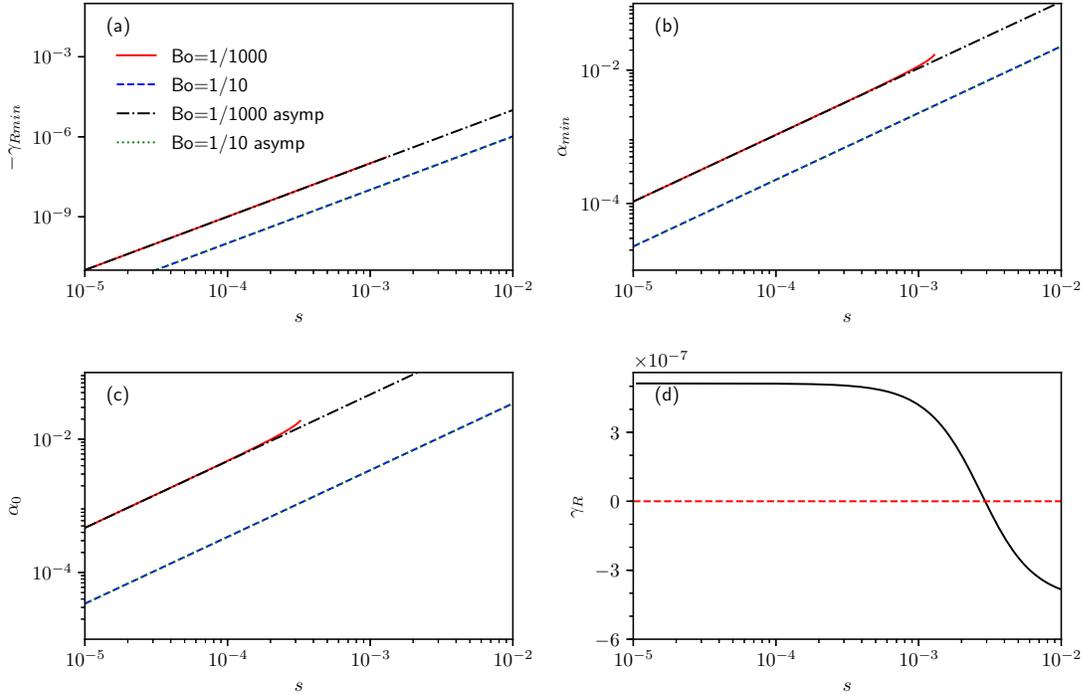}
\caption{(a) $\gamma_{R\text{min}}$, (b) $\alpha_{\text{min}}$, (c) $\alpha_{0}$,
for $\textrm{Bo}=-0.001$ and $\textrm{Bo}=-0.1$, along with their
small-$s$ asymptotics, and (d) $\gamma_{R}$ at $\alpha=0.01$ and
$\textrm{Bo}=-0.1$, vs $s$ for $n=2$ and $m=5$ (which is in the
$Q$ sector), with $\text{Ma}=1$.\label{fig:figgrminaminvss}}
\end{figure}

\begin{figure}
\centering \includegraphics[bb=0bp 0bp 288bp 216bp,clip,width=0.65\textwidth]{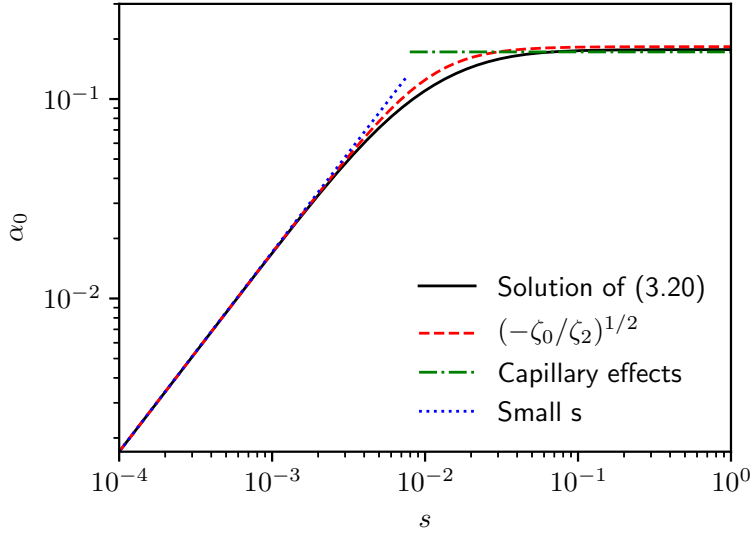}
\caption{Marginal wavenumber $\alpha_{0}$ vs the shear parameter $s$, along
with its asymptotics, at larger $s$, due to the capillary effects,
and at smaller $s$, due to the combined gravity-surfactant effects,
for $n=2$, $m=2$, $\text{Ma}=0.05$, and $\text{Bo}=-0.05$.\label{fig:figavssmalls}}
\end{figure}

In the analogous figure for the $Q$ sector, figure \ref{fig:figgrminaminvss},
the marginal wavenumber is the left endpoint of the interval of the
unstable wavenumbers, which is bounded away from the zero of the wavenumber
axis. There is a band of stable wavenumbers between this marginal
wavenumber and the zero, and inside it there is a minimum of the growth
rate, $\gamma_{Rmin}$, at the corresponding wavenumber $\alpha_{min}$;
their dependencies on $s$ are plotted in panels (a) and (b), respectively.
Correspondingly, panel (d) shows stability at the larger $s$, and
instability at the smaller $s$, since here, in the $Q$ sector, it
is the band of \emph{stable} wavenumbers that shrinks toward zero
as $s\downarrow0$. We call such cases, in which there is an interval
of unstable wavenumbers bounded away from zero, the mid-wave instability,
to distinguish them from the long-wave instability, in which the interval
of unstable wavenumbers is bordered by zero. We study the mid-wave
instability in detail below (see sections 5 and 6). 
By considering the formula for $\zeta_{2}$ (\ref{eq:zeta2}) for
sufficiently small $\textrm{Ma}$ and $\textrm{Bo}$, we see that
all the terms are negligible as compared to the last one (the capillary
term), and equation (\ref{eq:alpha0smalls1}), after being multiplied
by an appropriate factor, is interpreted as the instability term (\ref{eq:gamCsmallAlphaApprox})
being balanced by the capillary effect (corresponding to the term
$\alpha^{2}$ in $B=\textrm{Bo}+\alpha^{2}$, and arising from the
second term of equation (\ref{eq:alpha0smalls1}).) The resulting,
asymptotically $s$-independent, value of the marginal wavenumber,
as one can see at the larger $s$ in figure \ref{fig:figavssmalls},
is still small, consistent with the long-wave approximation. 
However, for the same fixed small values of $\textrm{Ma}$ and $\textrm{Bo}$,
at sufficiently small $s$, the last, capillary, term in $\zeta_{20}$
is negligible, and the stabilization near the marginal wavenumber
is due to non-capillary effects of the combined action of surfactants
and gravity. It is clear that the three corresponding terms in $\zeta_{20}$
are not zero only if both the Marangoni and Bond numbers are non-zero.
These (non-additively) combined surfactant-gravity effects are beyond
the lubrication approximation, and can be captured only by the post-lubrication
correction theory considered in \citet{Frenkel2016}. Figure \ref{fig:figavssmalls}
shows the numerical solution of the marginal-wavenumber equation (\ref{eq:MaEquation})
without using the long-wave asymptotics, along with the larger-$s$
(capillary) and small-$s$ (gravity- and surfactant-determined, non-lubrication)
approximations of the wavenumber given by the long-wave asymptotic
equation (\ref{eq:alpha0smalls1}). Excellent agreement is evident.

\subsubsection{Maximum growth rates}

As indicated earlier, a way to find $\gamma_{R\text{max }}$ and $\alpha_{\text{max}}$
is to solve equations (\ref{eq:MAXeqn}) and (\ref{eq:alphaeq:MAXeqn}).
For $s\downarrow0$, numerical computations suggest that $\alpha_{\text{max}}\propto s$
if $\text{Bo}\ne0$ (just like $\alpha_{0}$) and $\alpha_{\text{max}}\propto s^{2/3}$
if $\text{Bo}=0$, and that $\gamma_{R\text{max}}\propto s^{2}$ for
both $\text{Bo}\ne0$ and $\text{Bo}=0$, as one can see in figure
\ref{fig:fig4abcd}. We find the coefficients of these asymptotic
dependencies as follows:

For the case $\text{Bo}=0$, we write $\gamma_{R\text{max }}$ and
$s^{2}$ as functions of $\alpha_{\text{max}}$ to the two leading
orders, 
\begin{equation}
s^{2}\approx\phi_{1}\alpha^{3}+\phi_{2}\alpha^{4},\quad\gamma_{R\text{max}}\approx\psi_{1}\alpha_{\text{}}^{3}+\psi_{2}\alpha^{4},\label{eq:gammarsmalls}
\end{equation}
with indeterminate coefficients $\phi_{1}$, $\phi_{2}$, $\psi_{1}$
and $\psi_{2}$. We have to use two leading orders because the leading
order system for $\phi_{1}$ and $\psi_{1}$ turns out to be degenerate,
and only gives one relation between $\phi_{1}$ and $\psi_{1}$. The
other relation between $\phi_{1}$ and $\psi_{1}$ is found as the
solvability condition for the next order non-homogeneous system for
$\phi_{2}$ and $\psi_{2}$. The leading order of equation (\ref{eq:MAXeqn})
consists of terms that are proportional to $\alpha^{9}$. Therefore,
the terms which are nonlinear in $\gamma_{R}$ are discarded. This
yields 
\begin{equation}
(d\alpha^{6})\psi_{1}\alpha^{3}+(f\alpha^{6})\phi_{1}\alpha^{3}=0,\label{eq:grsmallsbo0eq}
\end{equation}
where $d=<Re(F_{1})>^{3}$ and $f=<Re(F_{1})><\Imag(F_{1})><\Imag(F_{0})>-F_{2}<\Imag(F_{0})>^{2}$.
Here the bracketed quantities are the coefficients of powers of $s$
and $\alpha$ in the leading order terms of the corresponding ``unbracketed''
coefficients (A24)-(A28): $<Re(F_{1})>=\frac{1}{3}n(m+n^{3})\text{Ma}$,
$<\Imag(F_{1})>=\frac{2}{3}n^{2}(n+1)(1-m)$, $<Re(F_{0})>=\frac{1}{36}n^{4}\text{Ma}$,
and $<\Imag(F_{0})>=\frac{1}{6}n^{2}(1-n^{2})\text{Ma}$. When obtaining
equation (\ref{eq:alphaeq:MAXeqn}) by differentiating with respect
to $\alpha$ at constant $\gamma_{R}$ and $s^{2}$, only the powers
of $\alpha$ inside the parentheses of equation (\ref{eq:gammarsmalls})
are differentiated, and this yields 
\[
(6d\alpha^{5})\psi_{1}\alpha^{3}+(6f\alpha^{5})\phi_{1}\alpha^{3}=0.
\]
So, the matrix of the coefficients of the linear homogeneous system
for $\phi_{1}$ and $\psi_{1}$ 
\[
M=\left[\begin{array}{cc}
f & d\\
6f & 6d
\end{array}\right]
\]
is singular, and the leading-order system yields the single relation
\begin{equation}
\phi_{1}=-\frac{d}{f}\psi_{1}.\label{eq:a1smalls}
\end{equation}
Therefore, we need to consider the next order of equation (\ref{eq:MAXeqn}),
proportional to $\alpha^{10}$. We obtain 
\begin{eqnarray}
(d\alpha^{6})\psi_{2}\alpha^{4}+(f\alpha^{6})\phi_{2}\alpha^{4} & = & -5(F_{2}<\Real(F_{1})>^{2}\alpha^{4})\psi_{1}^{2}\alpha^{6}\label{eq:a2b2eq1}\\
 &  & -(<\Real(F_{1})><\Imag(F_{1})>^{2}\alpha^{4})\psi_{1}\alpha^{3}\phi_{1}\alpha^{3}\nonumber \\
 &  & -(<\Real(F_{1})>^{2}<\Real(F_{0})>)\alpha^{10}.\nonumber 
\end{eqnarray}
Differentiating the quantities inside the parentheses with respect
to $\alpha$, the second equation for $\phi_{2}$ and $\psi_{2}$
is 
\begin{eqnarray}
(6d\alpha^{5})\psi_{2}\alpha^{4}+(6f\alpha^{5})\phi_{2}\alpha^{4} & = & -5(4F_{2}<\Real(F_{1})>^{2}\alpha^{3})\psi_{1}^{2}\alpha^{6}\label{eq:a2b2eq2}\\
 &  & -(4<\Real(F_{1})><\Imag(F_{1})>^{2}\alpha^{3})\psi_{1}\alpha^{3}\phi_{1}\alpha^{3}\nonumber \\
 &  & -10(<\Real(F_{1})>^{2}<\Real(F_{0})>)\alpha^{9}.\nonumber 
\end{eqnarray}
Equations (\ref{eq:a2b2eq1}) and (\ref{eq:a2b2eq2}) form a nonhomogeneous
linear system for $[\phi_{2},\;\psi_{2}]$ with the same matrix $M$.
The condition for the solution $[\phi_{2},\;\psi_{2}]$ to exist requires
that the right hand of the second equation is six times that of the
first equation, which yields after eliminating $\phi_{1}$ by equation
(\ref{eq:a1smalls}) the following equation for $\psi_{1}$ 
\[
\left(5F_{2}f-<\Real(F_{1})>^{2}<\Imag(F_{1})>^{2}\right)\psi_{1}^{2}=2f<\Real(F_{0})>.
\]
This determines $\psi_{1}$, and then from equation (\ref{eq:a1smalls}),
$\phi_{1}$, namely, 
\[
\phi_{1}=\left[\frac{8n^{2}(m+n^{3})^{6}}{3(n-1)(n+1)^{4}(n^{2}-m)\phi(16(m-1)^{2}(m+n^{3})^{2}+5(n-1)(n^{2}-m)\phi\psi)}\right]^{1/2}\text{Ma}^{3/2}
\]
and 
\[
\psi_{1}=\left[\frac{n^{4}(n-1)(n^{2}-m)\phi}{6(16n^{2}(m-1)^{2}(m+n^{3})^{2}+5(n-1)(n^{2}-m)\phi\psi)}\right]^{1/2}\text{Ma}^{1/2}.
\]
Returning to the independent variable $s$, the asymptotics 
\[
\gamma_{R\text{max}}=\frac{\psi_{1}}{\phi_{1}}s^{2},\;\alpha_{\text{max}}=\phi_{1}s^{2/3}
\]
are shown in figure \ref{fig:fig4abcd} along with the full dependencies
for a representative set of the parameter values.

For the case $\text{Bo}\ne0$, it is sufficient to consider only the
leading order of equations (\ref{eq:MAXeqn}) and (\ref{eq:alphaeq:MAXeqn})
(proportional correspondingly to $\alpha^{8}$ and $\alpha^{7}$)
to determine the coefficients $c_{1}$ and $d_{1}$ in the asymptotics
$s^{2}=c_{1}\alpha^{2}$ and $\gamma_{R\text{max}}=d_{1}\alpha^{2}$.
Since there are contributions from the terms of equations (\ref{eq:MAXeqn})
and (\ref{eq:alphaeq:MAXeqn}) with all powers of $\gamma_{R\text{max}}$,
the resulting system of two quartic equations for $c_{1}$ and $d_{1}$
can only be solved numerically. The small-$s$ asymptotics, 
\[
\gamma_{R\text{max}}=\frac{d_{1}}{c_{1}}s^{2},\;\alpha_{\text{max}}=c_{1}^{-1/2}s
\]
are shown in figure \ref{fig:fig4abcd} along with the full numerics.

We see that the cases $\text{Bo}=0$ and $\text{Bo}\ne0$ have different
powers of $s$ in the asymptotics for $\alpha_{\text{0}}$, and the
same is true for $\alpha_{\text{max}}$. Figure \ref{fig:fig4abcd}(c)
shows that as $\text{Bo}\downarrow0$, the interval of small $s$
for which $\alpha_{\text{0}}\propto s$ shrinks, and there is a crossover
to the $s^{1/2}$ behavior characteristic of $\text{Bo}=0$ for an
interval of larger (but still small) wavenumbers. Similarly, for $\alpha_{\text{max}}$
there is a crossover from $\alpha_{\text{max}}\propto s$ at the smallest
$s$ to the $s^{2/3}$ asymptotic characteristic of $\text{Bo}=0$
for an interval of larger wavenumbers.

These considerations clarify the transition from the instability at
$s\ne0$ to stability at $s=0$, and the relation between the different
powers in the $\alpha_{0}$ and $\alpha_{\text{max }}$ asymptotics
of the $\text{Bo}\ne0$ and $\text{Bo}=0$ cases.

\subsection{Instability thresholds in the different sectors and nearby asymptotic
behavior}

\label{sec:FiniteCaseLongwaveApproxCregions} In both the $R$ sector
$(1<m<n^{2})$ and the $Q$ sector, $(m>n^{2})$, the surfactant branch
(\ref{eq:gamSsmallAlphaApprox}) is stable for all $\text{Bo }$ and
the robust branch (\ref{eq:gamCsmallAlphaApprox}) is unstable if
$\Bo<\Bo_{cL}$, where, in view of equation (\ref{eq:gamCsmallAlphaApprox}),
the threshold value is 
\begin{equation}
\text{Bo}_{cL}=\frac{3\varphi(m-n^{2})}{4n^{3}(1-m)(n+m)}\text{Ma.}\label{eq:BoCritical}
\end{equation}
In the $R$ sector, the Marangoni effect is destabilizing, so $\textrm{B\ensuremath{o_{cL}}}>0$;
gravity renders the flow stable for $\text{Bo}>\text{Bo}_{cL}$, whereas
for $\text{Bo }<\text{Bo}_{cL}$, the flow is unstable. In the $Q$
sector (and in the $S$ sector as well), the Marangoni effect is stabilizing,
$\text{Bo}_{cL}<0$, and the gravity effect renders the robust branch
unstable when the (negative, destabilizing) $\text{Bo}<\text{Bo}_{cL}$.

\begin{figure}
\centering{}\includegraphics[clip,width=1\textwidth]{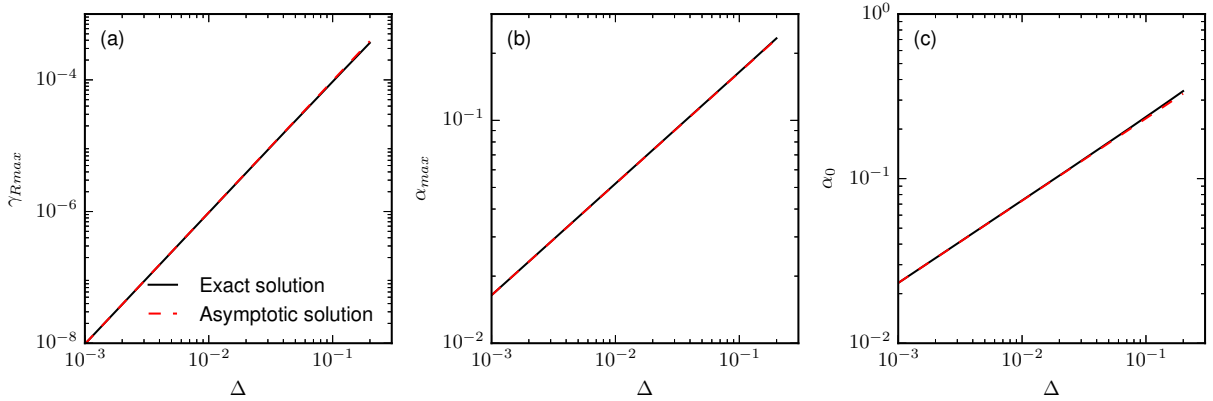} \caption{$\gamma_{R\text{max}},\alpha_{\text{max}}$ and $\alpha_{0}$ vs $\Delta$
for the same $n$, $m$ and $\text{Ma}$ as in figure \ref{fig:fig4abcd},
and $s=1$. The solid lines represent the full solutions, equation
(\ref{eq:MaEquation2}), and the dashed lines represent the asymptotics
given by (\ref{eq:MAapproxCmode}).\label{fig:figRTmaANDapp}}
\end{figure}

From equation (\ref{eq:BoCritical}) the ratio $\text{Bo}_{cL}/\text{Ma}$
is a function of $m$ and $n$ only, and its graph is a surface in
the $(n,m,\text{Bo}_{cL}/\text{Ma})$-space. This surface is plotted
in figure 3 of \citet{Frenkel2016}, and is discussed in detail there.
The window of unstable wavenumbers, $0<\alpha<\alpha_{0}$, shrinks
to zero as $\Bo\uparrow\Bo_{cL}$, so that the marginal wavenumber
$\alpha_{0}\downarrow0$ for both the $R$ and $Q$ sectors. To obtain
the asymptotic approximation for $\alpha_{0}$, we write the Bond
number as 
\begin{equation}
\text{Bo}=\text{Bo}_{cL}-\ \Delta\text{}\label{eq:BoDeltaFormula}
\end{equation}
with $\Delta\downarrow0$. Equation (\ref{eq:F2Re}) is substituted
into (\ref{eq:MAappEqn}) and when retaining the leading order terms
in $\Delta$ and $\alpha^{2}$ we find that $\zeta_{0}$ is proportional
to $\Delta$ and $\zeta_{2}$ is a cubic polynomial in $\text{Bo}_{cL}$
(and is independent of $\Delta$, to the leading order). The solution
is 
\begin{gather}
\alpha_{0}\approx\left[1+\beta_{1}\text{Bo}_{cL}+\beta_{3}\left.\text{Bo}_{cL}^{3}\right]^{-1/2}\Delta^{1/2}\text{}\right.\label{eq:MAapproxCmode}
\end{gather}
where the coefficients $\beta_{1}$ and $\beta_{3}$ are given by
equations (\ref{eq:beta1}) and (\ref{eq:beta3}) in appendix \ref{sec:Coefficients-of-equations}.
Note here that $\Ma$ has been written in terms of $\Bo_{cL}$ using
equation (\ref{eq:BoCritical}). If $\Bo_{cL}$ $\ll1$ (i.e., $\Ma\ll1$)
equation (\ref{eq:MAapproxCmode}) simplifies to 
\begin{equation}
\alpha_{0}\approx\Delta^{1/2}\text{.}\label{eq:MAapproxCmodeDeltaBo}
\end{equation}
We also find in the way described above the long-wave asymptotic dependences
\[
\alpha_{\text{max}}\propto\Delta^{1/2}\text{ and \ensuremath{\gamma_{R\text{max}}\propto\Delta^{2}}.}
\]
For example, the relative error of the asymptotic expression (\ref{eq:MAapproxCmodeDeltaBo})
for $n=m=2$, $s=1$, $\Bo=10^{-6}$, and $\Ma=10^{-6}$ to $\Ma=10$
is less than $10\%$ for $\Delta<0.2$. This is illustrated in figure
\ref{fig:figRTmaANDapp}, where $n=m=2$, $s=1$ and $\text{Ma}=1$.
The asymptotics for $\gamma_{R\text{max}}$, $\alpha_{\text{max}}$
and $\alpha_{0}$ near $\text{Bo}=\text{Bo}_{c}$ are practically
indistinguishable from the full numerical solutions. 
In the $S$ sector ($1<n<\infty$ and $0<m<1$), the robust branch
(\ref{eq:gamCsmallAlphaApprox}) is stable when $\text{Bo}>\text{Bo}_{cL}$,
the latter given by (\ref{eq:BoCritical}), and unstable otherwise.
However, equation (\ref{eq:gamSsmallAlphaApprox}) for the surfactant
branch does not contain the Bond number, and indicates instability.
Thus the surfactant mode is unstable for any $\text{Bo}$ provided
$\alpha$ is sufficiently small. However, it is easy to see that the
window of unstable wavenumbers shrinks to zero as $\text{Bo}\uparrow\infty$.
Indeed in this limit, equation (\ref{eq:MAappEqn}) reduces to 
\begin{equation}
s^{2}(n-1)(n+1)^{2}(m-1)+36n^{3}(n+m)\alpha^{2}\text{Bo}^{2}=0\text{.}\label{eq:MAappEqnSmodeBigBo}
\end{equation}
Hence the asymptotic formula for the marginal wavenumber is 
\begin{equation}
\alpha_{0}\approx\left[\frac{36s^{2}(n+1)^{2}(1-m)(n-1)}{n^{3}(n+m)}\right]^{1/2}\text{Bo}^{-1}\text{.}\label{eq:MAapproxSmode}
\end{equation}
For the $Q$ sector, the instability threshold (\ref{eq:BoCritical})
can be written in a different way: to state that (while the surfactant
branch (\ref{eq:gamSsmallAlphaApprox}) is stable for all $\text{Bo}$
and $\Ma$), the robust branch is stable if $\Ma$ exceeds a critical
Marangoni number, $\Ma_{cL}$ given by 
\begin{equation}
\text{Ma}_{cL}=\frac{4n^{3}(1-m)(n+m)}{3\varphi\left(m-n^{2}\right)}\text{Bo,}\label{eq:RTcritcalMa}
\end{equation}
which is the reciprocal of (\ref{eq:BoCritical}).

When $\Ma\uparrow\Ma_{cL}$, the marginal wavenumber is expressed
in terms of $\Delta_{M}=\Ma_{cL}-\Ma$. From equation (\ref{eq:MAappEqn}),
we obtain in the same way that we derived (\ref{eq:MAapproxCmode})
for the marginal wavenumber in the $R$ sector that 
\begin{equation}
\alpha_{0}\approx\left[M_{0}+M_{1}\text{Ma}_{cL}+M_{3}\text{Ma}_{cL}^{3}\right]^{-1/2}\Delta_{M}^{1/2}\text{.}\label{eq:RTmaApp}
\end{equation}
where the coefficients $M_{0}$, $M_{1}$ and $M_{3}$ are given by
(\ref{eq:M0})-(\ref{eq:M3}) in appendix B.

\subsection{Instabilities on the $\left(n,m\right)$-sector borders}

\label{sec:Instabilities on borders}The borders $m=1$, $m=n^{2}$,
and $n=1$ are considered separately because of singularities that
can occur in the expressions for the growth rates and the marginal
wavenumber derived in the previous sections for the $R$, $S$, and
$Q$ sectors.

\subsubsection{The $m=1$ border}

Consider first the case $m=1$ and $n\neq1$. In the long-wave limit,
$F_{1}^{2}\ll\left\vert F_{2}F_{0}\right\vert $ since $F_{1}^{2}\sim\alpha^{4}$
and $|F_{2}F_{0}|\sim\Ma\alpha^{3}$ (the truncated Taylor series
for such quantities are shown in Appendix C of \citet{schweiger2013gravity}).
Therefore, the roots to the dispersion equation (\ref{eq:QuadEqnGamma}),
are approximated by 
\begin{equation}
\gamma\approx\frac{1}{2F_{2}}\left(-F_{1}+(4F_{2}F_{0})^{1/2}\left[1+\frac{1}{2}\left(-\frac{F_{1}^{2}}{4F_{2}F_{0}}\right)\right]\right)\text{.}\label{eq:gammaIncrAPPm}
\end{equation}
Hence, the growth rates of the two branches are 
\begin{equation}
\gamma_{R}=\frac{-\operatorname{Re}(F_{1})+\operatorname{Re}(\sqrt{\zeta})}{2F_{2}}\label{eq:gamWithZm}
\end{equation}
where $\zeta$ is the discriminant of (\ref{eq:QuadEqnGamma}). To
leading order in $\alpha$, equation (\ref{eq:gamWithZm}) reduces
to 
\begin{equation}
\gamma_{R}\approx\frac{\operatorname{Re}(\sqrt{\zeta})}{2F_{2}}=\pm\frac{n\left[\left\vert n-1\right\vert (n+1)s\text{Ma}\right]^{1/2}}{2(n+1)^{2}}\alpha^{3/2}.\label{eq:GammaApproxSmallalpham}
\end{equation}
This result does not depend on the Bond number and is the same as
in FH and HF. It turns out that the next order correction, omitted
in the leading order expression, depends on both the Bond number and
the Marangoni number, and is proportional to $\alpha^{2}$. Note also
that (\ref{eq:GammaApproxSmallalpham}) is valid as $\alpha\downarrow0$
with the Marangoni number fixed but it is not valid as $\Ma\downarrow0$
with the wavenumber fixed. We will show below that for $m=1$, the
discriminant $\zeta$ in the expression for $\gamma_{R}$ is never
zero, and thus there are two branches of $\gamma_{R}$ that are continuous
at all parameter values and all $\alpha$, which we called the surfactant
branch and the robust branch. It is unclear from equation (\ref{eq:GammaApproxSmallalpham})
whether the positive growth rate corresponds to the surfactant branch
or the robust branch. Recall that, as $\Ma\downarrow0$, with $\alpha$
remaining finite, the identity of each branch is clear since, by definition,
the branch that vanishes in this limit is the surfactant branch. Starting
from there, each branch can be traced to the asymptotic region of
small $\alpha$ and finite $\Ma$ where equation (\ref{eq:GammaApproxSmallalpham})
is valid and thus the branches will be identified there.

The fact that there are two continuous branches of $\gamma(\alpha,\text{Ma)}$
(with the other parameters fixed and not shown explicitly) given by
(\ref{eq:QuadEqnGamma}) is seen as follows. As was discussed previously,
in section \ref{sec:FiniteCaseDerivationDispRelation} (see also Appendix
A), in any simply connected domain not containing 0 of the complex
$\zeta-$plane, there exist two distinct analytic branches of the
square root function, $f(\zeta)=\zeta^{1/2}$. The $\sqrt{\zeta}$
in the expression for $\gamma_{R}$, is a composite function of $(\alpha,\text{Ma)}$
through $\zeta(\alpha,\text{Ma}$). The discriminant $\zeta$ is a
single-valued continuous function of $(\alpha,\text{Ma)}$. It is
easy to see that it maps the first quadrant of the $(\alpha,\text{Ma)}$-plane
inside the upper half-plane $U$ of the $\zeta$-plane, which is a
simply connected domain not containing 0. Indeed, when $m=1$ ($n\neq1$
and $s\neq0$), then from equation (\ref{eq:F1Im}), $\textrm{Im}(F_{1})=0$,
and hence 
\begin{equation}
\operatorname{Im}(\zeta)=-4F_{2}\operatorname{Im}(F_{0}).\label{eq:IMzPOSmeq1-n-neq1}
\end{equation}
In view of $n>1$, we have $s_{\alpha n}>s_{\alpha}n$, and hence,
from equation (\ref{eq:F0Im}), $-\textrm{Im}(F_{0})>0$. Therefore,
equation (\ref{eq:IMzPOSmeq1-n-neq1}) yields $\operatorname{Im}(\zeta)>0$.
Since the upper half-plane $U$ of the $\zeta$-plane is a simply
connected domain not including 0, the square root function $\xi=f(\zeta)=\zeta^{1/2}$
in $U$ of $\zeta$ has two analytic branches. One of them maps $U$
onto the first quadrant of the $\xi-$plane, so that $\text{Re}(\sqrt{\zeta})>0$
for this branch, and thus $\text{Re}(\sqrt{(\zeta})$ is a \textit{positive}
continuous function of $(\alpha,\text{Ma})$. The other analytic branch
of $\xi=\zeta^{1/2}$ has its range entirely in the third quadrant
of the $\xi-$plane, so that $\text{Re}(\sqrt{\zeta})<0$ and thus
$\text{Re}(\sqrt{\zeta})$ is a \textit{negative} continuous function
of $(\alpha,\text{Ma})$ Thus, there is the one branch of $\text{Re}(\sqrt{\zeta})$
that is continuous and positive at all $(\alpha,\text{Ma )}$ and
the other branch of $\text{Re}(\sqrt{\zeta})$ that is continuous
and negative at all $(\alpha,\text{Ma)}$. (We note that for even
for arbitrary $m\ne0$, it readily follows that $\text{Im}(\zeta)>0$,
provided that $\text{Ma}\downarrow0$ and $\text{Bo}>0$, since then,
according to equations (\ref{eq:F1Re})-(\ref{eq:F0Im}), $F_{0}=0$,
$\text{Re}(F_{1})>0$ and $\text{Im}(F_{1})>0$.)

In the limit of $\text{Ma}\downarrow0$, the surfactant branch vanishes,
$\gamma_{R}=0$, which from equation (\ref{eq:gamWithZm}) means $\operatorname{Re}(\sqrt{\zeta})=\operatorname{Re}(F_{1})$.
Therefore, $sgn(\operatorname{Re}(\sqrt{\zeta}))=sgn(\operatorname{Re}(F_{1}))$,
where $sgn$ is the sign function. It is sufficient to consider here
only small wavenumbers, from an interval $[0,\alpha_{s}]$, by choosing
an arbitrary $\alpha_{s}$ such that $\alpha_{s}\ll1$ and $\alpha_{s}<|\Bo|$.
Then equation (\ref{eq:F1reApprox}) (with $\text{Ma}=0$) yields
$sgn(\operatorname{Re}(F_{1}))=sgn(\Bo)$, so that $sgn(\operatorname{Re}(\sqrt{\zeta}))=sgn(\Bo)$.
As was already established, each branch of $\Real(\sqrt{\zeta})$
has the same sign for all $(\alpha$,Ma$)$. Therefore, for the surfactant
branch, the relation $sgn(\operatorname{Re}(\sqrt{\zeta}))=sgn(\Bo)$
holds in the limit of $\alpha$ $\downarrow0$ as well. From equation
(\ref{eq:GammaApproxSmallalpham}), $sgn(\gamma_{R})=sgn(\operatorname{Re}(\sqrt{\zeta})$,
and then for the surfactant branch, $sgn(\gamma_{R})=sgn(\Bo)$. Thus,
the surfactant branch is unstable for $\Bo>0$, $\gamma_{R}\propto+\alpha^{3/2}$
and stable for $\Bo<0$, $\gamma_{R}\propto-\alpha^{3/2}$. Consequently,
the robust branch is stable (unstable) for $\Bo>0$ ($\Bo<0$). This
answers the question of identifying the stable and unstable modes
as belonging to the appropriate branches.

\begin{figure}
\centering \includegraphics[clip,width=0.55\textwidth]{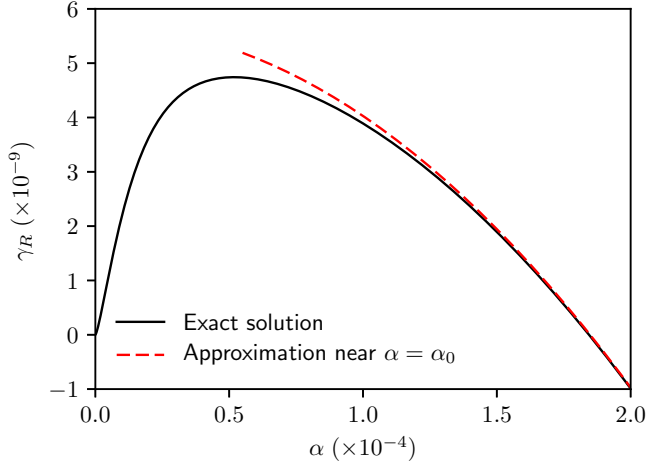} \caption{The exact dispersion curve (\ref{eq:QuadEqnGamma}) and the asymptotic
expression of the growth rate around the marginal wavenumber (\ref{eq:GammaApproxLargealphameq1})
of $\gamma_{R}$ for $m=1$, $n=2$, $s=1$, $\Ma=1$, and $\Bo=1000$.
\label{fig:Figm1gam_smallandlarge_a_app}}
\end{figure}

In certain limits it is possible to find a long-wave approximation
to $\gamma_{R}$ that captures the growth rate behavior close to the
marginal wavenumber $\alpha_{0}$. Assuming $\Bo\gg\Ma$, $\alpha^{2}\ll\textrm{Bo}$,
and $\Ma/\Bo^{2}\ll\alpha\ll1$, equation (\ref{eq:QuadEqnGamma})
can be simplified to yield, for the unstable surfactant branch, 
\begin{equation}
\gamma_{R}\approx\frac{27}{4}\frac{(n-1)^{2}(n+1)^{3}s^{2}\text{Ma}^{2}}{n^{5}\text{Bo}^{3}}-\frac{1}{4}\frac{n\text{Ma}}{(n+1)}\alpha^{2}\label{eq:GammaApproxLargealphameq1}
\end{equation}
which is valid for $\alpha\approx\alpha_{0}$. (Note that this equation
is not valid in the limit as $\alpha\downarrow0$; in the latter limit,
the leading order behavior is still given by (\ref{eq:GammaApproxSmallalpham})).
In figure \ref{fig:Figm1gam_smallandlarge_a_app} the growth rate
of the surfactant branch is plotted using (\ref{eq:QuadEqnGamma})
along with the asymptotic expression (\ref{eq:GammaApproxLargealphameq1}).
One can see the dashed line approximations approaches the full dispersion
curve as $\alpha\uparrow\alpha_{0}$. The long-wave $\gamma_{R}$
approximation (\ref{eq:GammaApproxSmallalpham}) is not plotted in
figure \ref{fig:Figm1gam_smallandlarge_a_app} but for the same parameter
values the error is less than $1\%$ when $\alpha$ $<1.4\times10^{-9}$.

An asymptotic expression for $\alpha_{0}$ is obtained by solving
for $\alpha$ equation (\ref{eq:GammaApproxLargealphameq1}) with
$\gamma_{R}=0$: 
\begin{equation}
\alpha_{0}\approx\frac{3s\left\vert n-1\right\vert (n+1)^{2}[3\text{Ma}]^{1/2}}{n^{3}\text{Bo}^{3/2}}.\label{eq:alpha0asy}
\end{equation}
The above expression is also obtained from the long-wave marginal
wavenumber equation (\ref{eq:MAappEqn}). This expression also suggests
that gravity is not completely stabilizing since $\alpha_{0}>0$ at
any positive finite value of $\Bo$. We had the similar result that
gravity, no matter how strong, cannot completely stabilize the Marangoni
instability for the $S$ sector.

\subsubsection{The case $n=1$}

Next, we consider the border $n=1$ with $m\neq1$. Just like the
$m=1$ and $n\neq1$ case, the imaginary part of the discriminant
$\zeta$, $\operatorname{Im}(\zeta)=2\operatorname{Re}(F_{1})\operatorname{Im}(F_{1})$,
is positive (or negative) for $m<1$ $($or $m>1)$, see (\ref{eq:F2Re})-(\ref{eq:F0Im}).
The growth rate for the robust mode is, from equation (\ref{eq:gamCsmallAlphaApprox}),
\begin{equation}
\gamma_{R}\approx-{\frac{\left(1+m\right)}{{m}^{2}+14\,m+1}}\left\{ \text{{Ma}}+\frac{1}{3}\text{Bo}\right\} {\alpha}^{2}\text{,}\label{eq:gamRn1border}
\end{equation}
but, since the coefficient of the $\alpha^{2}$ term in equation (\ref{eq:gamSsmallAlphaApprox})
becomes zero, we have for the surfactant branch, using equation (\ref{eq:ks})
with $n=1$, 
\begin{equation}
\gamma_{R}\approx-{\frac{1}{96}}\,{\frac{\left(1+m\right)}{{s}^{2}\left(m-1\right)^{2}}}\left\{ \,{\frac{1}{2}\,}\text{{Ma}}+{\frac{1}{3}}\,\text{{Bo}}\right\} \text{{Bo}}{\,}\text{{Ma}}{\alpha}^{4}\text{.}\label{eq:gamSn1border}
\end{equation}
For this case the robust and surfactant branches are long-wave stable
for $\Bo>0$. For $\Bo<0$ both branches are unstable if the magnitude
of $\Bo$ is sufficiently large. This occurs when the leading term
coefficients in (\ref{eq:gamRn1border}) and (\ref{eq:gamSn1border})
are positive, i.e. when $\Bo<-3\text{Ma}$ for (\ref{eq:gamRn1border}),
and $-3\text{Ma}/2<\text{Bo}<0$ for (\ref{eq:gamSn1border}).

\subsubsection{The $m=n^{2}$ border}

For the $m=n^{2}\neq1$ border, using the general equation (\ref{eq:gamGeneralEqnRSQsectors})
to obtain the growth rates to the leading orders, we find 
\begin{equation}
\gamma_{R}\approx-\left\{ \frac{n\text{Bo}}{12(n+1)}\right\} \alpha^{2}+\left\{ \frac{n\left(2\text{Ma}+n\text{Bo}-5\right)}{60(n+1)}\right\} \alpha^{4}\text{, }\label{eq:Midwave_gamC_smalla}
\end{equation}
and 
\begin{equation}
\gamma_{R}\approx-\left\{ \frac{\text{Ma}}{4(n+1)}\right\} \alpha^{2}.\text{}\label{eq:Midwave_gamS_smalla}
\end{equation}
We have kept two leading orders in equation (\ref{eq:Midwave_gamC_smalla})
because the $\alpha^{2}$ term vanishes for $\text{Bo}=0$. Equation
(\ref{eq:Midwave_gamS_smalla}) shows that the surfactant branch is
always stable, and this is consistent with HF in the limit $\Bo\rightarrow0$.
Also, in this limit the robust branch, equation (\ref{eq:Midwave_gamC_smalla}),
reproduces the corresponding HF result, their equation (4.13). Also,
for $\text{Bo}=0$, equation (\ref{eq:Midwave_gamC_smalla}) recovers
the long-wave dispersion relation found in FH.

Finally, for the $m=1$ and $n=1$ case, the solutions to the dispersion
equation (\ref{eq:QuadEqnGamma}) for arbitrary wavenumber are of
the form 
\begin{equation}
\gamma_{R}=\frac{-a\text{Ma}-b(\text{Bo}+\alpha^{2})\pm\lbrack a\text{Ma}-b(\text{Bo}+\alpha^{2})]}{2F_{2}\alpha^{4}}\text{,}\label{eq:gamPointCaseFull}
\end{equation}
where 
\[
a=\alpha^{2}(s_{\alpha}^{2}-\alpha^{2})(c_{\alpha}s_{\alpha}+\alpha)\text{ and }b=(s_{\alpha}^{2}-\alpha^{2})(c_{\alpha}s_{\alpha}-\alpha)\,\text{.}
\]
After substituting $F_{2}$, $a$, and $b$ into (\ref{eq:gamPointCaseFull}),
the growth rate for the robust branch is 
\[
\gamma_{R}=-\frac{(s_{\alpha}^{2}-\alpha^{2})(\text{Bo}+\alpha^{2})}{4\alpha(c_{\alpha}s_{\alpha}+\alpha)}\approx-\frac{1}{24}\left(\text{Bo}+\alpha^{2}\right)\alpha^{2}\text{ for }\alpha\ll1\text{,}
\]
and the growth rate for the surfactant branch is 
\[
\gamma_{R}=-\frac{\alpha(s_{\alpha}^{2}-\alpha^{2})\text{Ma}}{4(c_{\alpha}s_{\alpha}-\alpha)}\approx-\frac{1}{8}\text{Ma}\alpha^{2}\text{ for }\alpha\ll1\text{.}
\]
Note that the surfactant branch is always stable but the robust branch
is unstable if $\alpha^{2}<-\Bo$. Obviously, this only occurs if
$\Bo<0$.

\section{Arbitrary wavenumbers; mid-wave instability}

\label{sec:FiniteCaseMidWave}

In this section, results are given for arbitrary wavenumber, and comparisons
are made across all parameter sectors. First, the influence of gravity
on the maximum growth rate $\gamma_{R\max}$, the corresponding wavenumber
$\alpha_{\max}$ and the marginal wavenumber $\alpha_{0}$ in the
$R$, $S$, and $Q$ sectors are considered for fixed values of the
Marangoni number. Then similar results are given to show the influence
of surfactant for fixed values of the Bond number. Asymptotic results
are also discussed. 
\begin{figure}
\centering{}\includegraphics[bb=0bp 0bp 576bp 576bp,clip,width=0.95\textwidth]{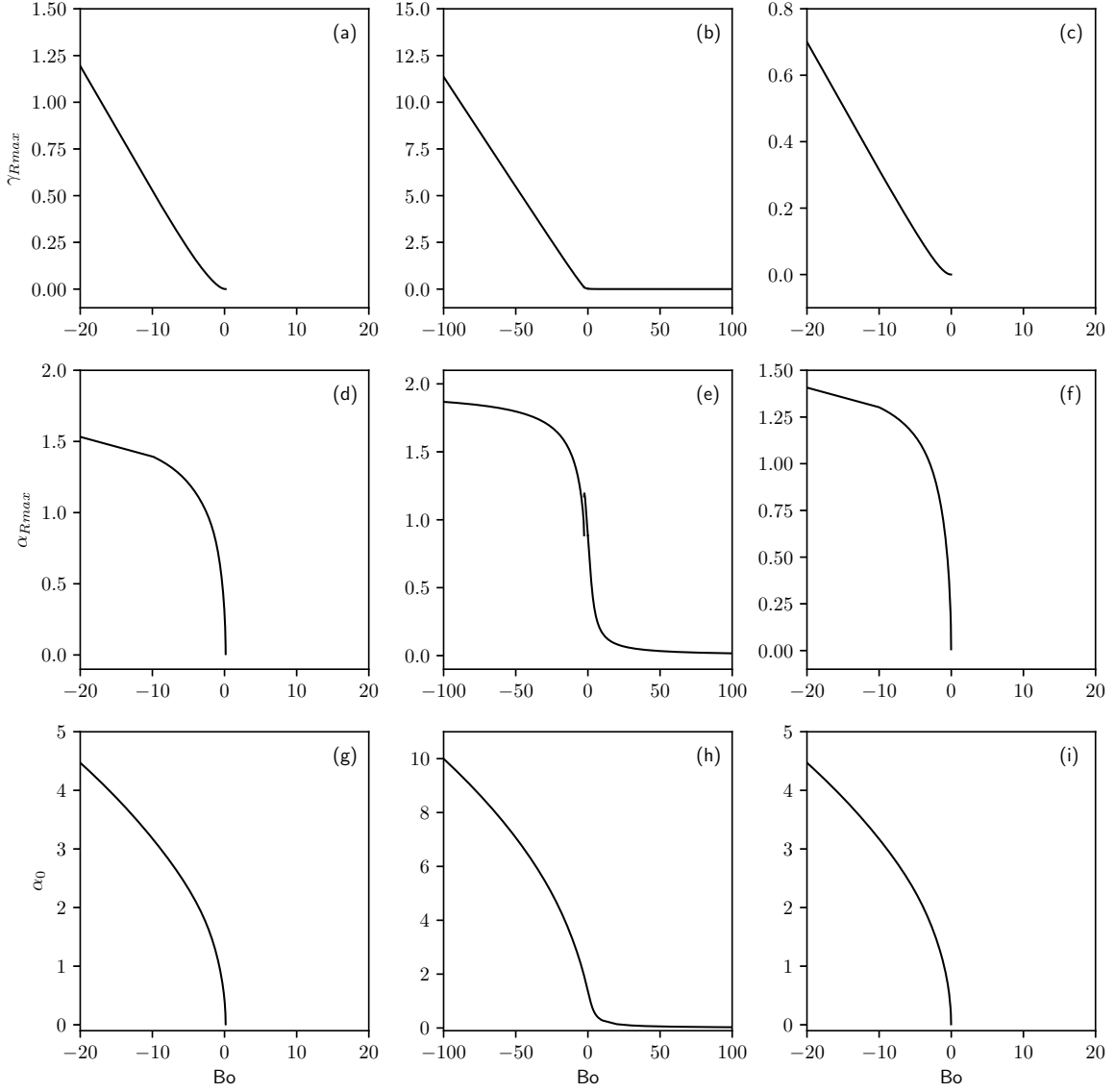}
\protect\caption{Dependence of the maximum growth rate $\gamma_{R\max}$, the corresponding
wavenumber $\alpha_{\max}$, and the marginal wavenumber $\alpha_{0}$
on $\textrm{Bp}$ in the $R$, $S$ and $Q$ sectors. Here $s=1$,
$\Ma=0.1$ and the values of the $(n,m)$ pairs for the $R$ (a,d,g),
$S$ (b,e,h), and $Q$ (c,f,i) sectors are $(2,2)$, $(2,0.5)$, and
$(2,5)$, respectively.\label{fig:Fig3maxmargRSQ} }
\end{figure}

\subsection{Effects of gravity}

\begin{figure}
\includegraphics[clip,width=0.95\textwidth]{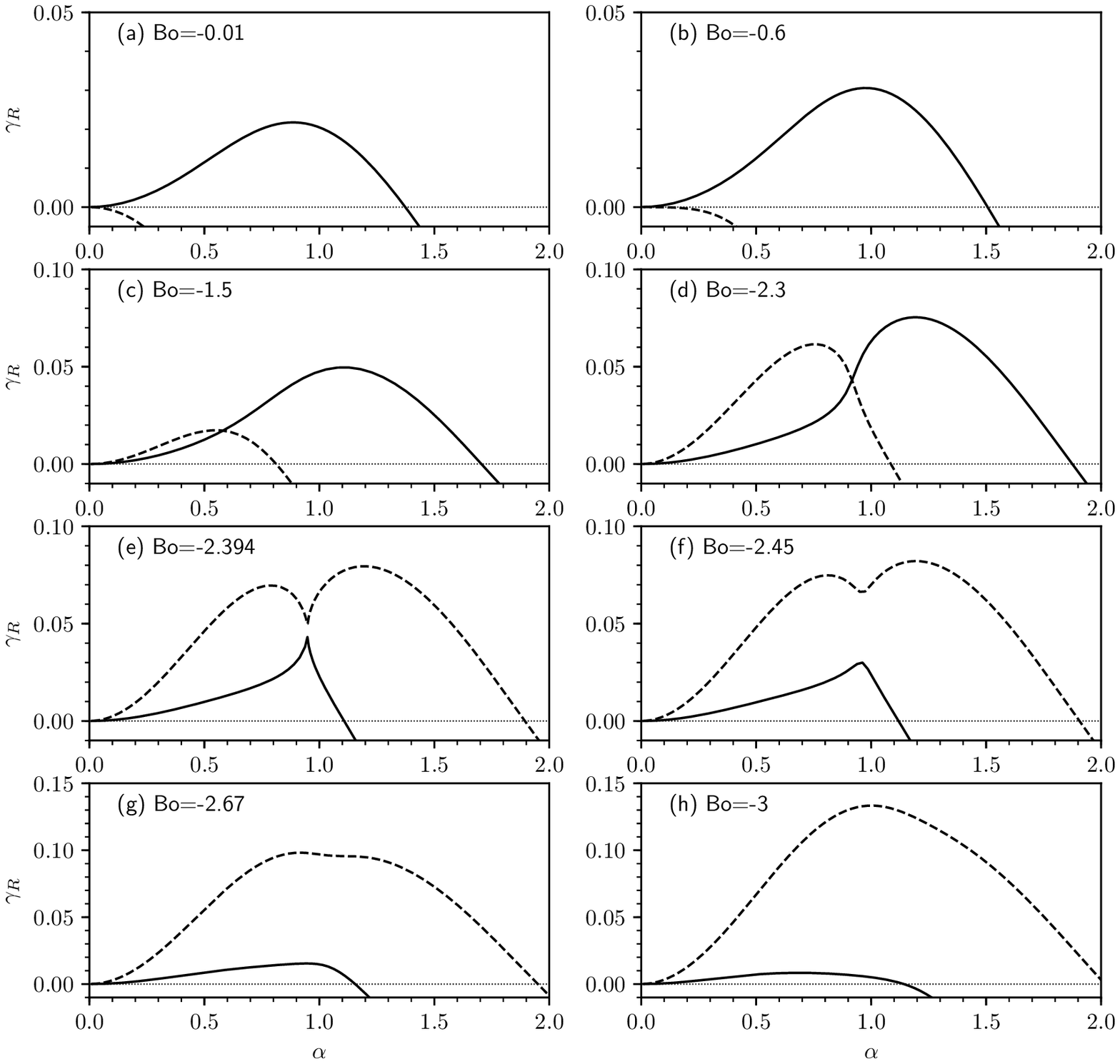} \protect\caption{Dispersion curves given by (\ref{eq:QuadEqnGamma}) in the $S$ sector
($n=2$, $m=0.5$) for selected values of $\Bo$ showing occurrence
of two local maxima and a jump in the global maximum. \ Here $s=1$
and $\Ma=0.1$. \label{fig:Fig_typical_MaxCrossingDCurve}}
\end{figure}

We first examine the influence of $\Bo$ on the maximum growth rate
$\gamma_{R\max}$, its corresponding wavenumber $\alpha_{\max}$,
and the marginal wavenumber $\alpha_{0}$. Figure \ref{fig:Fig3maxmargRSQ}
shows plots of $\gamma_{\max}$, $\alpha_{\max}$ and $\alpha_{0}$
for a representative $(n,m)$ pair from each of the three sectors
where panels (a, d, g), (b, e, h) and (c, f, i) represent the $R$,
$S$ and $Q$ sectors, respectively. In the $R$ sector, panels (a,
d, g) show that the system is unstable provided $\Bo$ does not exceed
a finite positive value $\Bo_{c}$ and that $\gamma_{R\max}$, $\alpha_{\max}$,
and $\alpha_{0}$ all decrease to zero as $\Bo\downarrow\Bo_{c}$.
These findings were also observed in the long-wave limit (see section
\ref{sec:FiniteCaseLongwaveApproxCregions}). This instability is
of the long-wave type even when the marginal wavenumber $\alpha_{0}$
is not small. However, for $m$ sufficiently close to $n^{2}$ but
still in the $R$ sector, there appears a ``mid-wave'' instability
(see figure 16 below), which is discussed below, in sections \ref{subsec:Surfactant-effects-in-Q-Sector}
and \ref{subsec:MA-Bo plane stability}. Panels (b, e, h) show the
surfactant branch is always unstable in the $S$ sector. The discontinuity
in the graph of $\alpha_{\max}$ in panel (e) is discussed below with
figure \ref{fig:Fig_typical_MaxCrossingDCurve}. In the $Q$ sector,
surfactants are completely stabilizing provided $\Bo>\text{Bo}_{c}$,
as shown in panels (c), (f) and (i). Note that $\Bo_{c}<0$ agrees
with the long-wave analysis (see equation (\ref{eq:RTcritcalMa})).
The discontinuity that can occur in the $S$ sector is displayed in
figure \ref{fig:Fig_typical_MaxCrossingDCurve}. Panel (a) shows that
for negligible $\Bo$, one branch is long-wave unstable and the other
one is stable. As the magnitude of $\Bo$ increases the previously
stable branch becomes unstable ($\Bo=-1$) and at some point the branches
cross ($\Bo=-1.5,-2.3$). Panel (e) shows that as $\left\vert \text{Bo}\right\vert $
continues to increase the crossing eventually disappears at which
point the upper branch has two local extrema. At some value of $\text{Bo}$,
the global maximum shifts from the right local extremum (as for $\Bo=-2.45$)
to the left local extremum (as for $\Bo=-2.67$). Finally, as $\Bo\downarrow-\infty$,
both branches are unstable in the long-wave manner, and feature a
single maximum.

\subsection{Effects of surfactants in the $R$ and $S$ sectors\label{subsec:Effects-of-surfactants in R and S}}

Here, we investigate, for a fixed value of $\Bo$ in the $R$ and
$S$ sectors, the Marangoni number $\Ma$ dependences of the maximum
growth rate $\gamma_{\max}$, the corresponding wavenumber $\alpha_{\max}$,
and the marginal wavenumber $\alpha_{0}$. The $Q$ sector turns out
to have somewhat different properties, which are discussed later (see
figure \ref{fig:FIG_RTgamNalphas2by4}). However, it is immediately
clear that in the $Q$ sector both branches are stable for $\Bo>0$
and fixed $\Ma$ (see panels (c), (f) and (i)) in figure \ref{fig:Fig3maxmargRSQ}.

\begin{figure}[ptb]
\includegraphics[clip,width=0.95\textwidth]{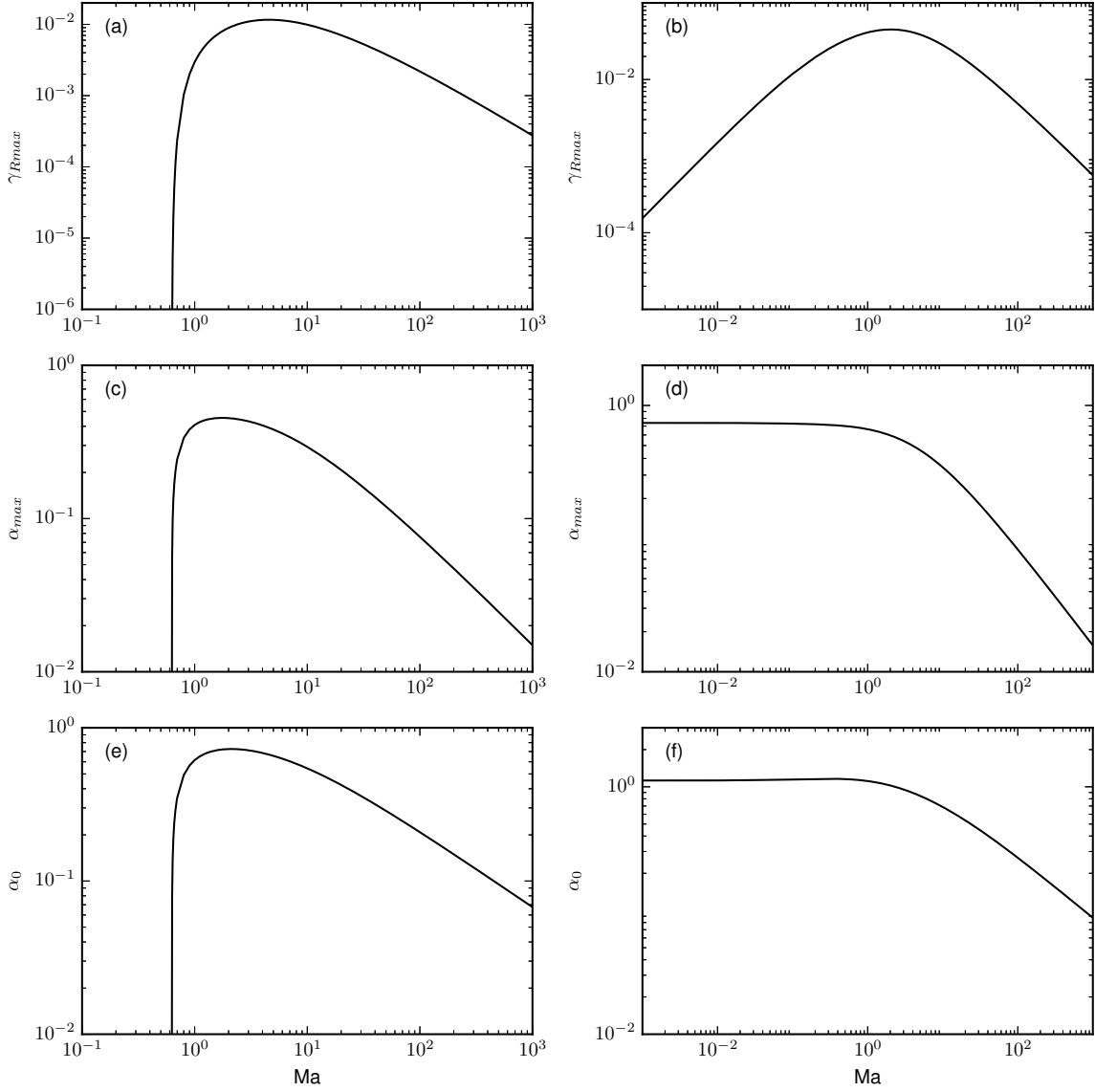} \protect\caption{(a,b) $\gamma_{R}{}_{\max}$, (c,d) $\alpha_{\max}$ and (e,f) $\alpha_{0}$
vs $\Ma$ for $\Bo=1.0$ in the $R$ sector (a,c,e) and $S$ sector
(b,d,f). Here $s=1$ and the values of the $(n,m)$ pairs in the $R$
and $S$ sectors are $(2,2)$ and $(2,0.5)$, respectively.\label{fig:FIG_Bo1_vs_Ma} }
\end{figure}

Panels (a) and (b) of figure \ref{fig:FIG_Bo1_vs_Ma} show that $\gamma_{R\max}$
attains a maximum at some $\text{Ma}=O(1)$ in both the $R$ and $S$
sectors, and that $\gamma_{R\max}\downarrow0$ as $\text{Ma}\uparrow\infty$.
Both $\alpha_{\max}$ and $\alpha_{0}$ also decrease to zero as $\text{Ma}\uparrow\infty$.
However, in the $R$ sector there is a threshold value of $\Ma$,
$\Ma_{cL}$, below which the flow is stable; while in the $S$ sector
the flow is unstable for all $\Ma>0$. Recall from the long-wave results
that $\Ma_{cL}(\Bo)$ is the inverse of $\Bo_{cL}(\Ma)$ (see equations
(\ref{eq:BoCritical}) and (\ref{eq:RTcritcalMa})). In the $S$ sector,
$\alpha_{\max}$ and $\alpha_{0}$ approach some non-zero constant
values and $\gamma_{R\max}\downarrow0$, showing no threshold value
of $\Ma$ for complete stabilization of the flow.

The small and large $\Ma$ asymptotics of $\alpha_{0}$ are discussed
next. Panels (e) and (f) suggest that $\alpha_{0}\downarrow0$ as
$\Ma\uparrow\infty$. By substituting equations (\ref{eq:k20app})
- (\ref{eq:k13app}) into the marginal-wavenumber equation (\ref{eq:MaEquation2}),
and keeping only the dominant $\Ma$ terms, the following expression
is obtained: 
\begin{equation}
\frac{n^{2}}{324}(n^{3}+m)^{2}\text{BoMa}\alpha^{2}+\frac{s^{2}}{108}\varphi(n-1)(n+1)^{2}(m-n^{2})=0,\label{eq:MAeqnMaInfin}
\end{equation}
from which 
\begin{equation}
\alpha_{0}\approx\frac{s(n+1)\sqrt{3\varphi(n-1)(n^{2}-m)}}{n(n^{3}+m)}\text{Bo}^{-1/2}\text{Ma}^{-1/2}.\label{eq:alpha0app0}
\end{equation}
This is consistent with the numerically-found behavior for $\alpha_{0}$
at large $\Ma$.

As $\Ma\downarrow0$, it is clear from panel (f) of figure \ref{fig:FIG_Bo1_vs_Ma}
that in the $S$ sector, $\alpha_{0}$ approaches some finite non-zero
value. Therefore, by keeping only the (dominant) linear $\Ma$ terms,
equation (\ref{eq:MaEquation2}) reduces to 
\begin{equation}
k_{11}+k_{13}B^{2}=0,\label{eq:reducedMaEqn2Ma0}
\end{equation}
where $k_{11}$ and $k_{13}$ depend on $\alpha$, as given by equations
(\ref{eq:k11}) and (\ref{eq:k13}). However, this equation must be
solved numerically for $\alpha_{0}$ since it is not necessarily small.
Some other asymptotics for $\alpha_{0}$ approaching zero in the $R$
sector were discussed above in subsection \ref{subsec:Marginal-wavenumbers}.
Panels (a), (b), (c) and (d) of figure \ref{fig:FIG_Bo1_vs_Ma} suggest
that $\gamma_{R\max}$ and $\alpha_{\max}\downarrow0$ as $\Ma\uparrow\infty$.
In the long-wave limit and for $\text{Ma}\gg1$, the linear and constant
terms of equation (\ref{eq:MAXeqn}), whose coefficients are proportional
to $\Ma^{2}$ and $\Ma^{3}$, are dominant, giving rise to the following
simplified equation for $\gamma_{R}$: 
\begin{equation}
\frac{1}{27}n^{3}(m+n^{3})(n^{2}-m)\alpha^{6}\text{Ma}\gamma_{R}-\frac{1}{108}(n-1)n^{4}(n^{2}-m)s^{2}\varphi\alpha^{6}\text{Ma}^{2}\approx0.\label{eq:gamRlargeMa}
\end{equation}
The latter gives 
\begin{equation}
\gamma_{R\max}\approx\frac{ns^{2}(n-1)(n+1)^{2}(m-n^{2})\varphi}{4(n^{3}+m)^{3}}\text{Ma}^{-1}\text{.}\label{eq:gamRmaxMaInfin}
\end{equation}
Because $\alpha^{6}$ appears in the simplified equation above, it
is convenient when solving for $\alpha_{\max}$ to subtract $\alpha$
times equation (\ref{eq:alphaeq:MAXeqn}) from six times equation
(\ref{eq:MAXeqn}) and obtain 
\begin{equation}
\frac{8}{27}(m-1)^{2}n^{5}(n+1)^{2}(n^{3}+m)s^{2}\alpha^{4}\text{Ma}\gamma_{R}-\frac{1}{162}n^{6}(n^{3}+m)^{2}\alpha^{8}\text{BoMa}^{3}\approx0.\label{eq:sub6times}
\end{equation}
Solving for $\alpha$ yields 
\begin{equation}
\alpha^{4}\sim48\frac{(m-1)^{2}(n+1)^{2}}{n(n^{3}+m)}s^{2}\text{Ma}^{-3}\text{Bo}^{-1}\gamma_{R}\text{.}\label{eq:equation2stars}
\end{equation}
Equation (\ref{eq:gamRmaxMaInfin}) is substituted into (\ref{eq:equation2stars}),
from which the following asymptotic expression for $\alpha_{\max}$
is obtained: 
\begin{equation}
\alpha_{\max}\approx\frac{\lbrack12\varphi(1-n)(m-n^{2})]^{1/4}(m-1)^{1/2}s}{(n^{3}+m)}\text{Ma}^{-3/4}\text{Bo}^{-1/4}\text{.}\label{eq:alphaMAXapp}
\end{equation}
\begin{figure}
\centering{}\includegraphics[clip,width=0.85\textwidth]{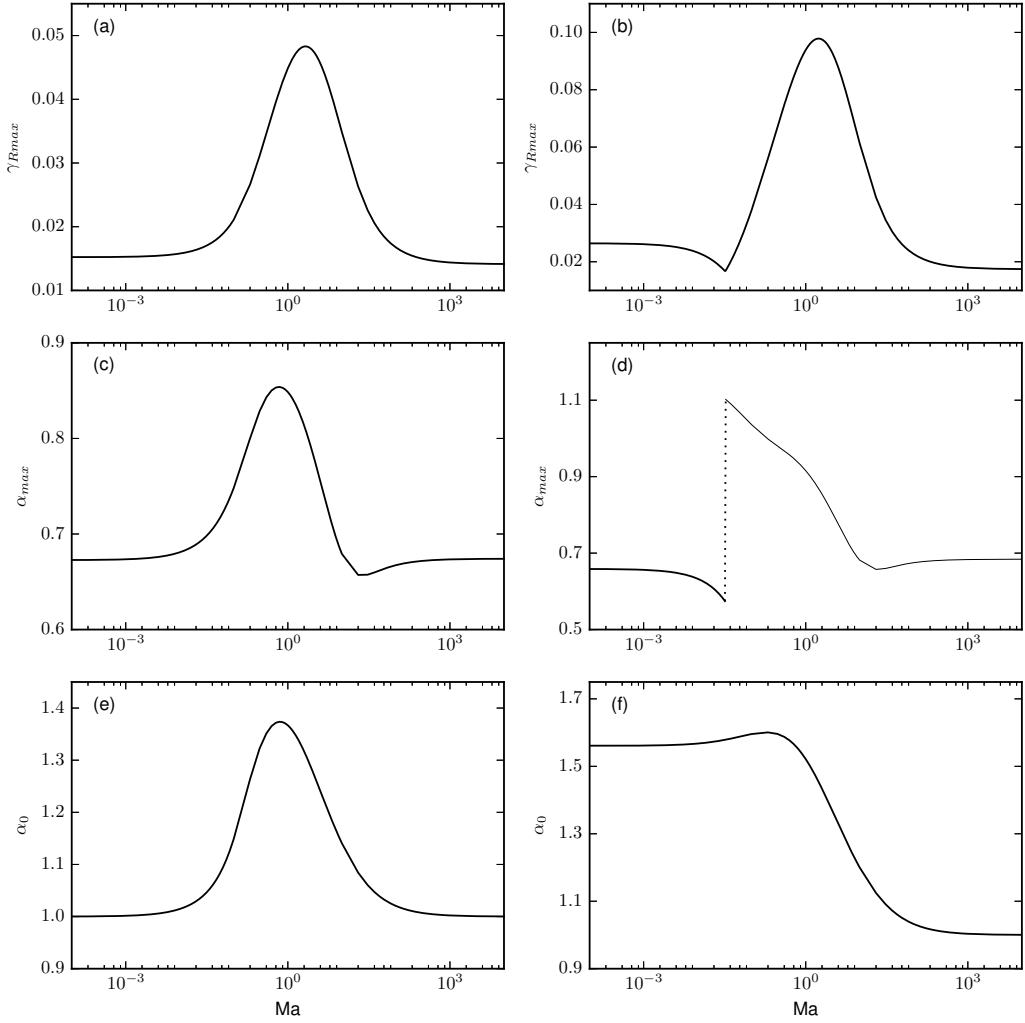}
\protect\caption{(a,b) $\gamma_{R}{}_{\max}$, (c,d) $\alpha_{\max}$ and (e,f) $\alpha_{0}$
as functions of $\Ma$ for $\Bo=-1.0$ in the $R$ sector (a,c,e)
and $S$ sector (b,d,f), for the same $s$ and $(n,m)$ points as
in figure \ref{fig:FIG_Bo1_vs_Ma}. \label{fig:FIG_Bon1_vs_Ma}}
\end{figure}
Panels (b) and (d) show that $\gamma_{R\max}\downarrow0$ and $\alpha_{\max}$
approaches some non-zero constant as $\Ma\downarrow0$. Therefore,
equation (\ref{eq:MAXeqn}) is approximately linear for $\gamma_{R}\ll1$,
$c_{10}\gamma_{R}\text{Ma}+c_{01}\approx0$ so that 
\begin{equation}
\gamma_{R}\approx-\frac{c_{01}}{c_{10}}\text{Ma}^{-1}\label{eq:gamRMa0Ssectors}
\end{equation}
where the $c_{ij}$ are independent of $\Ma$. An equation for $\alpha_{\max}$
is obtained by differentiating (\ref{eq:gamRMa0Ssectors}) with respect
to $\alpha$ and solving $d\gamma_{R}/d\alpha=0$ numerically for
$\alpha$, which is then substituted into (\ref{eq:gamRMa0Ssectors})
to obtain $\gamma_{R\max}$.

In contrast to the case shown in figure \ref{fig:FIG_Bo1_vs_Ma} for
$\Bo>0$, the flow is unstable for all $\Ma$ when $\Bo<0$ in either
the $R$ or $S$ sectors. Moreover, figures \ref{fig:FIG_Bon1_vs_Ma}
(a) and (b) also show that $\gamma_{R\max}$ has a global maximum
at $\Ma=O(1)$. However, in the $S$ sector $\gamma_{R\max}$ decreases
with increasing $\Ma$ for sufficiently small $\Ma$, up to $\Ma=\Ma_{0}$.
At $\Ma=\Ma_{0}$ there is a jump in $\alpha_{\max}$. This behavior
is due to the fact that the dispersion curve has two maxima, and at
this particular value of $\Ma$ there is a jump in the location of
the global maximum, similar to that shown in figure \ref{fig:Fig_typical_MaxCrossingDCurve}.
Figure \ref{fig:FIG_Bon1_vs_Ma} also shows that $\gamma_{R\max}$,
$\alpha_{\max}$ and $\alpha_{0}$ all approach some finite positive
constant in the limits $\Ma\uparrow\infty$ and $\Ma\downarrow0$
for both sectors.

Let us discuss the asymptotics of $\alpha_{0}$ with respect to the
Marangoni number for the case of $\text{Bo}<0$. Panels (e) and (f)
of figure \ref{fig:FIG_Bon1_vs_Ma} indicate that $\alpha_{0}$ asymptotes
to non-zero constants as both $\Ma\uparrow\infty$ and as $\Ma\downarrow0$.
The relevant values of $\alpha_{0}$ can be obtained as follows. For
$\Ma\uparrow\infty$, the dominant term in equation (\ref{eq:MaEquation2})
is the $\Ma^{3}$ term, and since $k_{13}\neq0$ this implies that
$\text{Bo}+\alpha^{2}\approx0$, or 
\begin{equation}
\alpha_{0}\approx\left\vert \text{Bo}\right\vert ^{1/2}.\label{eq:alpha0app}
\end{equation}
For $\Bo=-1$, $a_{0}\approx1$ which is consistent with the numerical
results shown in figures \ref{fig:FIG_Bon1_vs_Ma} (e) and (f). In
the limit Ma $\downarrow0$, equation (\ref{eq:MaEquation2}) reduces
to 
\begin{equation}
\left(k_{11}+k_{13}B^{2}\right)\text{Ma}B\approx0\text{.}\label{eq:eq:k11eq:k13eqn}
\end{equation}
In the $R$ sector, the solution $\alpha_{0}\approx\left\vert \text{Bo}\right\vert ^{1/2}$
is again obtained because $k_{13}$ is always positive and $k_{11}$
is the product of $(m-1)$ and a positive function, and thus $k_{11}>0$
in the $R$ sector. However, in the $S$ sector $k_{11}<0$, and $\alpha$
is a solution of $k_{11}+k_{13}$B$^{2}=0$ which is solved numerically
for $\alpha$. The solution is approximately $\alpha_{0}\approx1.56$
, and agrees with figure \ref{fig:FIG_Bon1_vs_Ma} (f).

Next, the asymptotics of $\gamma_{R\max}$ and $\alpha_{\max}$ in
the limit $\Ma\uparrow\infty$, and then in the limit $\Ma\downarrow0$,
(panels (a, b, c, d) of figure \ref{fig:FIG_Bon1_vs_Ma}) are discussed.
In this case, the terms proportional to $\Ma^{3}$ in equation (\ref{eq:MAXeqn})
yield 
\begin{equation}
c_{03}+c_{13}\gamma_{R}\approx0,\label{eq:maxapp}
\end{equation}
where the coefficients $c_{ij}$ correspond to the $\gamma_{R}^{i}\Ma^{j}$
terms in equation (\ref{eq:MAXeqn}). Therefore, 
\begin{equation}
\gamma_{R}\approx-\frac{c_{03}}{c_{13}}\approx-\frac{1}{2}\frac{\left(s_{\alpha}^{2}-\alpha^{2}\right)\left(s_{\alpha n}^{2}-\alpha^{2}n^{2}\right)\left(\text{Bo}+\alpha^{2}\right)}{\alpha\left(s_{\alpha}^{2}-\alpha^{2}\right)\left(s_{\alpha n}c_{\alpha n}+\alpha n\right)m+\alpha\left(s_{\alpha n}^{2}-\alpha^{2}n^{2}\right)\left(s_{\alpha}c_{\alpha}+\alpha\right)}\text{.}\label{eq:c03c13gameqn}
\end{equation}
Again, one must solve $d\gamma_{R}/d\alpha=0$ numerically for $\alpha_{\max}$
which in turn is substituted into equation (\ref{eq:c03c13gameqn})
to obtain $\gamma_{R\max}$.

\begin{figure}
\includegraphics[clip,width=0.95\textwidth]{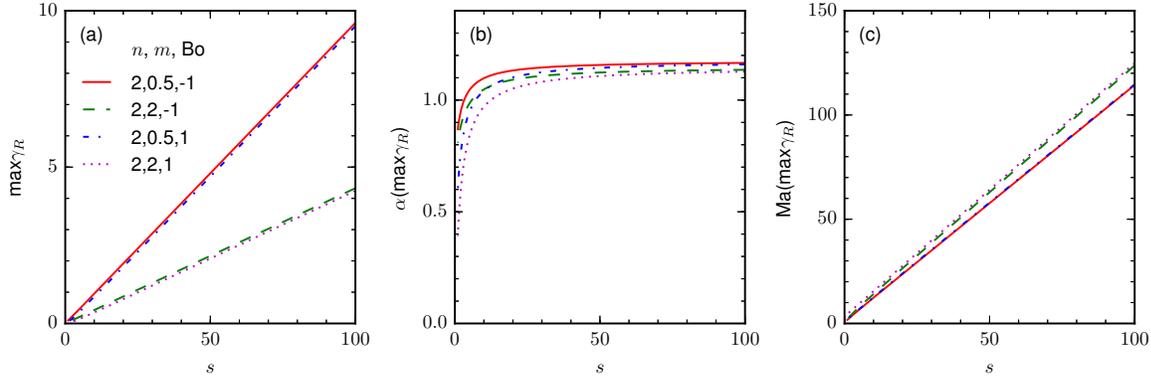} \protect\caption{The influence of $s$ on (a) the maximum of $\gamma_{R\text{max}}$
over all $\alpha$ and $\text{Ma}$ in the $R$ sector (at $n=2$
and $m=2$) and the $S$ sector (at $n=2,\;m=0.5$) for two different
values of $\text{Bo}$ as indicated in the legend. Panels (b) and
(c) show the corresponding $\alpha$ and $\text{Ma}$. \label{fig:fig13abc}}
\end{figure}

Figure \ref{fig:fig13abc} shows the results of varying the shear
parameter $s$. For any fixed $s$, the growth rate has a global maximum
over the $(\alpha,\textrm{Ma})$-plane, denoted $\max\gamma_{R}$.
We denote $\alpha(\max\gamma_{R})$ and $\textrm{Ma}(\max\gamma_{R})$
the values of the wavenumber and Marangoni number, respectively, at
which the growth rate attains its maximum, $\max\gamma_{R}$. These
quantities are plotted versus $s$ in figure \ref{fig:fig13abc},
for selected sampling points in the $R$ and $S$ sectors. 
We see that while in panels (a) and (c) the dependencies are linear,
and also practically independent of the Bond number, this does not
hold for the $\alpha(\max\gamma_{R})$ shown in panel (b); in particular,
in all four cases shown there, it stays almost constant (of magnitude
order 1) at large $s$ but falls off precipitously to zero as $s\downarrow0$.

In this subsection we only had to deal with the long-wave instability
because the values of $\text{Ma}$ considered are either sufficiently
large or sufficiently small, or the viscosity ratio was not sufficiently
close to the $R-Q$ boundary $m=n^{2}$. It turns out that for the
intermediate values of $\text{Ma}$ and the appropriate values of
$m$, even in the $R$ sector, a different type of instability, called
the ``mid-wave'' instability (HF), may happen. Its definition is
recalled in the next subsection where the $Q$ sector is considered,
since this instability is more prevalent there. Some results on the
mid-wave instability in the $R$ sector are found in section \ref{subsec:MA-Bo plane stability}
together with similar results for the $Q$ sector. In the $S$ sector,
the mid-wave instability sometimes coexists with the long-wave instability
of the robust mode. However, as far as we have observed, it is always
weaker than the long-wave instability of the surfactant branch there.
This is also discussed in section 6.

\subsection{Surfactant effects in the $Q$ sector \label{subsec:Surfactant-effects-in-Q-Sector}}

\begin{figure}
\includegraphics[bb=0bp 0bp 576bp 576bp,clip,width=0.95\textwidth]{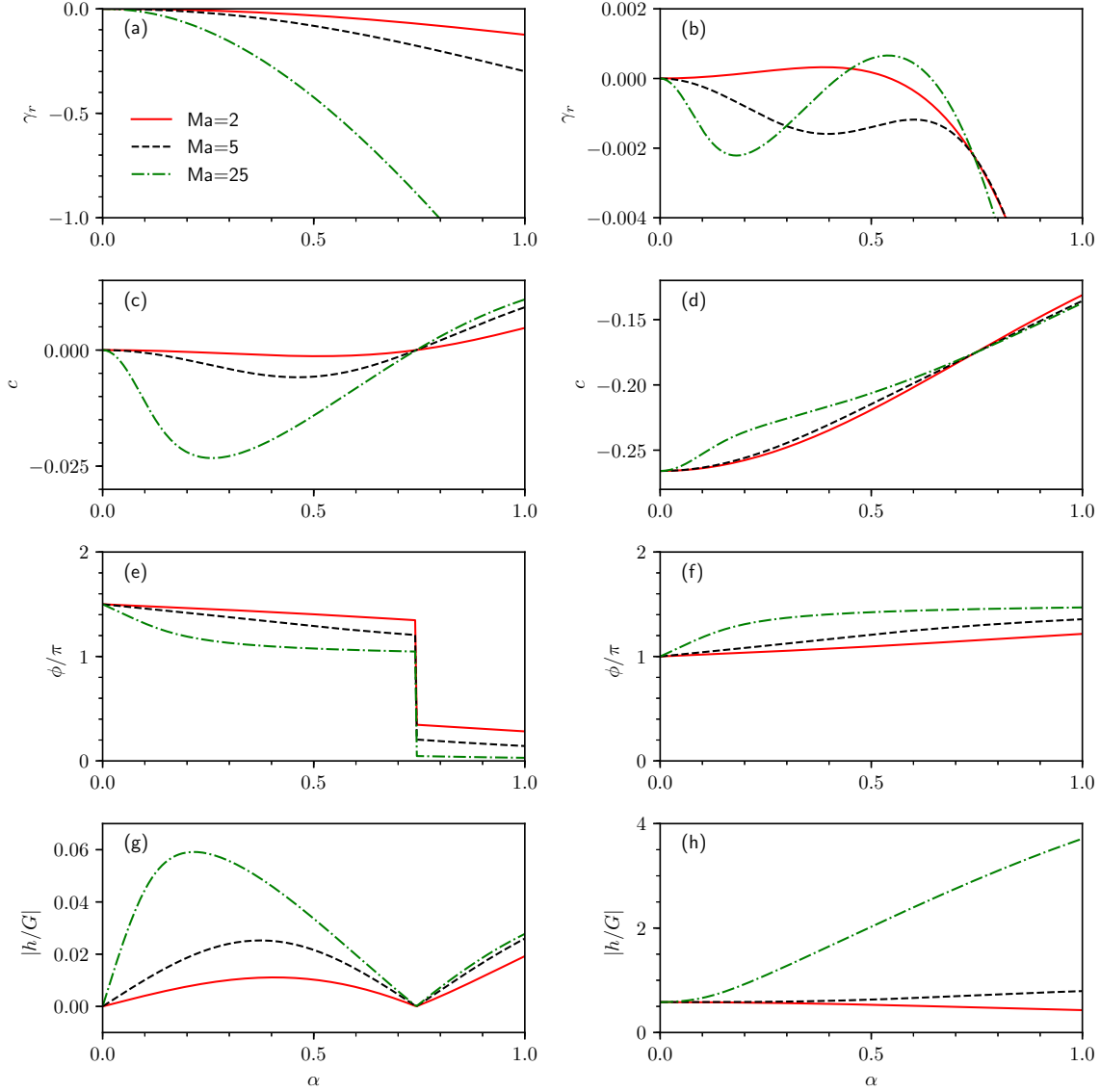}
\protect\caption{Curves for the four different functions of the wavenumber in the $Q$
sector ($n=2$, $m=5$) for $s=1$ and $\Bo=-0.45$. The stable mode
corresponds to the panels of the left-hand column, and the less stable
mode to the panels of the right-hand column. For the three values
of $\Ma$ given in the legend, panels (a) and (b) show the growth
rates, (c) and (d) show the wave velocities, (e) and (f) show the
interface-surfactant phase shifts, and (g) and (h) show the interface/surfactant
amplitude ratio. The transition from the long-wave instability to
stability to the mid-wave instability as $\Ma$ increases is evident
in panel (b).\label{fig:Fig_typical_midwaveDCurve}}
\end{figure}

It was shown in HF (for $\textrm{Bo}=0$) that for $\Ma>5/2$ and
$m>n^{2}$ ($Q$ sector), there is a mid-wave instability such that
$\gamma_{R}>0$ for a finite $\alpha$-interval bounded away from
$\alpha=0$. (Note that the mid-wave instability was called type I
in \citet{Cross1993} while the long-wave instability was called type
II). In order to investigate such an instability allowing for nonzero
Bond numbers, we introduce a critical Marangoni number, $\Ma_{cM}$
that corresponds to the onset (or the turnoff) of the mid-wave instability,
and let $\alpha_{cM}$ be the corresponding wavenumber. Thus, the
quantities $\Ma_{cM}$ and $\alpha_{cM}$ satisfy the equations $\gamma_{R}=0$
and $\partial\gamma_{R}/\partial\alpha=0$. In view of the quartic
equation (\ref{eq:MAXeqn}), $\Ma_{cM}$ and $\alpha_{cM}$ (for a
given $\Bo$) can be found by numerically solving simultaneously equation
(\ref{eq:MaEquation2}), which we write in the notation used in equation
(\ref{eq:alphaeq:MAXeqn}), 
\begin{equation}
C_{0}(\text{Ma},\alpha,\text{Bo})=0,\label{eq:C_0}
\end{equation}
along with 
\begin{equation}
\frac{\partial}{\partial\alpha}C_{0}(\text{Ma},\alpha,\text{Bo})=0.\label{eq:C_0' is zero}
\end{equation}
To illustrate the change of stability with $\Ma$, in the top panels
of figure \ref{fig:Fig_typical_midwaveDCurve}, the growth rate in
the $Q$ sector (for $n=2$ and $m=5$, at $s=1$) is plotted for
three selected values of the Marangoni number and $\text{Bo}=-0.45$.
The numerical results show that the instability is long-wave provided
$\text{Ma}<\text{Ma}_{cL}$ ($\approx2.28$ for the figure parameters). This
is then followed by a region of stability when $\Ma\in\left[\text{Ma}_{cL},\text{Ma}_{cM}\right]$,
where $\Ma_{cM}$ $\approx15.6$. For $\Ma_{cL}\,<\Ma<\Ma_{m}$, $\gamma_{R}$
decreases monotonically with $\alpha$ (so that there is no $\gamma_{R\text{max}}$;
such dispersion curves are not shown in the top right panel), but
starting from the $\Ma_{m}$ ($\approx3.70$), the local maximum $\gamma_{R\max}$
appears on the dispersion curves. So, the growth rate $\gamma_{R}$
has a local maximum $\gamma_{R\max}$ at some $\alpha_{max}>0~$ provided
$\Ma\geq\Ma_{m}$; and once $\Ma$ exceeds $\Ma_{cM}$, $\gamma_{R\text{max}}$
becomes positive, i.e., the mid-wave instability switches on. Note
that when $\Ma>\Ma_{cM}$ for at least some interval of $\text{Ma}$
corresponding to the mid-wave instability, there are two positive
marginal wavenumbers, one on the left at $\alpha=$ $\alpha_{0L}$
and another one on the right at $\alpha=$ $\alpha_{0R}$ so that
the interval of unstable wavenumbers is $\alpha_{0L}<\alpha<\alpha_{0R}$.
(Cases with both finite and infinite $\text{Ma}$ intervals of mid-wave
instability can be seen below in figure \ref{fig:Fig_noses}(a) and
are discussed in the last paragraph of section 6.3.)

Although the stability properties of the normal modes are fully given
by the dispersion curves (see panels (a) and (b) of figure \ref{fig:Fig_typical_midwaveDCurve}),
the normal modes have additional remarkable properties, such as the
phase speed, the phase difference between the co-traveling waves of
the interface and the surfactant, and the amplitude ratio of the interface
to the surfactant disturbances. As an example, these quantities are
plotted in figure \ref{fig:Fig_typical_midwaveDCurve} as functions
of the wavenumber $\alpha$. There, one notices a special value of
the wavenumber, $\alpha_{s}$, close to 0.7, at which the phase shift
of the decaying branch has a jump discontinuity. The wave speed at
$\alpha_{s}$ is zero for any $\Ma$, so all three curves intersect
at the same point $(\alpha_{s},0)$; similarly, the amplitude ratio
is zero, independent of $\text{Ma}$. For the other branch, in the
right panels (which, as panel (b) shows, goes, as $\text{Ma}$ increases,
from long-wave unstable, to stable and then to mid-wave unstable),
all three growth rates are equal at the same $\alpha_{s}$, and the
wave speeds are equal as well, but the amplitude ratios are non-zero
and different.

To explain these observations, note that the zero amplitude ratio
implies that if $h=0$ and $G\ne0$, then from the first equation
of (\ref{eq:dispEqnSystem}) $A_{12}=0$. Its solution, with the explicit
expression of $A_{12}$ from (\ref{eq:A12}), yields $\alpha_{s}$
in terms of $n$ and $m$ (but independent of $\Ma$). The second
equation of (\ref{eq:dispEqnSystem}) with $h=0$ yields $\gamma=-A_{22}$,
which by (\ref{eq:A22}), is real, negative, and proportional to $\text{Ma}$.
This agrees with the left upper panel of figure \ref{fig:Fig_typical_midwaveDCurve}.
The wave speed is zero because $\textrm{Im}(\gamma)=0$. The other
mode corresponds to the right panels of this figure, and must have
$h\ne0$. Since $A_{12}=0$ for $\alpha=\alpha_{s}$, we must have
$\gamma+A_{11}=0$. This implies $\text{Im}(\gamma)>0$, i.e. a negative
wave speed value, independent of $\text{Ma}$, corresponding to the
triple intersection in panel (d) of figure \ref{fig:Fig_typical_midwaveDCurve}.
The growth rate, $\gamma_{R}=-\text{Re(}A_{11})$, is seen to be negative
and independent of $\text{Ma}$, which explains the triple intersection
in panel (b). However, since $h\ne0$ for this branch, the amplitude
ratio is found to be 
\[
\left|\frac{h}{G}\right|=\left|\frac{A_{11}-A_{22}}{A_{21}}\right|.
\]
Only $A_{22}$ depends on $\text{Ma}$, and $|h/G|$ changes with
$\Ma$, so the three curves in figure \ref{fig:Fig_typical_midwaveDCurve}
go through different points at $\alpha=\alpha_{s}$.

Having noticed the existence of the normal modes in which the surfactant
is disturbed, $G\neq0$, but the interface is undisturbed, $h=0$,
the question arises if there exist some ``opposite'' modes, in which
only the interface, but not the surfactant is disturbed, so that $G=0$,
but $h\neq0$. We answer this question in Appendix \ref{sec:Are-there-normal}.
It turns out that such modes are possible, but only when $s=0$.

\begin{figure}
\includegraphics[clip,width=0.95\textwidth]{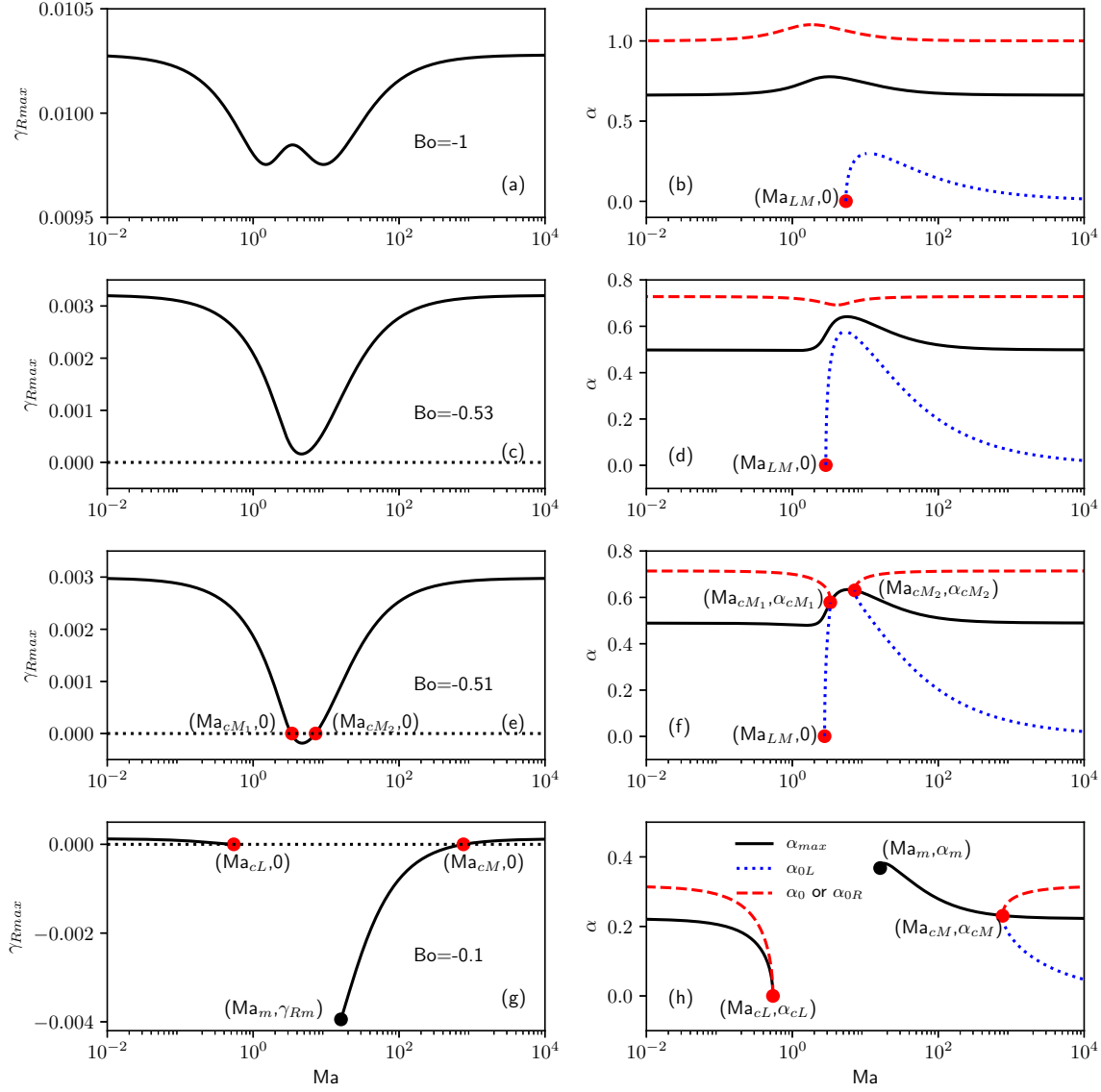} \protect\caption{Plots of $\gamma_{R\max}$ (left-hand panels) and corresponding $\alpha_{\max}$,
$\alpha_{0R}$, and $\alpha_{0L}$ (right-hand panels) vs $\Ma$ in
the $Q$ sector (here at $n=2$, $m=5$) for $s=1$ and the four indicated
values of $\Bo$. (For labeled points, see the text.)\label{fig:FIG_RTgamNalphas2by4}}
\end{figure}

In figure \ref{fig:FIG_RTgamNalphas2by4}, $\gamma_{R\text{max}}$,
$\alpha_{\text{max}}$ and $\alpha_{0}$ are plotted versus the Marangoni
number for $n=2$, $m=5$, $s=1$ and for four selected values of
$\text{Bo}$. If $\text{Bo}$ is sufficiently negative, as in panels
(a) and (c), then $\gamma_{R}>0$ for all $\text{Ma}$. 
For $\text{Ma}<\text{Ma}_{LM}$, the instability is long-wave, in
other words, there is no $\alpha_{0L}$, since its definition implies
that $\alpha_{0L}$ must be non-zero. However, a mid-wave instability
ensues when $\text{Ma}>\text{Ma}_{LM}$, and there appears $\alpha_{0L}>0$
(as in panels (b) and (d)). Initially, $\alpha_{0L}$ increases rapidly,
while $\alpha_{0R}$ decreases by a small amount, leading to the shrinkage
of the interval of unstable wavenumbers. After reaching a maximum,
$\alpha_{0L}$ decreases towards zero with increasing $\text{Ma}$
but never attains the zero value so that the instability does not
return to the long-wave type, and the interval of unstable wavenumbers
slowly expands. When $\text{Bo}=-0.51$ (see panels (e) and (f)),
the stability picture up to $\text{Ma}=\text{Ma}_{cM_{1}}$ is very
similar to that displayed in panels (b) and (d). The instability is
long-wave provided $\text{Ma}<\text{Ma}_{LM}$. Starting at $\text{Ma}=\text{Ma}_{LM}$,
corresponding to the lower left dot in panel (f), the long-wave instability
disappears, and the mid-wave instability mentioned previously emerges.
However, as $\text{Ma}$ continues to increase, the interval of unstable
wavenumbers quickly shrinks to a single, non-zero, $\alpha$ point,
indicated by the dot at $\text{Ma}=\text{Ma}_{cM_{1}}$. The flow
then becomes stable, with $\gamma_{R}<0$ for a range of Marangoni
numbers, $\text{Ma}_{cM_{1}}<\text{Ma}<\text{Ma}_{cM_{2}}$. Therefore,
in this range, $\alpha_{0L}$ and $\alpha_{0R}$ are non-existent,
but $\alpha_{\text{max}}$ is defined because $\gamma_{R}$ has a
local maximum at a nonzero $\alpha$. The mid-wave instability reappears
at $\text{Ma}_{cM_{2}}$, (see the right-most dot in panel (f)) starting
from $\gamma_{R}=0$, which corresponds to the right-hand intersection
point in panel (e). As $\text{Ma}$ increases beyond $\text{Ma}_{cM_{2}}$,
the interval of unstable wavenumbers expands in both directions. In
the final set of panels, (g) and (h), with $\text{Bo}=-0.1$, the
flow is stable, and $\gamma_{R\text{max}}$, $\alpha_{\text{max}}$,
and $\alpha_{0}$ do not exist, in the interval $\text{Ma}_{cL}\le\text{Ma}\le\text{Ma}_{m}$.
This is because $\gamma_{R}$ has no local maximum at any $\alpha>0$.
Note that, as with the previous set of panels, the flow is long-wave
unstable for $\text{Ma}<\text{Ma}_{cL}$ (i.e., to the left of the
left-most dot of panel (h)) and mid-wave unstable for $\text{Ma}>\text{Ma}_{cM}$
(to the right of the right-most dot).

Thus, we have observed here, for the first time, the existence of
another route to the mid-wave instability: the continuous transition
from long-wave instability (see the marked point $(\text{Ma}_{LM},0)$ 
in panel (f) of figure \ref{fig:FIG_RTgamNalphas2by4}). Only the other route, the onset of mid-wave
instability from stability, was present for the case of zero gravity
(see HF). In the former scenario, the mid-wave instability has a non-zero
growth rate and a final support interval from the very beginning.
A detailed investigation of the boundaries between the domains of
the mid-wave instability, long-wave instability and stability in the
$(\text{Ma},\text{Bo})$-plane appears below in section 6.

\begin{figure}
\includegraphics[clip,width=0.95\textwidth]{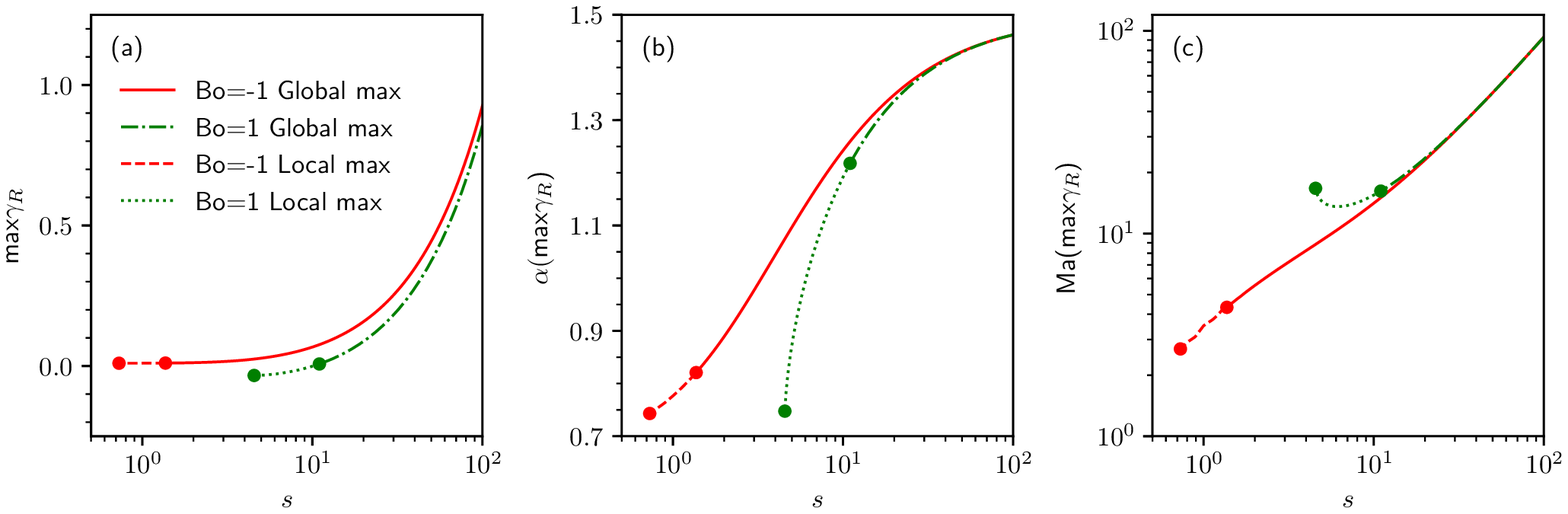} \protect\caption{The influence of $s$ on (a) the maximum of $\gamma_{R}$ over all
$\alpha$ and $\text{Ma}$ in the $Q$ sector (here at $n=2,\;m=5$)
for two different values of $\text{Bo}$, one positive and the other
one negative. Panels (b) and (c) show the values of $\alpha$ and
$\text{Ma}$ at which this maximum occurs. The global maxima of $\gamma_{Rmax}$
with respect to $\text{Ma}$, present at larger $s$, become local
maxima between the pairs of dots on each curve. At smaller $s$, to
the left of the end dot on each curve, there are neither global nor
local maxima.\label{fig:fig16abc}}
\end{figure}

Figure \ref{fig:fig16abc} shows the dependencies of $\text{max}\;\gamma_{R}$,
$\alpha(\text{max}\gamma_{R})$ and $\text{Ma}(\text{max}\gamma_{R})$
on the shear parameter $s$ in the $Q$ sector similar to those shown
in figure \ref{fig:fig13abc} for the other two sectors. We observe
that the existence of the global maximum in $\text{Ma}$ of the growth
rate maxima with respect to the wavenumber is less common in the $Q$
sector, especially for $\text{Bo}>0$. At smaller values of $s$,
the global maximum becomes a local one like the one in figure \ref{fig:FIG_RTgamNalphas2by4}(a).
This is indicated in figure \ref{fig:fig16abc} as the change from
the solid to the dashed curve at the negative $\text{Bo}$ and from
the dashed to the dotted one at the positive $\text{Bo}$. At still
smaller $s$, to the left of the end dot on each curve, there are
neither global nor local maxima.

\section{$\text{(Ma},\text{Bo)}$-plane stability diagrams\label{subsec:MA-Bo plane stability}}

\subsection{Regions of the long-wave and mid-wave instabilities}

\begin{figure}
\centering{}\includegraphics[bb=0bp 0bp 576bp 576bp,clip,width=0.95\textwidth]{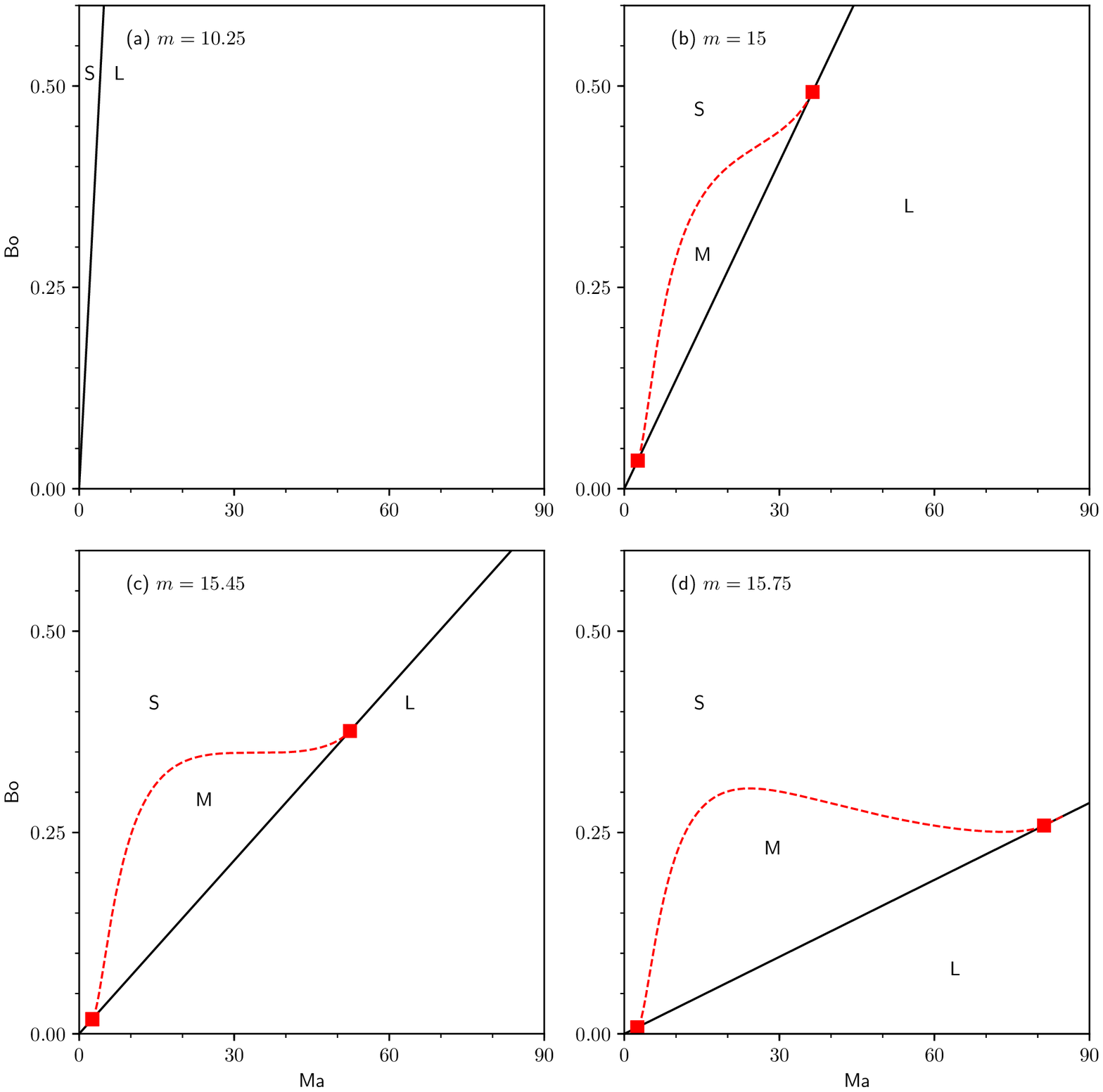}
\protect\caption{Stability diagrams in the $(\Ma,\Bo)$-plane showing the influence
of the viscosity ratio $m$ as $m\uparrow n^{2}$: (a) $m=10.25$;
(b)\ $m=15$; (c) $m=15.45$; and (d) $m=15.75$. The solid and dashed
curves represent long-wave and mid-wave instability boundaries respectively;
S, L, and M denote the stable, long-wave unstable, and mid-wave unstable
regions. Here $s=1$ and $n=4$. \label{fig:figMidwaveR1m_near_n2_n4s1}}
\end{figure}

Here we present a detailed account of the mid-wave instability changes
as the viscosity ratio is increased, starting from a value in the
$R$ sector, $1<m<n^{2}$, then crossing the $m=n^{2}$ border and
further growing in the $Q$ sector, $m>n^{2}$. In the $R$ sector,
the robust branch is long-wave unstable provided $\Bo<\Bo_{cL}$ where
$\Bo_{cL}(\Ma)$, as given by (\ref{eq:BoCritical}), is positive.
If $m<n^{2}$ and sufficiently far from the $m=n^{2}$ border, there
exists just one stability boundary, given by $\Bo=\Bo_{cL}$; it is
a straight line (starting at the origin) that separates the long-wave
unstable and stable regions, as shown in figure \ref{fig:figMidwaveR1m_near_n2_n4s1}(a).
As $m$ increases and gets sufficiently close to $m=n^{2}$, the onset
of a mid-wave instability is observed for certain intervals of $\Ma$
and $\Bo$. In panels (b) and (c), a mid-wave instability occurs provided
$\Bo_{cL}<\Bo\,<\Bo_{cM}$, for a finite interval of the Marangoni
numbers, $\Ma_{LM1}<\Ma<\Ma_{LM2}$, as the $\Bo_{cL}$ and $\Bo_{cM}$
curves ``intersect'' each other at $\Ma=\Ma_{LM1}$ and $\Ma=\ \Ma_{LM2}$.
The ``quasi-intersection'' points, marked in the figure as filled
squares, are the boundary points for the critical curve but are not
the critical points themselves: the critical wavenumber decreases
to zero as $\Ma\rightarrow\Ma_{LMj}$, but the zero value is prohibited
for a critical wavenumber. When $m$ is approaching ever closer to
$n^{2}$, at some $m$ the critical curve of the mid-wave instability
acquires a maximum and a minimum, such as the ones in panel (d). Clearly,
for each fixed $\Ma$ of the $\Ma$-interval $\Ma_{LM1}<\Ma<\Ma_{LM2}$,
there are three distinct $\Bo$-intervals: a semi-infinite interval
of stability $\Bo>\Bo_{cM}$; a finite interval of mid-wave instability
$\Bo_{cL}<\Bo<\Bo_{cM}$; and a semi-finite interval of long-wave
instability $\Bo<\Bo_{cL}$.

\begin{figure}
\centering{}\includegraphics[clip,width=0.95\textwidth]{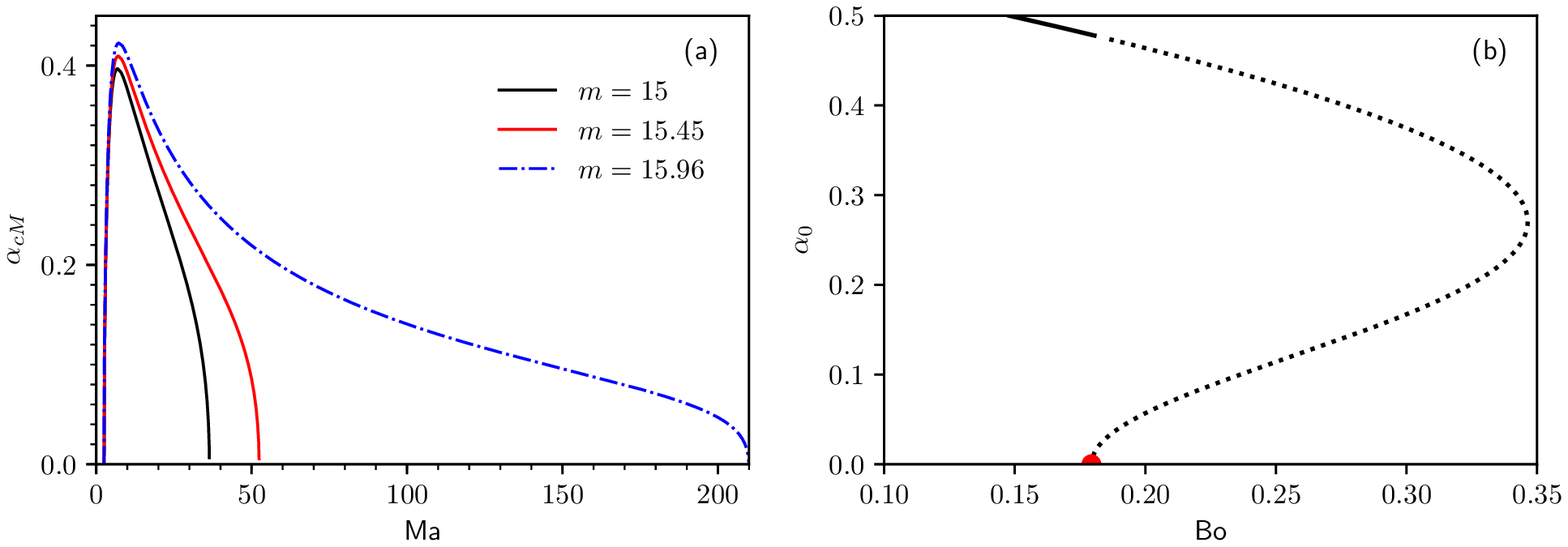}
\protect\caption{(a) The critical wavenumber $\alpha_{cM}$ versus Marangoni number
$\Ma$ for the same parameter value choices as in figure \ref{fig:figMidwaveR1m_near_n2_n4s1};
in particular, $n=4$ and $s=1$. (b) The marginal wavenumber $\alpha_{0}$
versus Bond number $\textrm{Bo}$ for $s=1$, $\textrm{Ma = 25}$,
$n=4$, and $m=15.45$. There is mid-wave instability in the region
bounded by the two semicircles on the horizontal axis, long-wave instability
to the left of this region, and stability to the right of this region.
\label{fig:figMidwaveAlphasR1m_near_n2_n4s1} }
\end{figure}

In figure \ref{fig:figMidwaveAlphasR1m_near_n2_n4s1}(a), the wavenumber
$\alpha_{cM}$ corresponding to $\Bo_{cM}$ is plotted versus $\Ma$
for the values of $m$ corresponding to panels (b) and (c) of figure
\ref{fig:figMidwaveR1m_near_n2_n4s1}, and also for $m=15.96$, which
is closer to the $m=n^{2}$ boundary value, $m=16$, than $m=15.75$
of figure \ref{fig:figMidwaveR1m_near_n2_n4s1}(d). With this, \ref{fig:figMidwaveAlphasR1m_near_n2_n4s1}(a)
suggests the hypothesis that in approaching the sector boundary, the
larger quasi-intersection value of $\textrm{Ma}$ tends to infinity.
The latter is in accordance with the stability diagram for the sector
boundary value $m=16$ (see figure \ref{fig:Fig_noses_16} below).
For all these cases, $\alpha_{cM}$ attains a maximum at an $\Ma$
such that $\Ma_{LM1}<\Ma<\Ma_{LM2}$. Figure \ref{fig:figMidwaveAlphasR1m_near_n2_n4s1}(b)
shows, for the parameters of figure \ref{fig:figMidwaveR1m_near_n2_n4s1}(c)
and $\textrm{Ma}=25$, that, as the Bond number grows, when it reaches
the value $\Bo_{cL}$, the long-wave instability changes into the
mid-wave one by the left endpoint of the interval of unstable $\alpha$
departing from the zero $\alpha$ point. The unstable $\alpha$ interval
continues to shrink from both ends, and finally becomes a single non-zero
$\alpha$ point at $\textrm{Bo=Bo}_{cM}$, the right-most point on
the curve. The maximum growth rate (not shown) decreases to zero at
this point, and there is stability for larger $\textrm{Bo}$, in agreement
with figure \ref{fig:figMidwaveR1m_near_n2_n4s1}(c). 
\begin{figure}
\includegraphics[clip,width=0.95\textwidth]{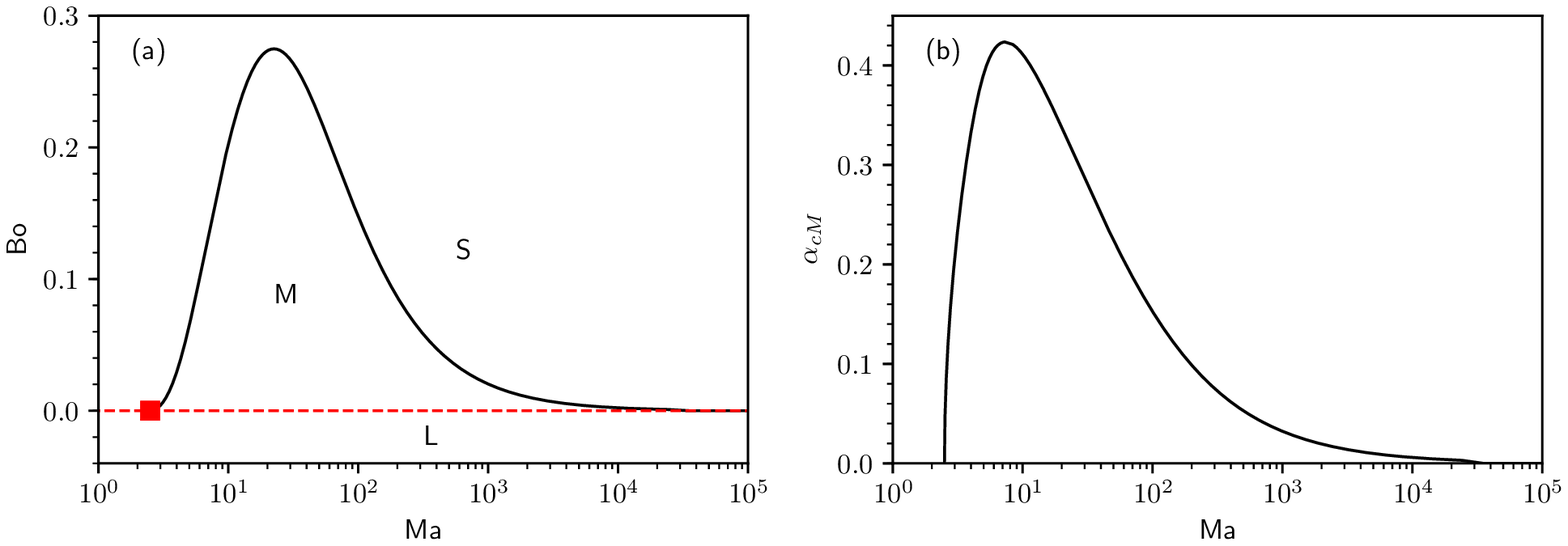} \protect\caption{(a)\ Stability diagram in the ($\text{Ma}$$,\text{Bo}$)-plane similar
to the ones shown in figure \ref{fig:figMidwaveR1m_near_n2_n4s1},
for a case where $m=n^{2}$ (here $m=16$) and (b) the corresponding
critical wavenumber, $\alpha_{cM}$. The end points have $\textrm{Ma}=5/2$.
Here $s=1$. \label{fig:Fig_noses_16}}
\end{figure}
On the $m=n^{2}$ border (e.g., for $m=16$, $n=4$), the robust branch
is long-wave unstable in the half-plane $\Bo<0$ (with the boundary
line $\Bo_{cL}=0$), as shown in figure \ref{fig:Fig_noses_16}. Along
the $\Ma$-axis ($\Bo=0$), the stability results of HF that show
the existence of a mid-wave instability for $\Ma>5/2$ are recovered:
$\Ma_{LM1}$ = 5/2 and $\Ma_{LM2}$ $=\infty$. Notably, $\Bo_{cL}(\Ma)\downarrow0$
as $\Ma\rightarrow\infty$. We also note that there is just a single
extremum, a maximum, on the critical curve. 
\begin{figure}
\includegraphics[bb=0bp 0bp 612bp 216bp,clip,width=0.95\textwidth]{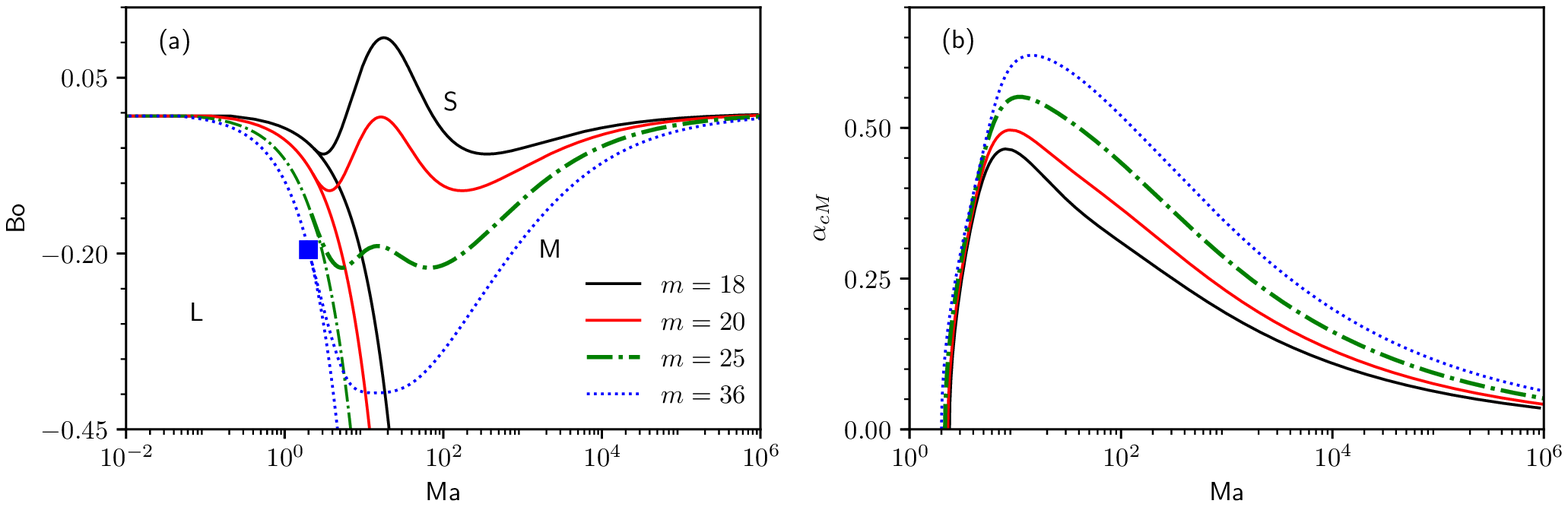}
\protect\caption{(a) Stability diagram showing the regions of mid-wave and long-wave
instability and stability defined by the curves $\Ma_{cL}$ and $\Ma_{cM}$
as $m$ increases in the $Q$ sector, and (b) the wavenumber corresponding
to $\Ma_{cM}$ for the indicated values of $m$. Here $n=4$ and $s=1$.
\label{fig:Fig_noses}}
\end{figure}
In the $Q$ sector, $\Bo_{cL}(\Ma)<0$, as given by (\ref{eq:BoCritical}).
We see the threshold curves $\textrm{Bo}=\text{Bo}_{cL}(\textrm{Ma)}$
in figure \ref{fig:Fig_noses}, for each value of $m$ represented
there; all the threshold curves have the ($\Bo$, $\Ma$)-origin as
their left-hand end (with linear scales on both axes, all the threshold
lines would start from the origin and have a negative slope). The
long-wave instability occurs below each threshold curve; the region
of long-wave instability is labeled with an ``L'' in the figure.
At some point on each L-threshold curve, the critical curve of the
mid-wave instability begins, going unbounded rightward, in the direction
of increasing $\Ma$; as $\Ma\uparrow\infty$, each critical curve
is asymptotic to $\textrm{Bo}=0$ (thus, in a difference with the
$R$ sector, but similar to the boundary between the $R$ and $Q$
sectors, the threshold line of the long-wave instability intersects
the critical curve of the mid-wave instability at a single point);
however, in contrast with the boundary between the $R$ and $Q$ sectors,
the critical curve approaches the axis $\textrm{Bo}=0$ from below.
Also, at the threshold-critical quasi-intersection, the $\Bo_{cL}(\Ma)$
increases as $\Ma\downarrow\Ma_{_{LM1}}$. Since there is still a
local maximum on the critical curve, just as there is one in the $R$
sector and on the inter-sector boundary, it follows that there must
be at least two local minima as well.

The mid-wave instability occurs below such a critical curve $\text{Bo}=\text{Bo}_{cM}$
(and above, or to the right of, the right-hand part ($\Ma>\Ma_{LM1}$)
of the corresponding threshold curve $\textrm{Bo}=\text{Bo}_{cL}(\textrm{Ma)}$).
This region is labeled with an ``M''. Above the critical curve,
as well as above the left-hand part ($\Ma<\Ma_{LM1}$) of the corresponding
threshold curve, the flow is stable. The critical curve is given by
a single-valued function $\textrm{Bo}=\textrm{Bo}_{cM}\textrm{(Ma)}$,
that is seen in figure \ref{fig:Fig_noses} to have two local minima
and a maximum in between them, provided the viscosity ratio $m$ is
below a certain value $m_{N}$. These two minima appear to occur at
the same value of $\Bo$, and as $m$ increases all three extrema
move downward, but the single maximum moves faster than the two minima.
Eventually, at $m=m_{N}$, the three extrema merge into a single minimum,
such as the one on the $m=36$ critical curve.

\begin{figure}
\includegraphics[bb=0bp 0bp 576bp 576bp,clip,width=0.8\textwidth]{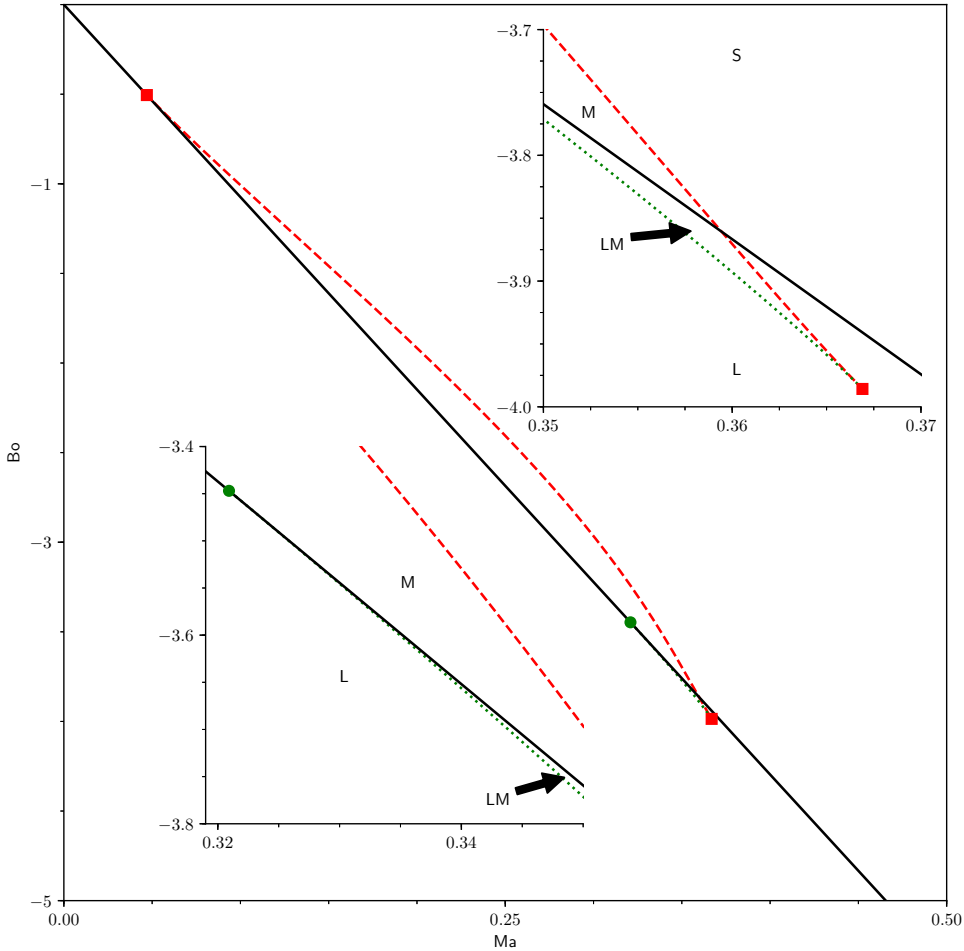}
\caption{(a) Stability diagram of the less ustable mode for $s=1$, $n=10$,
and $m=0.1$. The long-wave instability is present below the solid
line and absent above it, while the mid-wave instability is present
between the dashed curve, and either the solid line line or the dotted
curve. The top and bottom insets zoom in on the regions near the two
lower pairwise intersections, marked by the square and the circle,
respectively.}
\label{fig:S midwave_mp1n10s1} 
\end{figure}

\begin{figure}
\includegraphics[bb=0bp 0bp 576bp 576bp,clip,width=0.95\textwidth]{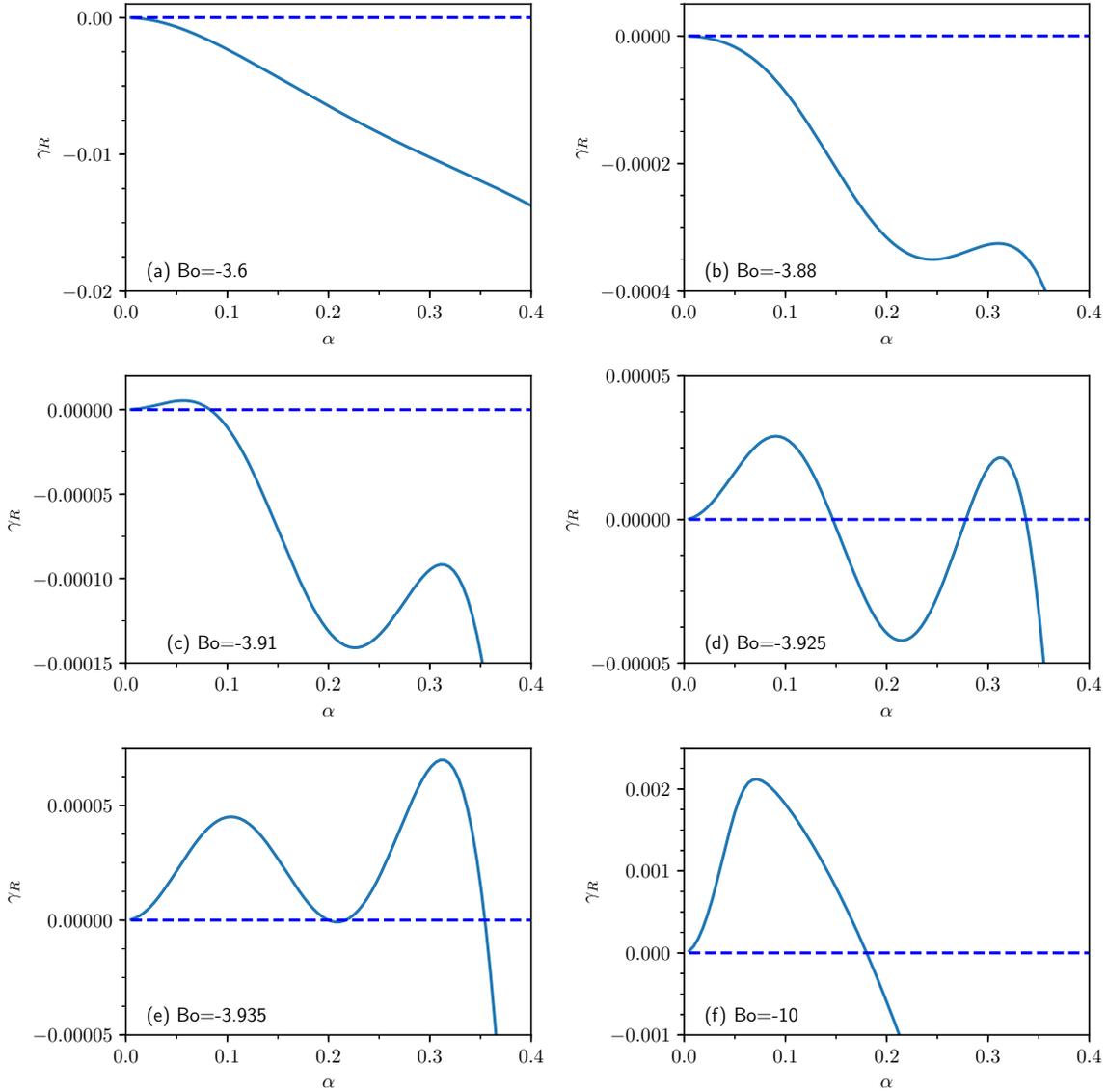}
\caption{Dispersion curves for the robust mode in the $S$- sector. Here $n=10$,
$m=0.1$, $s=1$, and $\text{Ma}=0.363$. The values of $\text{Bo}$
are as indicated in each panel.}
\label{fig:S dispersion curves-1} 
\end{figure}

In the $S$ sector, as was mentioned at the end of section 5.2, the
mid-wave instability occurs for the robust mode, although it is overshadowed
by the long-wave instability of the surfactant mode. It is illustrated
in figure \ref{fig:S midwave_mp1n10s1} for the parameter values indicated
there. 
The zoom-in, the upper inset, shows that, in contrast with the other
sectors, the critical curve does not end at its intersection with
the threshold curve of the linear instability, but continues below
the intersection, until it meets another critical curve. On the latter
curve, each point corresponds to a dispersion curve having zero growth
rate at a local minimum (as will be illustrated in the next figure).
The bottom inset of figure \ref{fig:S midwave_mp1n10s1} is a zoom-in
near the quasi-intersection point of the lower critical curve and
the threshold line, marked by a small circle, located at $\text{Ma}$
slightly above 0.32 and $\text{Bo}$ slightly above -3.45. The quasi-intersection
point of the upper critical curve and the threshold line, marked by
a small square, is located at $\text{Ma}$ slightly above 0.046 and
$\text{Bo}$ slightly below -0.5. Figure \ref{fig:S dispersion curves-1}
illustrates the change of the dispersion curves of the robust mode
for the same values of $n$, $m$ and $s$ as in figure \ref{fig:S midwave_mp1n10s1},
and $\text{Ma}$ fixed at 0.363 for a decreasing sequence of $\text{Bo}$
values corresponding to moving in the upper inset of figure \ref{fig:S midwave_mp1n10s1}
from the domain of stability (\ref{fig:S dispersion curves-1}(a))
to long-wave instability (\ref{fig:S dispersion curves-1}(b)) to
the domain of coexisting long-wave and mid-wave instabilities (\ref{fig:S dispersion curves-1}(c))
to the lower critical curve (corresponding to the zero minimum in
figure \ref{fig:S dispersion curves-1}(d)) and finally to the domain
of long-wave instability (see panels (e) and (f) of figure \ref{fig:S dispersion curves-1}).
The mid-wave instability starts at a certain $\text{Bo}$ between
those of panels (b) and(c) as the maximum, which is negative in panel
(b), grows through the zero to positive values as in panel (c) near
$\alpha=0.3$. In this process both intervals of (co-existing) long-wave
instability and mid-wave instability expand, until they coalesce which
corresponds to the snapshot shown in panel (d). Also, in going from
panel (c) to panel (d), the local minimum increases from negative
to zero value, and becomes positive, as in panel (e). Finally, this
minimum disappears, and the dispersion has a single maximum, see panel
(f).

\begin{figure}
\includegraphics[clip,width=0.95\textwidth]{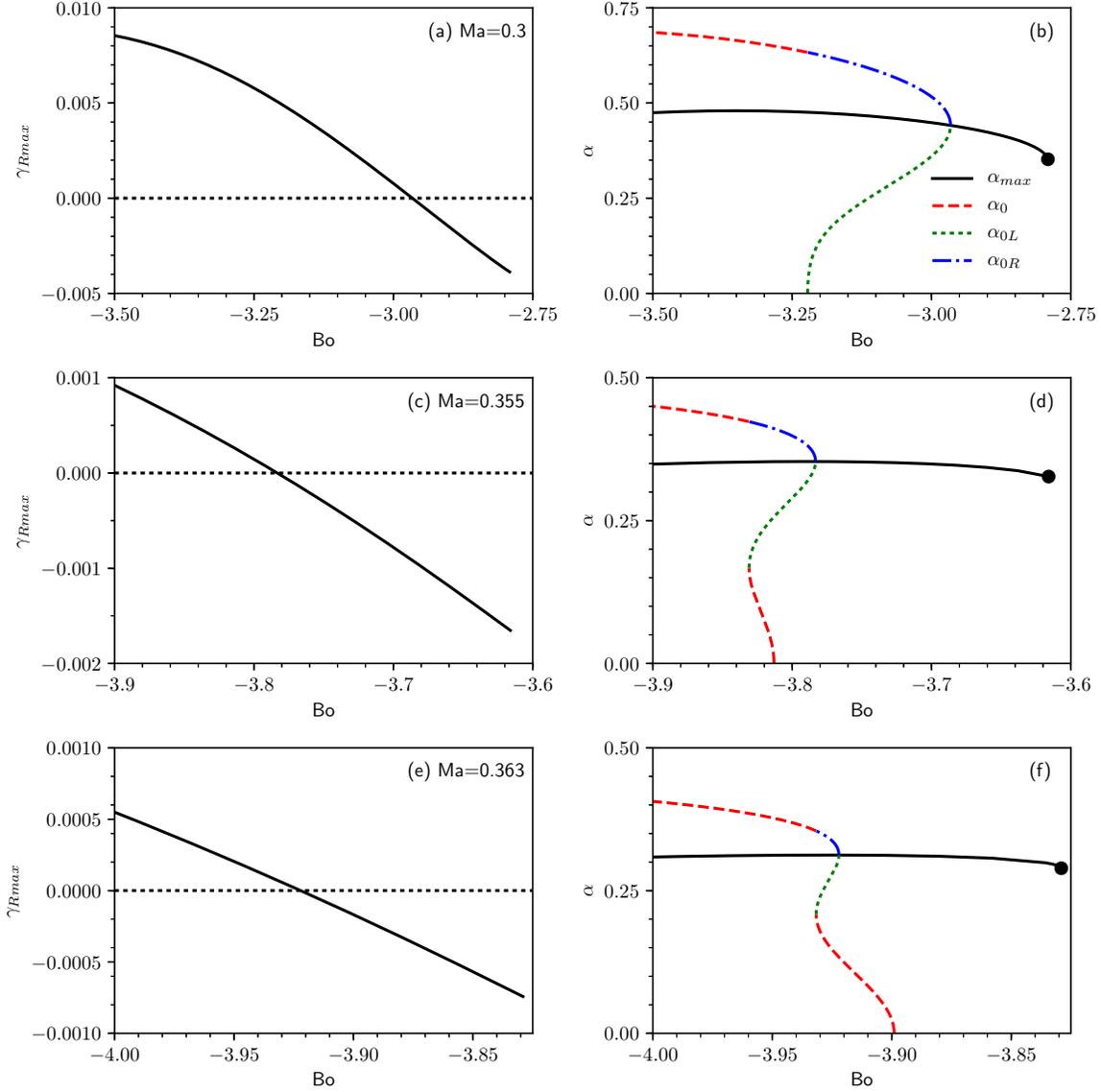} \caption{Plots of $\gamma_{R\max}$, corresponding $\alpha_{\max}$, and $\alpha_{0}$,
vs $\text{Bo}$ in the $S$ sector for the indicated values of $\textrm{Ma}$.
Here $n=10$, $m=0.1$ and $s=1$.Note that when there are two local
maxima on the dispersion curves, $\gamma_{Rmax}$ shown here corresponds
to the right-hand maximum even if it is smaller than the left one.}
\label{fig:S midwave Bo-alpha} 
\end{figure}

\begin{figure}
\centerline{\includegraphics[clip,width=0.6\textwidth]{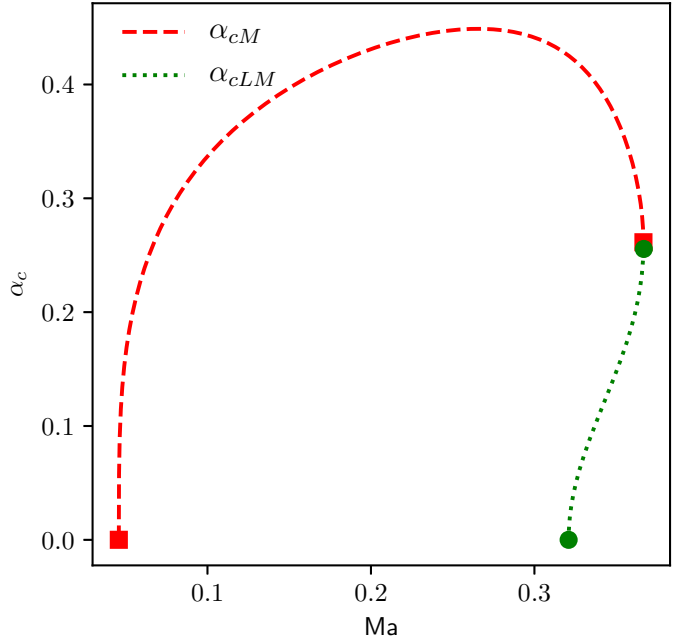}}
\caption{Critical $\alpha$ versus $\text{Ma}$ corresponding to the previous
figure.}
\label{fig: S critical alpha v Ma} 
\end{figure}

Figure \ref{fig:S midwave Bo-alpha} shows the salient features of
the dispersion curves, such as the maximum growth rate, $\gamma_{R\text{max}}$,
the corresponding wavenumber, $\alpha_{\text{max}}$, and the marginal
wavenumbers, $\alpha_{0}$, $\alpha_{0L}$ and $\alpha_{0R}$, as
continuous functions of the Bond number for three different values
of the Marangoni number. In particular, figure \ref{fig:S dispersion curves-1}
corresponds to panels (e) and (f) of figure \ref{fig:S midwave Bo-alpha}.
For smaller values of the Marangoni number, such as $\textrm{Ma}=0.355$
in panels (c) and (d), which are to the left of the intersection of
the (maximum) critical curve and the threshold curve, the mid-wave
instability emerges before the long-wave instability as the value
of $\text{Bo}$ becomes more negative (see the upper inset of figure
\ref{fig:S midwave_mp1n10s1}). For a small range of $\text{Bo}$,
both long-wave and mid-wave instabilities can coexist (indicated by
the label ``LM'' in the upper inset of figure \ref{fig:S midwave_mp1n10s1}).
This is then followed by a completely long-wave unstable regime. For
still smaller Ma, such as $\textrm{Ma}=0.3$, in panels (a) and (b),
we observe the emergence of the mid-wave instability, which, subsequently,
turns into a long-wave instability, similar to figure \ref{fig:figMidwaveAlphasR1m_near_n2_n4s1}(b).
Figure \ref{fig: S critical alpha v Ma} is the plot of the critical
wavenumber corresponding to the two critical curves in the preceding
figure. It shows, similar to the analogous figures for the other two
sectors, that the critical wavenumber, $\alpha_{c}$, approaches zero
at the quasi-intersection points. It also reveals that the rate of
change of the critical wavenumber approaches infinity at the common
point of the two critical curves. Using small wavenumber expansions
as described in section \ref{subsec:Asymptotics-of-the critical curves near boundaries},
we obtained the cubic equation given by (\ref{eq:cubic for Bo}) below,
and solved it numerically to verify that at the left quasi-intersection
point $\textrm{Ma=0.0458}$, and at the other one $\textrm{Ma=0.321}$.

\subsection{Asymptotics of the critical curves near their boundaries\label{subsec:Asymptotics-of-the critical curves near boundaries} }

\subsubsection{General considerations\label{subsec:General-considerations 6-2-1}}

It should be possible to establish the asymptotic behavior of the
critical curves near their boundaries, in particular, the sense of
the curve inclination at a finite boundary point, a priori, using
only minimal numerical information. This, as already was indicated
above, leads to certain conclusions about the number and sense of
possible extrema, that in their turn facilitate the complete determination
of the curve extrema. Near any finite quasi-intersection point, for
both $R$ and $Q$ sectors, we look for the critical point coordinates
in the form of generic power expansions 
\begin{equation}
\text{Ma}=\text{Ma}_{0}+\alpha^{2}\text{Ma}_{2}+\alpha^{4}\text{Ma}_{4}+...\label{eq:crit Ma expans}
\end{equation}
and 
\begin{equation}
\text{Bo}=\text{Bo}_{0}+\alpha^{2}\text{Bo}_{2}+\alpha^{4}\text{Bo}_{4}+...,\label{eq:crit Bo expans}
\end{equation}
where, to simplify notations, $\Ma_{0}$ stands for $\Ma_{LMj}$ (with
$j=1,2$), etc. We substitute these expansions into the critical curve
equations (\ref{eq:C_0}) and (\ref{eq:C_0' is zero}) and require
the collected coefficients of each power to vanish. Since the point
($\Ma_{0}$, $\Bo_{0}$) lies on the threshold curve of the long-wave
instability, we have $\Bo_{0}=\kappa\Ma_{0}$, where $\kappa$ is
the coefficient of $\Ma$ in equation (\ref{eq:BoCritical}). Because
of this relation, the leading orders $\alpha^{6}$ in (\ref{eq:C_0})
and $\alpha^{5}$ in (\ref{eq:C_0' is zero}) are satisfied identically.
The next order system, given by the orders $\alpha^{8}$ in (\ref{eq:C_0})
and $\alpha^{7}$ in (\ref{eq:C_0' is zero}), is 
\begin{eqnarray*}
k_{206}\text{Ma}_{2}+k_{116}\text{Bo}_{2} & = & r_{1},\\
6k_{206}\text{Ma}_{2}+6k_{116}\text{Bo}_{2} & = & 8r_{1},
\end{eqnarray*}
where the coefficients $k_{pqr}$ are functions of $(m,n,s)$ given
in Appendix B, and $r_{1}$ is a cubic polynomial in $\text{Ma}_{0}$
whose coefficients are known combinations of $k_{pqr}$, and which
lacks the quadratic term (cf. the discussion around equation (\ref{eq:cubic discrim-eq0})).
The consistency of this system requires that $r_{1}=0$, which is
a cubic equation for $\text{Ma}_{0}$. Clearly $\text{Bo}_{2}=-(k_{206}/k_{116})\text{Ma}_{2}$,
which simplifies to 
\begin{equation}
\text{Bo}_{2}=\kappa\text{Ma}_{2}.\label{eq:bo2 is kappa times ma2}
\end{equation}
The cubic equation for $\text{Ma}_{0}$ can be examined using the
well known Cardano's formula and the underlying theory for the case
with real coefficients.

The inclination of a critical curve at any quasi-intersection point
is 
\begin{equation}
d\text{Bo}/d\text{Ma}=(d\text{Bo}/d\alpha)/(d\text{Ma}/d\alpha).
\end{equation}
Using (\ref{eq:crit Ma expans}) and (\ref{eq:crit Bo expans}) we
get $d\text{Bo}/d\text{Ma}=\text{Bo}_{2}/\text{Ma}_{2}=\kappa$, where
we have used (\ref{eq:bo2 is kappa times ma2}). Thus, at the boundary
point, the critical curve is tangent to the threshold curve through
that quasi-intersection point.

\subsubsection{The $R$ sector finite critical curves and the threshold for their
existence \label{subsec:R sector critical curves}}

We find that in the $R$ sector the cubic equation for $\Ma_{0}$
has two distinct positive roots, corresponding to the two quasi-intersection
points, for $m$ greater than some threshold value $m_{d}$, and one
non-physical negative root. For $m=m_{d}$, the two positive roots
merge into a single double root, which means that the interval of
mid-wave instability shrinks to a single point, so that there is no
mid-wave instability for $m<m_{d}$. If the cubic equation is written
in the form $\text{Ma}_{0}^{3}+p\text{Ma}_{0}+q=0$, the condition
for the double root is that a certain discriminant is zero, or $27q^{2}+4p^{3}=0$,
whose solution for given $n$ and $s$ is $m_{d}$, the threshold
value above which the mid-wave instability exists. For example, when
$n=4$ and $s=1$, as in \ref{fig:figMidwaveR1m_near_n2_n4s1}, $m_{d}=10.2783$.
This value of $m$ is between those for the panels (a) and (b), as
it should be. Thus, one can predict also the location of the boundaries
of the critical curves in the $R$ sector. A somewhat different way
for this, leading to a cubic equation for $\Bo$, is as follows. A
more explicit form of the system (\ref{eq:C_0})-(\ref{eq:C_0' is zero})
is 
\begin{equation}
C_{0}(\textrm{Ma},\;\alpha,\;\textrm{Bo})=\textrm{Ma}(A_{1}+A_{2}\text{Ma}+A_{3}\text{Ma}^{2})=0,\label{eq:q_1is0}
\end{equation}
\begin{equation}
\frac{\partial C_{0}}{\partial\alpha}=\textrm{Ma}(A_{1}'\text{}+A_{2}'\text{Ma}+A_{3}'\text{Ma}^{2})=0\label{eq:q_2is0}
\end{equation}
where 
\begin{equation}
A_{1}=k_{11}B+k_{13}B^{3},\;A_{2}=k_{20}+k_{22}B^{2},\;A_{3}=k_{31}B,\label{eq:A_jdefinition}
\end{equation}
and the prime stands for the $\alpha-$derivative. Since $\text{Ma}>0$,
we divide equations (\ref{eq:q_1is0}) and (\ref{eq:q_2is0}) by $\text{Ma}$,
and then the system consists of two quadratic equations, from which
we obtain two different linear equations for $\text{Ma}$, one by
eliminating the quadratic term, and the other by eliminating the zero-power
term. The solvability condition, obtained by equating the two expressions
for $\text{Ma}$, is 
\begin{equation}
(A_{1}A_{3}^{'}-A_{1}^{'}A_{3})^{2}-(A_{1}A_{2}^{'}-A_{1}^{'}A_{2})(A_{2}A_{3}^{'}-A_{2}^{'}A_{3})=0.\label{eq:Ma crit solvability cond}
\end{equation}
Since $\alpha\downarrow0$ near a boundary point, we use the small-$\alpha$
expansions to find, to the leading order, the standard-form cubic
equation 
\begin{equation}
\textrm{Bo}^{3}g_{3}+\textrm{Bo}g_{1}+g_{0}=0,\label{eq:cubic for Bo}
\end{equation}
where the coefficients are defined as $g_{0}=k_{116}k_{206}^{2}$,
$g_{1}=k_{206}(k_{118}k_{206}-k_{116}k_{208})$, and $g_{3}=k_{116}^{2}k_{318}-k_{116}k_{206}k_{228}+k_{206}^{2}k_{138}$.
(One can see from the expressions for $k_{pqr}$ that here $g_{0}>0$
and $g_{3}>0$.) This cubic equation can be written in the standard
form $\text{Bo}^{3}+p_{1}\text{Bo}+q_{1}=0$, with $p_{1}=g_{1}/g_{3}$
and $q_{1}=g_{0}/g_{3.}$ The viscosity value $m_{d}(n,s)$ satisfies
the double-root condition 
\begin{equation}
27q_{1}^{2}+4p_{1}^{3}=0,\label{eq:cubic discrim-eq0}
\end{equation}
which is essentially the same equation as the one found above using
a different approach, where no explicit expressions were shown for
$p$ and $q$ (in fact, it is clear from relation (\ref{eq:bo2 is kappa times ma2})
that $p_{1}=\kappa^{2}p$ and $q_{1}=\kappa^{3}q$).

Consider the asymptotics as $s\uparrow\infty$. Note that $p_{1}\propto s^{2}$
and $q_{1}\propto s^{2}$. Hence, (\ref{eq:cubic discrim-eq0}) simplifies
to $p_{1}=0$, which implies $g_{1}=0$ (provided $g_{3}\neq0$),
and then, since $k_{206}>0$, it follows that $k_{206}k_{118}=k_{208}k_{116}$.
Expanding, this equation involves $m$ and $n$ only: 
\[
(1/2)\varphi(m-n^{2})(m(n(8n-3)+7)+n(n(7n-3)+8))
\]
\begin{equation}
=(n+1)(n+m)(m^{2}(n(3n+8)+3)-4mn^{2}(n^{2}-1)-n^{4}(n(3n+8)+3)),\label{eq:s large}
\end{equation}
where $\varphi$ is given by (\ref{eq:phi}). For $n\uparrow\infty$,
we look for solutions in the form $m\sim\chi n^{2}$ with $0<\chi<1$.
The leading order is proportional to $n^{8}$, yielding $9\chi^{2}-4\chi-1=0$.
The only acceptable solution is $\chi=(2+\sqrt{13})/9\approx0.623$.
Note that even for $s=1$ and $n=4$, our (mentioned above) result
$m_{d}=10.2783$ implies $m_{d}/n^{2}\approx0.643$ (cf. the asymptotic
value $0.623$).

If $s\uparrow\infty$ but $n\downarrow1$, it turns out that no appropriate
solutions exist for $m_{d}$. Then the curve $m=m_{d}(n)$ should
intersect the sector boundary $m=1$ at some finite $n=n_{0}>1$.
Substituting $m=1$ into (\ref{eq:s large}), we obtain the following
equation for $n_{0}$: $(n-1)^{4}-16n^{2}=0$, which has a single
acceptable solution, $n_{0}=3+\sqrt{8}$. 
\begin{figure}
\centerline{\includegraphics[clip,width=0.7\textwidth]{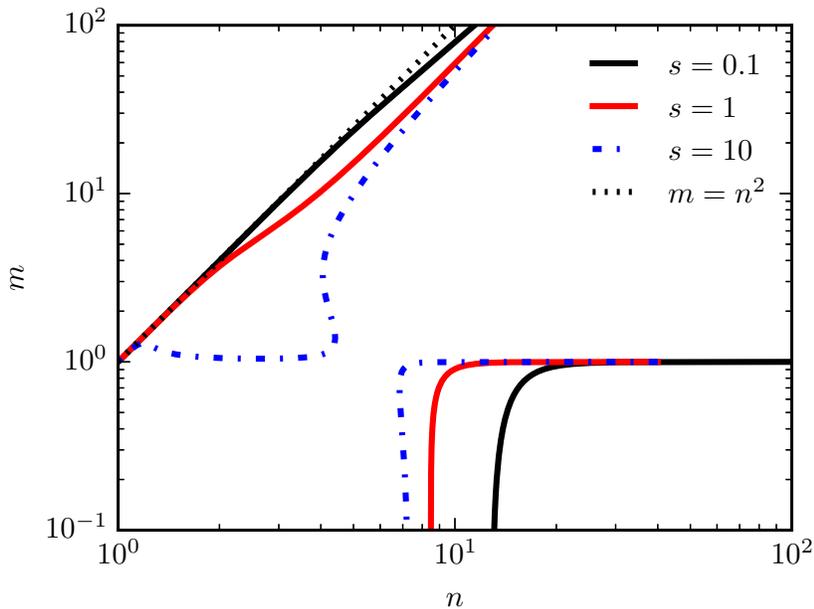}}
\caption{Numerical solutions of equation (\ref{eq:cubic discrim-eq0}) for
the representative values of $s$ given in the legend.}
\label{fig:fig20rt} 
\end{figure}
Consider now the asymptotic case $s\downarrow0$. Here, equation (\ref{eq:cubic discrim-eq0})
simplifies to the leading order equation $q_{1}=0$, and thus its
numerator is also zero. But this contradicts the fact, mentioned above,
that it is strictly positive. Therefore, there is no mid-wave instability
for sufficiently small base shear.

Fixing the value of $s$, we solve numerically equation (\ref{eq:cubic discrim-eq0})
for the solution curve $m=m_{d}(n)$. In figure \ref{fig:fig20rt},
we show these solution curves for several representative values of
$s$, ranging from small, to medium, to large. For large and small
values of $s$, numerical solutions can be verified with analytical
asymptotics. It is difficult to get numerical solutions for very large
$s$. In particular, we obtain the point ($n=n_{0}$, $m=1$) which
is approached when $s\uparrow\infty$ by the $m_{d}$ curves of the
$R$ and $S$ sectors (the upper and lower branches in figure \ref{fig:fig20rt}.

As was established in the last paragraph of section \ref{subsec:General-considerations 6-2-1},
at any boundary point of a critical curve, the latter is tangent to
the threshold curve through that quasi-intersection point of the two
curves. Hence, since in the $R$ sector the threshold curves have
positive slopes (see figure \ref{fig:figMidwaveR1m_near_n2_n4s1}),
the same holds for the critical curves near their boundary points.
This means that the critical function $\text{Bo}_{cM}(\text{Ma})$
is increasing near its boundary points. Therefore, if there is a maximum,
then there must be a minimum between this maximum and the right-end
quasi-intersection point. It transpires that as $m$ rises through
a certain threshold value $m_{t}$, such a maximum and a minimum appear
at some ($\textrm{Ma, Bo}$). The latter is an inflection point on
the $m=m_{t}$ critical curve, where the tangent is horizontal. We
call it an ``extrema bifurcation point'' (EBP; see figure \ref{fig:figMidwaveR1m_near_n2_n4s1}
(c)). The EBPs, in both $R$ and $Q$ sectors, are discussed in detail
below, in section \ref{subsec:Extrema-bifurcation-points}.

\subsubsection{The $Q$ sector semi-infinite critical curves and their asymptotic
behavior}

Turning next to the $Q$ sector, the cubic equation for $\text{Ma}_{0}$
has a single positive root and two non-physical complex conjugate
roots. The physical root corresponds to the single ``quasi-intersection''
points in figure \ref{fig:Fig_noses}(a). Since the threshold curve
has $\text{Bo}=\text{Bo}_{cL}(\text{Ma})<0$ and for the critical
curve $\text{Bo}\to0$ as $\text{Ma}\to\infty$, it is clear that
the critical curve of the mid-wave instability lies above this threshold
curve of the long-wave instability. This conclusion agrees with figure
\ref{fig:Fig_noses}.

For the $Q$ sector, the fact of the shared direction with the threshold
curve at the boundary point of the critical curve, $d\text{Bo}/d\text{Ma}=\text{Bo}_{2}/\text{Ma}_{2}=\kappa$,
(see the last paragraph of section \ref{subsec:General-considerations 6-2-1}),
implies that the function $\text{Bo}_{cM}(\text{Ma})$ is decreasing
near the (single) quasi-intersection point. For $\text{Ma}\uparrow\infty$,
postulating, from numerical results, that $\text{Bo}\rightarrow0$
and also $\alpha\rightarrow0$, we look for asymptotics $\text{Ma}=c_{1}\alpha^{-\zeta}$
(with $c_{1}\neq0$) and $\text{Bo}=d_{1}\alpha^{\xi}$ (with $d_{1}\neq0$),
where $\zeta$ and $\xi$ are positive, and substitute this into the
system of equations (\ref{eq:C_0}) and (\ref{eq:C_0' is zero}).
In more detail, these equations are (\ref{eq:MaEquation2}), which
for convenience is divided by $\text{Ma}^{2}$, and its partial derivative
with respect to $\alpha$. Considering the first of these equations,
it is clear that the second term is much smaller than the first one
and the fourth and fifth terms are negligible in comparison with the
third one. Thus, at leading order, the third term must balance the
first one: 
\begin{equation}
k_{20}+k{}_{31}\text{Ma}B=0.\label{eq:c_1-1}
\end{equation}
Since the $k_{20}\propto\alpha^{6}$, and $k_{31}\propto\alpha^{8}$,
it follows that the product $\text{Ma}B\propto\alpha^{-2}$. Since
$B=\text{Bo}+\alpha^{2}$ (which, clearly, entails that $\partial B/\partial\alpha=2\alpha$),
one can see that necessarily $\xi=2$. This can be proved by showing
that the assumption of $\xi<2$ or $\xi>2$ leads to a contradiction
in the system consisting of (\ref{eq:c_1-1}) and 
\begin{equation}
k'_{20}+k'_{31}\text{Ma}B+k_{31}\text{Ma}(2\alpha)=0\label{eq:c_1}
\end{equation}
has the power $9-\zeta$ which is greater than 5 (since $-\zeta+\xi=-2$,
so $\zeta=2+\xi<4$). Therefore, the last term of (\ref{eq:c_1})
is negligible compared to the other terms, which are clearly of power
$-5$. The first equation of the system yields $k_{206}+k_{318}c_{1}d_{1}=0$,
and the second equation becomes $6k_{206}+8k_{318}c_{1}d_{1}=0$,
which is clearly contradictory for $c_{1}d_{1}\neq0$.

If we assume that $\xi>2$ then $B=\alpha^{2}$ to leading order.
The first equation of the system yields $k_{206}+k_{318}c_{1}=0$
and the second equation $6k_{206}+8k_{318}c_{1}+2k_{318}c_{1}=0$.
This system again has only the unacceptable solution $c_{1}=0$. Thus,
we are left with $\xi=2$, and therefore $\zeta=4$. The system for
$c_{1}$ and $d_{1}$ is now 
\[
k_{206}+k_{318}(1+d_{1})c_{1}=0
\]
and 
\[
3k_{206}+k_{318}(5+4d_{1})c_{1}=0.
\]
Eliminating $k_{206}$ from the last two equations yields $d_{1}=-2$.
Then 
\[
c_{1}=\frac{k_{206}}{k_{318}}=\frac{3(n-1)(m-n^{2})(n+1)^{2}}{n^{2}(n^{3}+m)^{2}}>0.
\]
Therefore, $\text{Bo}=c_{1}^{1/2}d_{1}\text{Ma}^{-1/2}<0$. This is
in excellent quantitative agreement with the numerical results documented
in figure \ref{fig:Fig_noses}(a).

\subsection{Local extrema of the critical curves}

As figure \ref{fig:Fig_noses} shows, in the $Q$ sector there is
a local maximum on the critical curve for $m$ sufficiently close
to $n^{2}$, just as there is one at $m=n^{2}$, the boundary between
the $R$ and $Q$ sectors (see figure \ref{fig:Fig_noses_16}). Taking
into account that $\text{Bo}_{cM}(\text{Ma})$ is increasing at large
$\text{Ma}$ (as it is negative and goes up to zero in the limit of
infinitely increasing $\text{Ma}$), we conclude that there must be
at least two local minima on the critical curve, which is also in
agreement with the numerical results shown in figure \ref{fig:Fig_noses}(a).
For sufficiently large $m$, however, the critical curve is seen numerically
to have just a single minimum.

At any extremum, be it in the $R$ or the $Q$ sectors, we have 
\begin{equation}
\frac{d\text{Bo}}{d\text{Ma}}=0.\label{eq:dBo/dMa is 0}
\end{equation}
Also, since substituting the solutions $\textrm{Bo(Ma)}$ and $\alpha(\textrm{Ma})$
of the system of equations (\ref{eq:C_0}) and (\ref{eq:C_0' is zero})
for the critical curve into the left-hand side of equation (\ref{eq:C_0})
makes it true for all $\textrm{Ma}$, the total $\textrm{Ma}$-derivative
of the left-hand side must be zero, i.e. 
\[
\frac{\partial C_{0}}{\partial\text{Ma}}+\frac{\partial C_{0}}{\partial\alpha}\frac{d\alpha}{d\text{Ma}}+\frac{\partial C_{0}}{\partial\text{Bo}}\frac{d\text{Bo}}{d\text{Ma}}=0.
\]
For the extremum, in view of equations (\ref{eq:C_0' is zero}) and
(\ref{eq:c_1-1}), this leads to the third equation in addition to
(\ref{eq:C_0}) and (\ref{eq:C_0' is zero}): 
\begin{equation}
\frac{\partial C_{0}}{\partial\textrm{Ma}}=0.\label{eq:dC_0/dMa is 0}
\end{equation}
Thus the system of the three quadratic equations for the extrema points
is 
\begin{equation}
\textrm{\ensuremath{A_{1}}+\ensuremath{A_{2}\text{Ma}}+\ensuremath{A_{3}\text{Ma}^{2}}}=0,\label{eq:q_11s0}
\end{equation}
\begin{equation}
A_{1}'+A_{2}'\text{Ma}+A_{3}'\text{Ma}^{2}=0,\label{eq:q_21s0}
\end{equation}
\begin{equation}
A_{1}+2A_{2}\text{Ma}+3A_{3}\text{Ma}^{2}=0.\label{eq:q_3is0}
\end{equation}
Subtracting (\ref{eq:q_11s0}) from (\ref{eq:q_3is0}), we get the
linear equation 
\begin{equation}
A_{2}+2A_{3}\text{Ma}=0,\label{eq:linearforMa}
\end{equation}
which can be solved for $\textrm{Ma}$ in terms of the other variables,
provided that $A_{3}\neq0$, i.e., since $k_{31}>0$, that $B\neq0$.
On the other hand, another linear equation for $\textrm{Ma}$ is obtained
by eliminating the quadratic terms by linearly combining the quadratic
equations (\ref{eq:q_11s0}) and (\ref{eq:q_3is0}), 
\begin{equation}
2A_{1}+A_{2}\text{Ma}=0.\label{eq:linearforMa-1}
\end{equation}
This can also be solved for $\textrm{Ma}$ in terms of the other variables
(provided that $A_{2}\neq0$; also, it is easy to see that one has
to assume that $B\neq0$ in order to have a nonzero $\textrm{Ma}$).
The solvability condition of the over-determined system of the two
linear equations for $\textrm{Ma}$, (\ref{eq:linearforMa}) and (\ref{eq:linearforMa-1}),
is 
\begin{eqnarray}
{\cal D} & = & 0,\label{eq:discriminant in terms of Ajs-1}
\end{eqnarray}
where we have introduced the notation 
\begin{eqnarray}
{\cal D} & = & A_{2}^{2}-4A_{1}A_{3},\label{eq:discriminant in terms of Ajs}
\end{eqnarray}
which is independent of $\textrm{Ma}$.

One has to distinguish the cases $B\ne0$ and $B=0$. For $B\ne0$,
the solution of equation (\ref{eq:linearforMa}) is 
\begin{equation}
\text{Ma}=-\frac{A_{2}}{2A_{3}}.\label{eq:Ma in terms of Ajs}
\end{equation}
Substituting this into the quadratic equation (\ref{eq:q_2is0}),
we have a system of two transcendental equations for $B$ and $\alpha$,
which can be written in the following form: 
\begin{eqnarray}
{\cal D} & = & 0,\label{eq:discriminant of q_1 is 0}\\
{\cal D}' & = & 0.\label{eq:discriminant prime is 0}
\end{eqnarray}
In the $R$-sector, two solutions, a maximum and a minimum, are found
by solving numerically the system of equations, (\ref{eq:discriminant of q_1 is 0})
and (\ref{eq:discriminant prime is 0}), and then finding $\textrm{Ma}$
from (\ref{eq:Ma in terms of Ajs}). In the $Q$ sector, this gives
a single solution, which is a maximum for $m<m_{N}$, and a minimum
for $m>m_{N}$.

Also, there are, in a certain interval of viscosity ratios, solutions
with $B=0$. In this case, the solvability condition (\ref{eq:discriminant in terms of Ajs-1})
implies 
\begin{equation}
k_{20}=0,\label{eq:k_20 is 0}
\end{equation}
which yields the wavenumber. Then the Bond number is determined uniquely
as 
\begin{equation}
\text{Bo}=-\alpha^{2}.\label{eq:Bo is -alpha^2}
\end{equation}
The quadratic equation (\ref{eq:q_21s0}), with the now known $\alpha$
and $\textrm{Bo}$, gives two distinct solutions for the Marangoni
number if the discriminant $\zeta$ is positive, where 
\begin{equation}
\zeta=A_{2}^{'2}-4A_{1}^{'}A_{3}^{'}.\label{eq:zeta is q_2 discriminant}
\end{equation}
For the case at hand we have the simplified relations $A_{2}^{'}=k_{20}^{'}$,
$A_{1}^{'}=2\alpha k_{11}$, and $A_{3}^{'}=2\alpha k_{31}$. Thus,
the solutions are 
\begin{equation}
\text{Ma}=\frac{-k_{20}'\pm\sqrt{k_{20}'^{2}-16\alpha^{2}k_{11}k_{31}}}{4k_{31}\alpha},\label{eq:Ma is solution of q_2}
\end{equation}
corresponding to the two minima on the critical curves in the $Q$
sector (see figure \ref{fig:Fig_noses}).

\begin{figure}
\centerline{\includegraphics[bb=0bp 0bp 468bp 280bp,clip,width=0.7\textwidth]{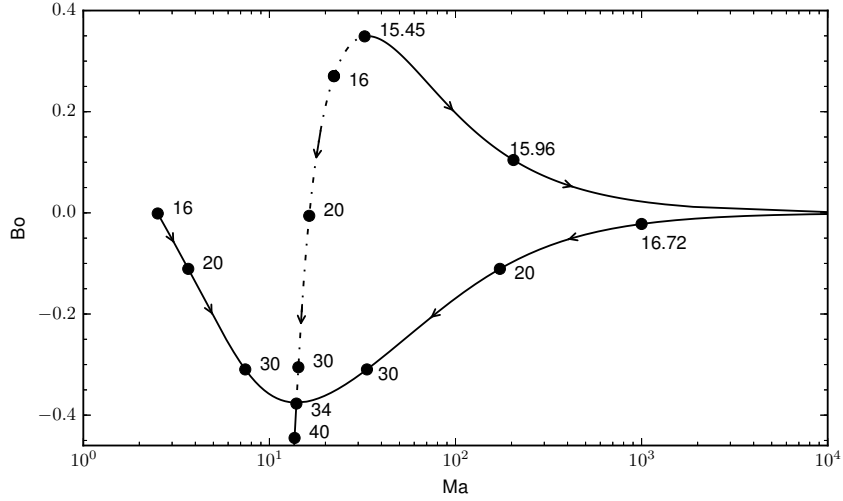}}
\protect\caption{The trajectories of the extrema of the critical curves $\text{Ma}=\text{Ma}_{cM}(\text{Bo})$
in the ($\text{Ma}$, $\text{Bo}$)-plane (for $n=4$ and $s=1$;
see figures \ref{fig:figMidwaveR1m_near_n2_n4s1}(c) and (d), \ref{fig:Fig_noses_16}(a),
and \ref{fig:Fig_noses}(a)) as $m$ changes starting in the $R$
sector and increasing through the $R$ sector and after that, for
$m>16$, the $Q$ sector. In the $R$ sector, for $15.45<m<16$, there
is one maximum, the dashed curve, and one minimum, the solid curve,
for $16<m<34.31$, there is one maximum between two minima, and finally,
for $m>34.31$ there is one minimum. The arrows indicate the increase
of $m$ and the dots correspond to the displayed values of $m$ next
to them. The minimum moves to infinite $\textrm{Ma}$ as $m\to16$,
from either side. \label{fig:fig__rtmovingnose}}
\end{figure}

Figure \ref{fig:fig__rtmovingnose} shows the trajectories of the
extrema in the $(\text{Ma},\text{Bo)}$-plane for $n=4$ as the viscosity
ratio $m$ increases, starting from $m=15.45$, in the $R$ sector,
reaching the $Q$ sector at $m=16$, and continuing to increase in
the $Q$ sector. Consistent with the stability diagrams shown in figures
\ref{fig:figMidwaveR1m_near_n2_n4s1} and \ref{fig:Fig_noses_16},
there are two extrema, a maximum and a minimum, for $m<16$, which
is in the $R$ sector, provided $m>m_{t}=15.45$. In the $Q$ sector,
there are three extrema as long as $m<m_{N}$. At $m=m_{N}$, the
three extrema, one of them a maximum and two of them minima, collapse
together into a single minimum, which then persists through the $Q$
sector. (Recall that we term this point the extrema bifurcation point
(EBP)).

In connection with the non-monotonic character of critical curves
that have multiple local extrema, we note the following. In panels
(e) and (f) of figure \ref{fig:FIG_RTgamNalphas2by4} (where $n=2$),
we see that as $\textrm{Ma}$ increases, the long-wave instability
gives way to stability at $\text{Ma}=\text{Ma}_{cL}$, which persists
up to $\text{Ma}=\text{Ma}_{cM}$, at which point the mid-wave instability
sets in, further persisting for all larger $\textrm{Ma}$. For short,
we symbolically describe this sequence of Ma-intervals with different
stability types as L-S-M, (where L indicates the long-wave instability,
S denotes stability, and M stands for the mid-wave instability). The
same stability interval sequence is obtained from figure \ref{fig:Fig_noses}
(where $n=4$) if, e.g., we fix $m=25$ and $\textrm{Bo}=-0.1$, and
go rightwards parallel to the $\Ma$-axis. However, different sequences
occur for other sets of parameters. For example, at $m=20$ and $\textrm{Bo}=-0.05$,
we observe the sequence L-S-M-S-M; at $m=36$ and $\textrm{Bo}=-0.3$,
the sequence is L-M-S-M; and at $m=25$ and $\textrm{Bo}=-0.2$, we
have the longer sequence L-M-S-M-S-M. It appears that for any $\textrm{Bo}<0$,
any sequence starts with L and ends with M. In contrast, for positive
$\Bo$, e.g., at $\textrm{Bo}=0.05$ and $m=18$, we have a S-M-S
sequence of $\Ma$-intervals.

\subsection{Extrema bifurcation points\label{subsec:Extrema-bifurcation-points}}

\subsubsection{The EBP in the Q sector}

We turn now to the problem of equations determining the extrema bifurcation
points. In this section, we consider those in the $Q$-sector, while
those in the $R$-sector are examined in the following section.

\begin{figure}
\centerline{\includegraphics[bb=0bp 0bp 288bp 216bp,clip,width=0.7\textwidth]{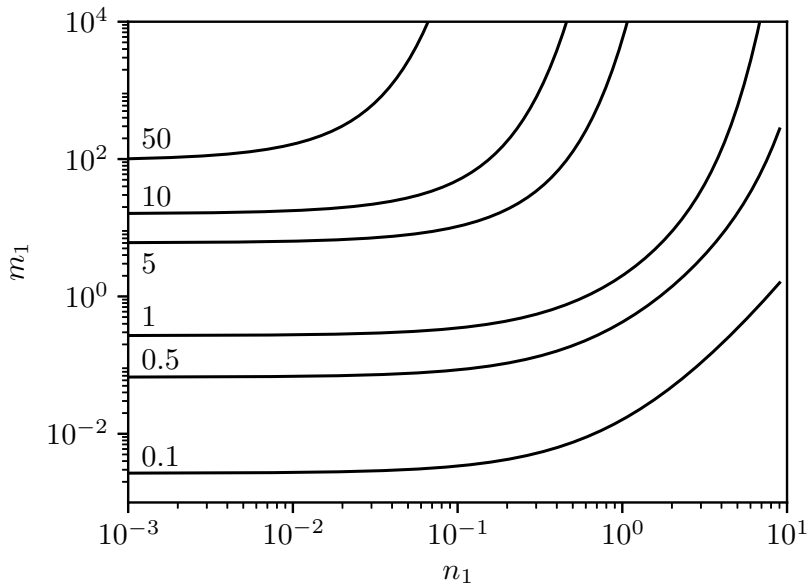}}
\caption{Level curves of $\alpha(n_{1},\;m_{1})$ for the extrema bifurcation
points in the $Q$ sector. The numbers next to the curves are the
corresponding values of $\alpha$. \label{fig:Level-curves-of alpha in Q}}
\end{figure}

Clearly, the bifurcation point of the two minima and one maximum at
$m=m_{N}$ in the $Q$ sector, which has $B=0$ (inherited, by continuity,
from the $B=0$ property of the two minima existing at the smaller
$m$), must satisfy, in addition to equations (\ref{eq:C_0}), (\ref{eq:C_0' is zero})
and (\ref{eq:dC_0/dMa is 0}), the condition that the $\text{Ma}$
values of the two minima coalesce to a double root. It is clear from
equation (\ref{eq:Ma is solution of q_2}) that this means 
\begin{equation}
k_{20}'^{2}-16\alpha^{2}k_{11}k_{31}=0.\label{eq:D1 is 0}
\end{equation}
As we already noted above, the latter corresponds to the discriminant
(\ref{eq:zeta is q_2 discriminant}) being zero, so that the two $\text{Ma}$
solutions of (\ref{eq:q_21s0}) for the two minima merge into just
a single one. For the extrema bifurcation points in the $Q$ sector,
it is convenient to use new variables $n_{1}=n-1$ and $m_{1}=(m-n^{2})/(n-1)$so
that the $Q$ sector corresponds to the entire first quadrant. For
any given ($n_{1},m_{1}$), as we already mentioned above, we can
find the other properties of the EBP as follows: First, $\alpha$
is determined by solving equation (\ref{eq:k_20 is 0}) (which can
be simplified, yielding that the quantity within the curly-bracket
of $k_{20}$ in (\ref{eq:k20}), denoted by ${\cal C}$, must vanish).
This dependence $\alpha(n_{1},m_{1})$ is shown as the contour plot
in figure \ref{fig:Level-curves-of alpha in Q}. 
We observe this unique solution for the extended region of realistic
$(n_{1},m_{1})$. For small $n_{1}$ and $\alpha$, we find that asymptotically
$m_{1}=(4/15)\alpha{}^{2}$, independent of $n_{1}$, so that the
level curves of $\alpha$ intersect the vertical axis at different
heights. With corrections, the equation of the level curves at $\alpha\ll1$
and $n_{1}\ll1$ is $m_{1}=4/15\alpha^{2}(1+5n_{1}/2+2n{}_{1}^{2})$,
where we have suppressed the terms which have powers of $\alpha^{2}$
higher than one or powers of $n_{1}$ higher than two. Keeping the
two correction terms in the formula is necessary for predicting the
flip of the sign of the level curve curvature as one switches between
the linear and log scales of the $n_{1}$-axis.

Having found $\alpha$ from equation (\ref{eq:k_20 is 0}), the Bond
number $\text{Bo}(n_{1},m_{1})$ is given by equation (\ref{eq:Bo is -alpha^2})
with $\alpha=\alpha(n_{1},m_{1})$. Next, equation (\ref{eq:D1 is 0}),
after being divided through by $s^{2}$, is a linear equation in $s^{2}$,
and gives $s(n_{1},m_{1})$, provided the derivative of $k_{20}$
with respect to $\alpha$ is negative. In view of $k_{20}=0$, it
is enough to require that ${\cal C}'<0$. There is a strong evidence
that the latter is indeed the case. At fixed $n_{1}$ and $m_{1}$,
${\cal C}>0$ and growing at sufficiently small $\alpha$; asymptotically,
we find that ${\cal C}=\frac{1}{9}\varphi n^{2}\left(n_{1}+2\right)n_{1}m_{1}\alpha^{8}$.
This factor attains a (positive) maximum and then monotonically decreases;
asymptotically, at $\alpha\uparrow\infty$, we have ${\cal C}=-\frac{1}{32}\left(m+1\right)\exp(4\alpha n+2\alpha)$.
Since ${\cal C}$ is decreasing at its zero, it is clear that ${\cal C}'<0$
at that $\alpha$, and this is confirmed by numerics. In addition,
for a few fixed values of $n_{1}$, we computed $\alpha(n_{1},m_{1})$
and the corresponding ${\cal C}'$ up to large values of $m_{1}$,
e.g. $m_{1}=10^{6}$, and this always showed ${\cal C}'<0$. We find
analytically that the large-$m_{1}$ log-log asymptotic slope is $\text{d}\ln(-{\cal C}')/\text{d}\ln m_{1}=4(n_{1}+1)/n_{1}$
(with the logarithmic asymptotics $\text{d}\alpha/\text{d\ensuremath{\ln}}m_{1}=(1/2)n_{1}$).
These asymptotic results are in excellent agreement with the numerical
computations (which are not shown).

Finally, $\text{Ma}(n_{1},m_{1})$ of the EBP is given by (\ref{eq:Ma is solution of q_2})
with discriminant equal to zero, so that 
\begin{equation}
\text{Ma}=\frac{-k_{20}'}{4k_{31}\alpha}.\label{eq:Ma is solution of q_2-1}
\end{equation}
For example, there is a solution that corresponds to the EBP in figure
\ref{fig:fig__rtmovingnose}, with $n=4$, $s=1$, $m=m_{N}=34.31$,
$\Ma=13.97$, $\Bo=-0.375$, and $\alpha=0.61$. These values are
also consistent with figure \ref{fig:Fig_noses}. Note that in figure
\ref{fig:Level-curves-of alpha in Q}, $\alpha(n_{1},m_{1})$ has
no external parameters. The same is true of the other EBP dependencies:
$\text{Bo}(n_{1},m_{1})$,; $s(n_{1},m_{1})$; and $\text{Ma}(n_{1},n_{1})$.
\begin{figure}
\centering{}\includegraphics{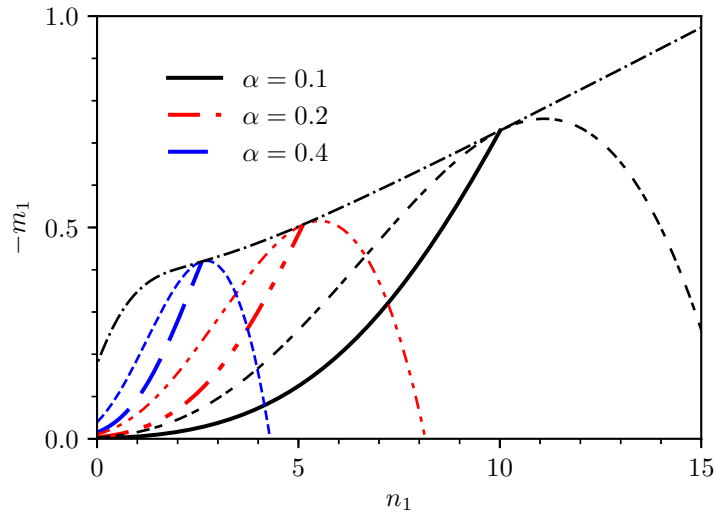} \caption{Level curves of $\alpha(n_{1},\;m_{1})$ for the extrema bifurcation
points in the $R$ sector. (For other curves, see the text.)\label{fig:Level-curves-of alpha(n1,m1)}}
\end{figure}

\subsubsection{The EBP in the $R$ sector}

It was mentioned above, at the end of section \ref{subsec:R sector critical curves},
that one maximum and one minimum appear at the EBP on the critical
curve in the $R$ sector corresponding to a threshold value $m_{t}$
of $m$. It is clear that for this EBP 
\[
\frac{d^{2}\text{Bo}}{d\text{Ma}^{2}}=0
\]
along the critical curve. (This bifurcation point of extrema corresponds
to the inflection point with the horizontal tangent line in figure
\ref{fig:figMidwaveR1m_near_n2_n4s1}(c).) Here $\text{Bo}(\text{Ma})$
is one of the two functions defined implicitly by the system (\ref{eq:C_0})
and (\ref{eq:C_0' is zero}), where the other implicit function is
$\alpha(\textrm{Ma})$. By the well-known formula for the derivative
of an implicit function we have 
\[
\frac{d\text{Bo}}{d\text{Ma}}=\frac{\partial(C_{0},C'_{0})}{\partial(\alpha,\text{Ma})}\left[\frac{\partial(C_{0},C'_{0})}{\partial(\text{Bo},\alpha)}\right]^{-1}.
\]
We differentiate this expression with respect to $\text{Ma}$, taking
into account (\ref{eq:C_0}), (\ref{eq:C_0' is zero}) and (\ref{eq:dC_0/dMa is 0}).
As a result, we obtain a fourth equation of the system for the bifurcation
point of the extrema: 
\[
\left(\frac{\partial^{2}C_{0}}{\partial\alpha\partial\textrm{Ma}}\right)^{2}-\frac{\partial^{2}C_{0}}{\partial\alpha^{2}}\frac{\partial^{2}C_{0}}{\partial\textrm{Ma}^{2}}=0,
\]
which is, more explicitly, given by 
\begin{equation}
2\textrm{Ma}(A_{2}+3A_{3}\textrm{Ma})(A_{1}''+A_{2}''\text{Ma}+A_{3}''\text{Ma}^{2})-(A_{1}'+2A_{2}'\text{Ma}+3A_{3}'\text{Ma}^{2})^{2}=0.\label{eq:q_2is0-1}
\end{equation}
We note that $\text{Bo}>0$ in the $R$ sector and therefore $B\ne0$.
The four equations, (\ref{eq:discriminant of q_1 is 0}), (\ref{eq:discriminant prime is 0}),
(\ref{eq:Ma in terms of Ajs}), and (\ref{eq:q_2is0-1}), are solved
numerically. As an example, for $n=4$ and $s=1$, we find $\alpha=0.21$,
$\text{Bo}=0.35$, $\text{Ma}=33.81$ and $m=m_{t}=15.45$. These
numbers are consistent with figure \ref{fig:figMidwaveR1m_near_n2_n4s1}(c)
and figure \ref{fig:fig__rtmovingnose}.

Similarly to the $Q$ sector procedure used above, for the four EBP
equations in the $R$ sector, an algebraic reduction is possible where
a single equation is used to solve for one variable, and then the
three other parameters of the EBP are found (with given values of
$n$ and $m$). For this, we note the algebraic identity $k_{22}{}^{2}-4k_{13}k_{31}=0$.
Hence, equation (\ref{eq:discriminant of q_1 is 0}) can be written,
explicitly showing the $s$ and $B$ dependencies, as 
\begin{equation}
k_{20s}^{2}s^{2}+2B^{2}k_{sB}=0\label{eq:sB1}
\end{equation}
and equation (\ref{eq:discriminant prime is 0}) is written as 
\begin{equation}
k_{20s}k_{20s}'s^{2}+B^{2}k_{sB}'+4B\alpha k_{sB}=0.\label{eq:sB2}
\end{equation}
Here we have defined the quantity $k_{sB}$ as $k_{sB}=k_{20s}k_{22}-2k_{11s}k_{31}$,
where $k_{20s}$ and $k_{11s}$ are defined by $k_{20s}=k_{20}/s^{2}$
and $k_{11s}=k_{11}/s^{2}$. The last two displayed equations are
linear equations for $s^{2}$, so $s^{2}$ is obtained explicitly
in terms of the quantities $\alpha$, $m$, $n$, and $B$. Moreover,
the solvability condition of the over-determined system of the two
linear equations for $s^{2}$ yields (after dividing through by $Bk_{20s}$)
a linear equation for $B$, 
\begin{equation}
B(k_{20s}k_{sB}'-2k_{20s}'k_{sB})+4k_{20s}\alpha k_{sB}=0,\label{eq:linear for B}
\end{equation}
whose coefficients depend on $\alpha$, $m$, and $n$. Solving it
(provided that the coefficient of $B$ is nonzero) yields $B$ in
terms of $\alpha$, $m$, and $n$; using this expression in equation
(\ref{eq:sB1}), we obtain $s^{2}$ in terms of $\alpha$, $m$, and
$n$, and then, from equation (\ref{eq:Ma in terms of Ajs}), an expression
for $\text{Ma }$ in terms of $\alpha$, $m$, and $n$. Substitute
these expressions into (\ref{eq:q_2is0-1}) to obtain an equation
containing $\alpha$, $m$, and $n$, which can be numerically solved
for $\alpha$ giving it as a function of $n_{1}$ and $m_{1}$. Then,
for the given values of $n_{1}$ and $m_{1}$, we find sequentially
$B$, $s^{2}$, and $\textrm{Ma}$, in that order, using the linear-equations
solutions for them described above. Thus, for given $n_{1}$ and $m_{1}$,
we determine all the parameters, $\alpha$, $B$, $s$, and $\textrm{Ma}$,
of the corresponding EBP in the $R$ sector. The level curves of $\alpha$
are seen in figure \ref{fig:Level-curves-of alpha(n1,m1)} as the
monotonically rising curves. Through the upper point of each curve
passes the level curve, with the same value of $\alpha$, of the $\alpha$
function which makes identically vanish the coefficient $B_{d}$ of
$B$ in equation (\ref{eq:linear for B}). It is clear that the envelope
of the family of level curves for $B_{d}=0$ is the locus of the upper
ends of the level curves for the EBPs. (The envelope curve shown in
the figure was obtained by solving the system $B_{d}=0$ and $\partial B_{d}/\partial\alpha=0$.)
When approaching the envelope curve, the values of $B$ grow to infinity.
The EBP level curves can be formally continued above the envelope
curve, but lead to unphysical negative values of $B$ and $\text{Ma}$.
(Note that the $R$ sector is completely mapped into the region of
the $(n_{1},-m_{1})$-plane bounded above by the line $-m_{1}=n_{1}+2$,
corresponding to $m=1$; however, this line is outside the range of
figure \ref{fig:Level-curves-of alpha(n1,m1)}.)

\section{Summary and discussion}

\label{sec:Conclusions}

In this paper, we have considered the linear stability of two immiscible
viscous fluid layers flowing in the channel between two parallel plates
that may move steadily with respect to each other driving a Couette
flow. The combined effects of gravity and an insoluble surfactant
monolayer at the fluid interface were examined for certain flows such
that the effect of inertia on their stability properties is negligible.
The bulk velocity components satisfy linear homogeneous equations
with constant coefficients. Therefore, their general solution, in
the standard normal-mode analysis, is available with a few undetermined
constants. The latter are determined, by the plate and interfacial-balance
boundary conditions, in terms of the interface deflection and surfactant
disturbance amplitudes. This yields a system of two algebraic linear
homogeneous equations for the latter two amplitudes. Nontrivial solutions
of this algebraic eigenvalue problem exist only if the increment $\gamma$,
the complex ``growth rate,'' satisfies a quadratic equation whose
coefficients are known functions of the wavenumber $\alpha$, the
Marangoni number $\text{Ma}$, the Bond number $\text{Bo}$, the viscosity
ratio $m$, the aspect ratio $n$, and the interfacial shear parameter
$s$. The two solutions of this dispersion equation were shown to
yield two continuous increment branches, defined almost everywhere
in the wavenumber-parameters space (with a ``branch cut'' hypersurface
excluded from it), and their real parts, the two continuous growth
rate branches, were analyzed to infer conclusions concerning the stability
of the flow. Similar to FH and subsequent papers, we call one of the
branches the ``robust'' branch, as it is present even when $\Ma=0$,
and we call the other one, that vanishes as $\Ma\downarrow0$, the
``surfactant'' branch. Thus, we have explicit formulas allowing
us to readily compute the growth rates of instability for any given
input values of the wavenumber and the five parameters of the problem.

In the long-wave analysis of FH, three open sectors in the part of
the ($n$, $m$)-plane given by $n\geq1$ and $m\geq0$, categorizing
the stability of the system without gravity ($\Bo=0$), were identified:
the $Q$ sector, ($m>n^{2}$), where both modes are stable; the $R$
sector, ($n^{2}>m>1$), where only the robust branch is unstable;
and the $S$ sector, ( $0<m<1$), where only the surfactant mode is
unstable. The same long-wave sectors were found to be relevant for
non-zero $\Bo$ in the lubrication theory of FH17. In the present
paper, by using the long-wave asymptotics for the coefficients of
the quadratic dispersion equation, we corroborate the lubrication
approximation results of FH17 for the instability thresholds. In the
$S$ sector, the surfactant mode remains unstable for all $\Bo$,
that is for arbitrarily strong stabilizing gravity; while in the $R$
sector the growth rate of the robust branch is unstable provided $\textrm{Bo}$
is below some positive threshold value $\textrm{Bo}_{c}$. In the
$Q$ sector, both branches remain stable for Bo $\geq0$, but the
robust branch is long-wave unstable for the smaller values of $\Ma$
(while mid-wave unstable for larger values of $\Ma$, so that there
are longer waves that are stable, as discussed below), when $\textrm{Bo}$
is below some negative $\textrm{Bo}_{c}$. We have obtained the long-wave
marginal wavenumbers and extremum growth rates which depend on the
two main orders of the growth rate expression and were not considered
in FH17. In particular, the small-$s$ behavior of the marginal wavenumber
was obtained from the asymptotic form of our general equation for
the marginal wavenumber. We have established that in the $R$ sector
there are parametric situations in which the stabilizing effects,
responsible for the emergence of the marginal wavenumber, are due,
instead of the capillary forces, as is usual for larger $s$, to the
nontrivial combined action of gravitational and surfactant forces.

We also obtained the asymptotic small-$s$ behavior of the (long-wave)
growth rate maximum and its corresponding wavenumber, which yielded
different power laws for the cases of zero and non-zero Bond numbers.
The asymptotic behavior in nearing the instability thresholds in the
different sectors was established as well.

The long-wave instabilities at the different borders between the three
($n,m$)-sectors were analyzed, such as the $S-R$ one, $m=1$. For
the latter case, it was not clear from the small-wavenumber expression
for the growth rates, equation (\ref{eq:GammaApproxSmallalpham}),
whether the unstable mode belonged to the surfactant branch, or, alternatively,
to the robust one. We used complex analysis to show that there are
indeed two separate branches of the growth rate function continuous
for all wavenumbers and all the values of the parameters, one of the
branches everywhere positive and the other one everywhere negative.
The surfactant branch is easily identified near the wavenumber axis
in the wavenumber-Marangoni number space, as the one of the two branches
which vanishes in this limit of Marangoni number approaching zero,
and it turns out to be positive or negative for positive or negative
Bond number, respectively. The same is then true in the alternative
limit, the wavenumber approaching zero at a finite Marangoni number,
(corresponding to equation (\ref{eq:GammaApproxSmallalpham})), since
the branches keep their signs everywhere, and in particular the surfactant
branch of the growth rate has the same sign near the $\textrm{Ma}$-axis
as its sign near the $\alpha$-axis. In this way, we established that
the unstable mode, corresponding to the positive sign in equation
(\ref{eq:GammaApproxSmallalpham}), belongs to the surfactant (robust)
branch for positive (negative) Bond numbers (and the stable mode belongs
to the other branch, in each case).

For cases of arbitrary, (not necessarily small) wavenumbers, we still
have explicit formulas for the stability quantities of interest, albeit
more complicated and therefore, in general, studied numerically. It
was found that in the $S$ sector and in the $R$ sector sufficiently
far from the $Q$ sector, as well as in the $Q$ sector for sufficiently
small Marangoni numbers, the dominant-mode instability has a long-wave
character, in the sense that the left endpoint of the interval of
unstable wavenumbers is zero. Otherwise, in particular in the $Q$
sector, for sufficiently large Marangoni numbers, the 'mid-wave' instability
may occur, in which the interval of unstable wavenumbers is bounded
away from zero. These two situations were considered in turn. An interesting
phenomenon, the dispersion-curve reconnection, was observed in the
$S$ sector. Both branches are unstable for sufficiently negative
values of Bond number, and, as $\textrm{Bo}$ decreases further, the
robust-mode dispersion curve starts to cross the other dispersion
curve at a single intersection point. Later in this process, at some
sufficiently large value of $\left\vert \text{Bo}\right\vert $, the
four parts of the two curves emanating from the intersection point
recombine and detach, forming two new, non-intersecting, continuous
curves, with the upper curve having two local maxima, of unequal heights.
Then, as the Bond number decreases further, a jump in the global maximum
may occur, as the shorter local maximum grows and eventually overcomes
the other local maximum (figure \ref{fig:Fig_typical_MaxCrossingDCurve}).

The long-wave instability was studied with respect to gravity effects,
as indicated by the dependencies of the characteristic dispersion
quantities on the Bond number, figure \ref{fig:Fig3maxmargRSQ}, and,
in the $S$ and $R$ sectors, with respect to the surfactant effects,
as expressed in the dependencies on the Marangoni number, figure \ref{fig:FIG_Bo1_vs_Ma}.
For the small and large values of these parameters, the relevant wavenumbers
may be small, allowing for simpler asymptotics. Even when the limits
of the characteristic dispersion quantities are not small, we sometimes
get simplified equations which are easier to solve numerically, or,
occasionally, even approximate analytic expressions, such as equation
(\ref{eq:alpha0app}).

In the $R$ and $S$ sectors, at a fixed Bond number, the long-wave
growth rate has a maximum at certain finite values of the wavenumber
and the Marangoni number. We have observed, numerically, that both
the maximum growth rate and its Marangoni number, grow linearly with
the shear parameter $s$, starting from zero, while the corresponding
wavenumber, which starts from zero as well, grows very fast at first,
but then remains almost constant at larger $s$ (figure \ref{fig:fig13abc}).
Similar dependencies take place in the $Q$ sector as well (figure
\ref{fig:fig16abc}).

The mid-wave instability turns out to emerge in two distinct ways
(as a control parameter increases): it starts either from a stability
stage, which we call the true onset of the mid-wave instability, or,
alternatively, from a long-wave instability stage. The latter occurs
when the left endpoint of the interval of the unstable wavenumbers,
which is zero for the long-wave instability, starts moving away from
zero (as shown in figure \ref{fig:figMidwaveAlphasR1m_near_n2_n4s1}(b)),
the maximum growth rate remaining positive all along. In the alternative
scenario of the onset of the mid-wave instability, the maximum growth
rate is equal to zero at a certain positive wavenumber, for which,
therefore, the marginal wavenumber equation holds. But in view of
the maximum, the partial derivative of the growth rate equals zero
as well. Thus, we have a system of two equations, whose solution gives
the critical values of the Marangoni number and the wavenumber asa
function of the Bond number, for arbitrarily fixed values of the remaining
three parameters. We follow, as the viscosity ratio is increased in
the $R$ sector toward its border with the $Q$ sector, the emergence
of the critical curve, and its consequent change, in the Marangoni
number-Bond number plane (figure \ref{fig:figMidwaveR1m_near_n2_n4s1}).
The critical curve has its two endpoints on the threshold curve of
the long-wave instability. The latter is rightward-increasing in the
$R$ sector, horizontal at the boundary with the $Q$ sector (figure
\ref{fig:Fig_noses_16}), and a decreasing curve in the $Q$ sector
(figure \ref{fig:Fig_noses}). The right-side endpoint of the critical
curve moves away to infinity as we cross into the $Q$ sector. The
critical wavenumber is small near a critical curve endpoint, and so
one can look for the critical solutions in the form of asymptotic
power series. This gives rise to a cubic equation for the endpoint
locations. Analysis of this equation leads to conclusions which are
in agreement with the numerical observations, such that the critical
curve in the $R$ sector exists only above a certain value of the
viscosity ratio and has two endpoints, while there is just one single
endpoint in the $Q$ sector. In all cases, the critical curve at its
end point is tangent to the long-wave threshold curve. We also obtain
and solve equations for the extrema of the critical curve, obtaining
conclusions that agree with the numerical results. In the $R$ sector,
there is a certain value of the viscosity ratio below which the critical
curve has no extrema, but above which it has exactly two extrema:
a maximum and a minimum. The latter disappears into the right-side
infinity at the boundary with the $Q$ sector, and so we have just
one extremum at this boundary, a maximum. Moving into the $Q$ sector
as the viscosity ratio increases, there are at first one maximum in
between two minima on the critical curve. These extrema coalesce into
a single minimum at a certain value of the viscosity ratio $m$, and
this minimum persists for the larger values of $m$.

As we go from an arbitrary critical point to a critical extremum,
one more constraint is added, which decreases the number of free parameters
by one. The 'extrema bifurcation points', at which the number of extrema
changes, correspond to another reduction of the number of free parameters.
Thus, for given $n$ and $m$, they determine all the other values:
the wavenumber, Marangoni number, Bond number, and the shear parameter
of the corresponding extrema bifurcation point (figure \ref{fig:Level-curves-of alpha(n1,m1)}).
Thus, figures \ref{fig:Fig_typical_midwaveDCurve}, \ref{fig:FIG_RTgamNalphas2by4},
\ref{fig:Fig_noses}, \ref{fig:fig__rtmovingnose} and \ref{fig:Level-curves-of alpha(n1,m1)}
represent different levels of information about the stability properties.
Namely, going from one of the figures to the next, in the given order,
the description gets more refined. On the other hand, the amount of
data in the description decreases, in a certain sense. Figure \ref{fig:Fig_typical_midwaveDCurve}
gives the growth rates at every wavenumber, but all the parameters
are fixed at certain values. So, out of the seven quantities, $\alpha$,
$\gamma_{R}$,$\Ma$,$\Bo$, $m$, $n$ and $s$, six are independent
variables, and just one quantity is a dependent variable. Thus, these
data make up a six-dimensional hypersurface in the seven-dimensional
space. Figure \ref{fig:FIG_RTgamNalphas2by4} corresponds to some
five independent variables determining the values of the other two
quantities, thus resulting in a five-dimensional manifold of data.
Figure \ref{fig:Fig_noses} corresponds to a four-dimensional manifold,
figure \ref{fig:fig__rtmovingnose} implies a three-dimensional manifold
of data, and figure \ref{fig:Level-curves-of alpha(n1,m1)} corresponds
to a two-dimensional manifold parameterized by the independent variables
$m$ and $n$, whose values determine $\alpha$, $\gamma_{R}$, $\Ma$,
$\Bo$, $s$, (where $\gamma_{R}=0$ since our consideration here
is confined to the critical conditions of mid-wave instability.) The
envelope curve in figure \ref{fig:Level-curves-of alpha(n1,m1)} corresponds
to a one-dimensional curve in the seven-dimensional space of the relevant
quantities. Finally, for the inflection point of the envelope curve
in figure \ref{fig:Level-curves-of alpha(n1,m1)}, there are no independent
variables, and all seven quantities are uniquely determined.

There is the mid-wave instability of the robust branch in the $S$
sector too, albeit the long-wave instability of the surfactant branch
is the stronger of the two there. In the $(\textrm{Ma,Bo})$-plane,
in the vicinity of the threshold line of the long-wave instability,
in addition to the more usual critical mid-wave curve which consists
of the points that correspond to dispersion curves with zero maximum
growth rate, there is, below the latter, another critical mid-wave
curve, consisting of the points corresponding to dispersion curves
with zero minimum growth rate (see figure \ref{fig:S midwave_mp1n10s1}).
Correspondingly, as the Bond number decreases (to bigger-magnitude
negative values), it is possible that at some point after the onset
of the mid-wave instability, the long-wave instability starts, whose
wavenumber interval is initially small and does not intersect the
mid-wave interval of unstable wavenumbers. The coexistence of the
mid-wave and the long-wave instabilities lasts until their intervals
coalesce, corresponding to the critical curve of zero minimum growth
rates in the $(\textrm{Ma,Bo})$-plane. After this coalescence, there
is just one long-wave continuous interval of the unstable wavenumbers,
with the dispersion curve having two positive local maxima of the
growth rate at first, but just one single maximum eventually, at the
most negative Bond number values. For another range of Marangoni number,
an alternative scenario is possible, which differs from the one described
above solely in that the long-wave instability starts first and the
mid-wave one at the smaller (more negative) values of the Bond number.
The consequent coalescence into purely long-wave instability is the
same in both scenarios (figures \ref{fig:S midwave Bo-alpha} and
\ref{fig:S dispersion curves-1}).

\appendix

\section{The Continuous Branches of the Growth Rate Function\label{sec:On-the-Continuous-Branches}}

Recall that the two distinct analytic branches of the function $\sqrt{\zeta}$
exist in any simply connected domain in the complex plane that does
not contain the origin ($\zeta=0$). As was mentioned in the text,
it may happen for the discriminant $\zeta$ of the dispersion relation
that $\zeta=0$ for some values of $\alpha$ and the parameters. This
implies the two real equations, $\textrm{Re}(\zeta)=0$ and $\textrm{Im}(\zeta)=0$.
The imaginary part of $\zeta$ (\ref{eq:DiscZ}) is 
\begin{align}
\operatorname{Im}(\zeta) & =\frac{s}{\alpha^{5}}\left(k_{Ma}\text{Ma}+\frac{k_{b}}{\alpha^{2}}\left(\text{Bo}+\alpha^{2}\right)\right)\label{eq:infinIMz}
\end{align}
with the coefficients here 
\begin{eqnarray*}
k_{Ma} & = & (m-1)(\alpha n+\alpha n^{2}-n^{2}s_{\alpha}c_{\alpha}-s_{\alpha n}c_{\alpha n})\\
 &  & \times\left(m(s_{\alpha}^{2}-\alpha^{2})(\alpha n+s_{\alpha n}c_{\alpha n})+(\alpha+s_{\alpha}c_{\alpha})(s_{\alpha n}^{2}-\alpha^{2}n^{2})\right)\\
 &  & +2(s_{\alpha n}-ns_{\alpha})(ns_{\alpha}+s_{\alpha n})(-\alpha^{4}(-1+m)^{2}n^{2}+(mc_{\alpha n}s_{\alpha}+c_{\alpha}s_{\alpha n})^{2}\\
 &  & +\alpha^{2}(-c_{\alpha}n^{2}m^{2}+n(-c_{\alpha}^{2}n+m(-2+mns_{\alpha}^{2}))+s_{\alpha n}^{2}))
\end{eqnarray*}
and 
\begin{eqnarray*}
k_{b} & = & (m-1)(n(\alpha+\alpha n-c_{\alpha n}s_{\alpha})-c_{\alpha n}s_{\alpha n})(\alpha^{3}n(m+n)+s_{\alpha}s_{\alpha n}(mc_{\alpha n}s_{\alpha}+c_{\alpha}s_{\alpha n})\\
 &  & -\alpha^{2}(c_{\alpha n}^{2}s_{\alpha}+mc_{\alpha n}s_{\alpha n})-\alpha(mns_{\alpha}^{2}+s_{\alpha n}^{2}))
\end{eqnarray*}
As we mentioned before, the two equations $\text{Re}(\zeta)=0$ and
$\text{Im}(\zeta)=0$ define a manifold of codimension two in the
$\alpha$-parameter space. This manifold is analogous to a multivalued-function
branch point in the complex plane. We consider the trace of this ``branch
manifold'' in the three-dimensional space of $(\alpha,\;\text{Ma},\;\text{Bo})$,
with the rest of the parameters fixed, as follows. Solving $\operatorname{Im}(\zeta)=0$
for Marangoni number yields 
\begin{equation}
\text{Ma}=-\frac{k_{b}}{\alpha^{2}k_{Ma}}(\text{Bo}+\alpha^{2})\text{.}\label{eq:MaWhenIM-eq-0}
\end{equation}
Note that not all values of $($Bo$,\text{Ma})$ are appropriate here
because $\Ma$ must be positive.

\begin{figure}
\includegraphics[clip,width=0.95\textwidth]{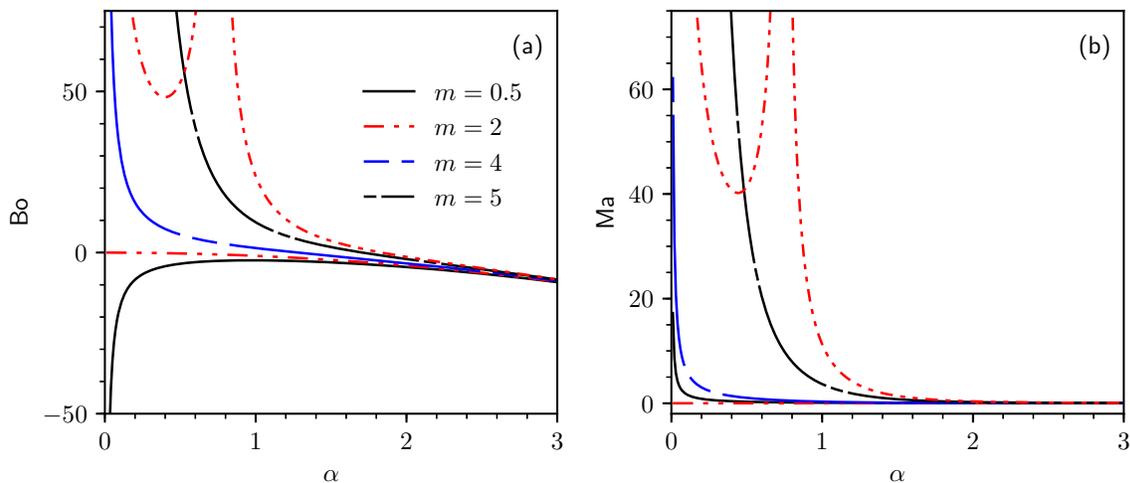} \caption{The curves (a) $\text{Bo}=\text{Bo}_{m}(\alpha)$ and (b) $\text{Ma}=\text{Ma}_{m}(\alpha)$
such that the discriminant $\zeta=0$ are plotted for the values of
viscosity ratio $m$ indicated in the legend. \label{fig:zetaeq0-BomandMamvsalpha}}
\end{figure}

Similarly to the above expression for $\text{Im}(\zeta)$, we obtain
\[
Re(\zeta)=\frac{1}{\alpha^{10}}\left(K_{20}\text{Ma}^{2}+K_{02}\left(\text{Bo}+\alpha^{2}\right)^{2}+K_{11}\text{Ma}(\text{Bo}+\alpha^{2})+K_{00}\right),
\]
where 
\begin{eqnarray*}
K_{20} & = & \frac{1}{4}\alpha^{4}(\alpha n(\alpha^{2}(m+n)+\alpha c_{\alpha n}s_{\alpha}-sms_{\alpha}^{2})-mc_{\alpha n}(s_{\alpha}^{2}-\alpha^{2})s_{\alpha n}-(\alpha+c_{\alpha}s_{\alpha})s_{\alpha n}^{2})^{2},\\
K_{02} & = & \frac{1}{4}(\alpha^{3}n(m+n)+s_{\alpha}s_{\alpha n}(mc_{\alpha n}s_{\alpha}+c_{\alpha}s_{\alpha n})-\alpha^{2}(c_{\alpha n}^{2}s_{\alpha}+mc_{\alpha n}s_{\alpha n})-\alpha(mns_{\alpha}^{2}+s_{\alpha n}^{2}))^{2},\\
K_{11} & = & \frac{1}{2}\alpha^{2}(m(s_{\alpha}^{2}-\alpha^{2})(s_{\alpha n}^{2}-\alpha^{2}n^{2})\left((\alpha+c_{\alpha}s_{\alpha})(c_{\alpha n}s_{\alpha n}-\alpha n)+(c_{\alpha}s_{\alpha}-\alpha)(\alpha n+c_{\alpha n}s_{\alpha n})\right)\\
 &  & +(c_{\alpha}^{2}s_{\alpha}^{2}-\alpha^{2})(s_{\alpha n}^{2}-\alpha^{2}n^{2})^{2}\\
 &  & +m^{2}(s_{\alpha}^{2}-\alpha^{2})^{2}(c_{\alpha n}^{2}s_{\alpha n}^{2}-\alpha^{2}n^{2})+2(s_{\alpha}^{2}-\alpha^{2})(s_{\alpha n}^{2}-\alpha^{2}n^{2})(\alpha^{4}(m-1)^{2}n^{2}\\
 &  & -(mc_{\alpha n}s_{\alpha}+nc_{\alpha}s_{\alpha})^{2}+\alpha^{2}(m^{2}c_{\alpha n}^{2}+n(nc_{\alpha}^{2}+m(2-mns_{\alpha}^{2}))-s_{\alpha n}^{2}))),\\
K_{00} & = & -s^{2}\alpha^{6}(m-1)^{2}(-\alpha n(1+n)+c_{\alpha n}^{2}s_{\alpha}+c_{\alpha n}s_{\alpha n})^{2}.
\end{eqnarray*}
To solve the system $\operatorname{Re}(\zeta)=0$ and $\operatorname{Im}(\zeta)=0$
for $\Ma$ and $\Bo$ as functions of $\alpha$ (with $s$, $m$,
and $n$ fixed), equation (\ref{eq:MaWhenIM-eq-0}) is substituted
into $\operatorname{Re}(\zeta)$ which yields 
\begin{equation}
\operatorname{Re}(\zeta)=AB^{2}+C=0,\label{eq:RezABC}
\end{equation}
where $B=\Bo+\alpha^{2}$, and $A$ and $C$ do not depend on $\Ma$:
\[
A=\frac{1}{\alpha^{10}}\left(\frac{k_{b}^{2}}{\alpha^{4}k_{Ma}^{2}}K_{20}+K_{02}-\frac{k_{b}}{\alpha^{2}k_{Ma}}K_{11}\right),\;C=\frac{K_{00}}{\alpha^{10}}.
\]
Therefore, $\Bo=\Bo(\alpha)$, where 
\begin{equation}
\text{Bo}(\alpha)=-\alpha^{2}\pm\sqrt{-\frac{C}{A}}\text{.}\label{eq:BoACal}
\end{equation}
Substituting (\ref{eq:BoACal}) for $\Bo$ into equation (\ref{eq:MaWhenIM-eq-0})
yields $\Ma$ such that $\zeta=0$ for a given $\alpha$. Only the
unique value $\Bo=\Bo(\alpha)$ that yields $\Ma=\Ma(\alpha)$
$>0$ is admitted here. In figures \ref{fig:zetaeq0-BomandMamvsalpha}(a)
and (b) curves $\Bo=\Bo(\alpha)$ and $\Ma=\Ma(\alpha)$ are
plotted for various values of $m$. One can see that $\Ma\uparrow\infty$
in the limit $\alpha\downarrow0$ for all $m$. In this limit, $\Bo\uparrow\infty$
for $m>1$, but $\Bo\uparrow-\infty$ for $m<1$. At $\alpha\uparrow\infty$,
for all $m$, $\text{Bo}\sim-\alpha^{2}$ and $\Ma\downarrow0$. There
are no points where the discriminant is zero for $m=1$, as was shown
in the main text for all parameter values (formally, in the figure,
we get $\Ma(\alpha)=0$ and $\Bo(\alpha)=-\alpha^{2}$). This
indicates that the branch manifold consists of at least two pieces,
and perhaps more than two, some with $m>1$ and others with $m<1$.
The same fact is reflected in the infinite discontinuities of the
curves in the figure at finite values of $\alpha$, which take place
provided $m>n^{2}$.

Also, if we consider the ($\alpha$, $\Ma$)-plane, with all the other
parameters fixed, including $\text{Bo}$, corresponding to a horizontal
line in figure \ref{fig:zetaeq0-BomandMamvsalpha}(a), there will
be at most two branch points in the ($\alpha,\Ma$)-plane since any
horizontal line there intersects any curve at no more than two points.
Therefore, in some sufficiently narrow infinite strip whose left boundary
is the (vertical) $\Ma$-axis, the discriminant is non-zero at all
its points, and so there are two continuous branches, in agreement
with the long-wave results in the main text. These results also show
no intersections of the two dispersion curves (when the wavenumbers
are small enough), which means that $\operatorname{Re}\sqrt{\zeta}$
is non-zero in a sufficiently narrow strip bordering the $\Ma$-axis.
The equation $\operatorname{Re}\sqrt{\zeta}=0$ implies that $\zeta$
is real (and negative). We have solved this equation for $\Ma$ as
a function of $\alpha$ at fixed values of $\Bo$ (and the other parameters),
and every resulting curve in the ($\alpha,\Ma$)-plane indeed lies
entirely outside some strip bordering the $\Ma$-axis.

Regarding the entire ($\alpha,\Ma$)-plane, if we remove from it the
branch points together with the infinite rays emanating from each
branch point to the right and going parallel to the $\alpha$-axis,
then in the remaining domain the discriminant is nowhere zero, and
thus there are two continuous branches of the growth rate in this
domain, smooth in $\alpha$ at each point that they are defined.

Next, we note that the horizontal line $\Bo=0$ in panel (a) of figure
\ref{fig:zetaeq0-BomandMamvsalpha} intersects every curve whose $m>1$.
So, even in the absence of gravity, there may be intersections of
the two dispersion curves. As $\Ma$ is varied, these intersections
disappear at some $\Ma$, with the reconnection of the curve parts
lying to the right of the ``marginal intersection'' point and consequent
separation of the two ``renovated'' dispersion curves. This happens
in the ranges of wavenumbers when both branches are stable, which
was not noted in HF.

\begin{figure}
\centerline{\includegraphics[clip,scale=0.75]{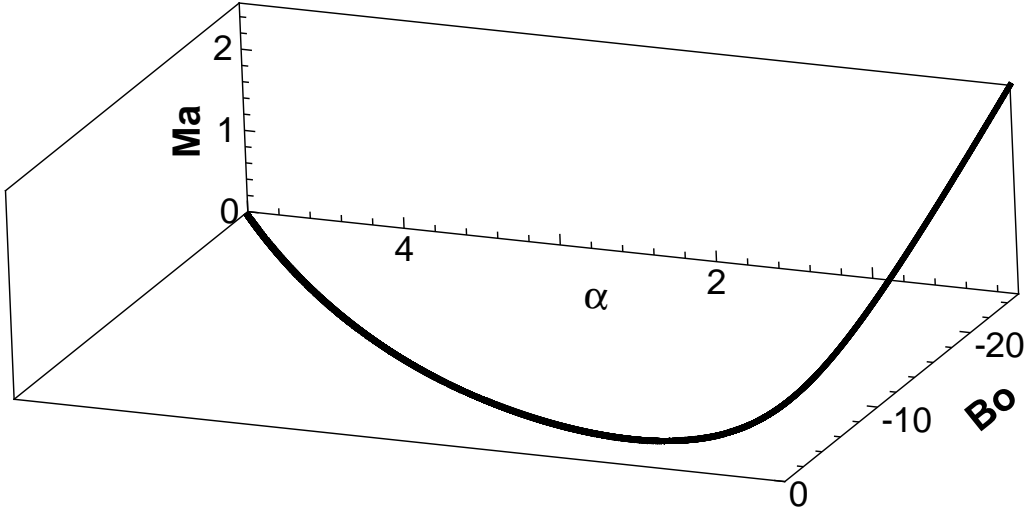}} \caption{The zero discriminant curve corresponding to the $m=2$ projection
curves shown in panels (a) and (b) of figure \ref{fig:zetaeq0-BomandMamvsalpha}.
\label{fig:The-zero-discriminant3d}}
\end{figure}

Figure \ref{fig:The-zero-discriminant3d} shows, as an example, the
curve in the three-dimensional space which corresponds to the two
dash-dotted, $m=2$, curves of figure \ref{fig:zetaeq0-BomandMamvsalpha}
. The coordinate box there is shown with its front, top, and right
faces removed for a better view. The curve of zero discriminant starts
at the back top right vertex and steadily goes downward and to the
left simultaneously twisting first toward the viewer and then backward,
until it ends at the back bottom left vertex.

Next, we demonstrate that there is always a strip $\mathcal{D}_{s}=\{0<\alpha<\alpha_{s},\Ma>0\}$
where $\zeta\neq0$. Indeed, it appears in figure \ref{fig:zetaeq0-BomandMamvsalpha}(a)
that any horizontal line $\Bo=\Bo_{f}$ intersects any of the graphs
of $\Bo=\ \Bo_{m}(\alpha)$ at no more than three points. If there
are no intersections then the value of $\alpha_{s}$ is chosen completely
arbitrarily. Otherwise, $\alpha_{s}$ must be smaller than the smallest
$\alpha$ of the intersection points. For the purpose of this paper,
the existence of $\mathcal{D}_{s}$ (and thus of the two branches
of the growth rate) is sufficient with any small but finite $\alpha_{s}$.
The existence of $\alpha_{s}$ is shown analytically for small values
of $\alpha$.

\section{Coefficients of equations \label{sec:Coefficients-of-equations}}

\setcounter{equation}{0} The coefficients $A_{11}$, $A_{12}$, $A_{21}$,
and $A_{22}$ of equation (\ref{eq:dispEqnSystem}) are: 
\begin{align}
\text{Re}(A_{11}) & =\left(m(s_{\alpha}^{2}-\alpha^{2})(s_{\alpha n}c_{\alpha n}-\alpha n)+(s_{\alpha n}^{2}-\alpha^{2}n^{2})(s_{\alpha}c_{\alpha}-\alpha)\right)\frac{1}{2\alpha^{5}F_{2}}B,\label{eq:ReA11}\\
\text{Im}(A_{11}) & =-\frac{(m-1)s}{\alpha^{2}F_{2}}\left(n^{2}(s_{\alpha}c_{\alpha}-\alpha)+s_{\alpha n}c_{\alpha n}-\alpha n\right),\\
A_{12} & =\left(s_{\alpha n}^{2}-\alpha^{2}n^{2}-mn^{2}(s_{\alpha}^{2}-\alpha^{2})\right)\frac{\text{Ma}}{2\alpha^{2}F_{2}},\label{eq:A12}\\
\text{Re}(A_{21}) & =\left(s_{\alpha n}^{2}-\alpha^{2}n^{2}-mn^{2}(s_{\alpha}^{2}-\alpha^{2})\right)\frac{1}{2\alpha^{2}F_{2}}B,\label{eq:ReA21}\\
\text{Im}(A_{21}) & =\left(m\left((s_{\alpha}c_{\alpha}-\alpha)(s_{\alpha n}c_{\alpha n}+\alpha n)+c{}_{\alpha n}^{2}(s_{\alpha}^{2}-\alpha^{2})+\alpha^{2}n^{2}s_{\alpha}^{2}\right)\right.\nonumber \\
 & \left.+(s_{\alpha}c_{\alpha}+\alpha)(s_{\alpha n}c_{\alpha n}+\alpha n)+c_{\alpha}^{2}(s_{\alpha n}^{2}-\alpha^{2}n^{2})+\alpha^{2}s_{\alpha n}^{2}\right)\frac{s}{\alpha^{3}F_{2}},\label{eq:ImA21}\\
A_{22} & =\left(m((s_{\alpha}^{2}-\alpha^{2})(s_{\alpha n}c_{\alpha n}+\alpha n))+(s_{\alpha n}^{2}-\alpha^{2}n^{2})(s_{\alpha}c_{\alpha}+\alpha)\right)\frac{\text{Ma}}{2\alpha^{3}F_{2}}.\label{eq:A22}
\end{align}
The coefficients $k_{20}$, $k_{11}$, $k_{31}$, $k_{22}$ and $k_{13}$
that appear in equation (\ref{eq:MaEquation2}) are 
\begin{align}
k_{20} & =\frac{s^{2}}{4\alpha^{6}}\left(s_{\alpha n}^{2}-n^{2}s_{\alpha}^{2}\right)\left\{ (m-1)\left(s_{\alpha n}c_{\alpha n}-\alpha n+{n}^{2}\left(s_{\alpha}c_{\alpha}-\alpha\right)\right)\right.\nonumber \\
 & \times\left[m\left({\,s_{\alpha}^{2}-{\alpha}^{2}}\right)({s_{\alpha n}c_{\alpha n}+\alpha\,n})+(s_{\alpha n}^{2}-{\alpha}^{2}n^{2})\left({\alpha+s_{\alpha}c_{\alpha}}\right)\right]\nonumber \\
 & -\left(s_{\alpha n}^{2}-{s}_{\alpha}^{2}n^{2}\right)\left[m^{2}\left({\,s_{\alpha}^{2}-{\alpha}^{2}}\right)\left({c_{\alpha n}^{2}}+{\alpha}^{2}{n}^{2}\right)\right.\nonumber \\
 & \left.\left.+2m(n^{2}\alpha^{4}-n\alpha^{2}+{s_{\alpha}c_{\alpha}s_{\alpha n}c_{\alpha n})}+\left({c_{\alpha}^{2}}+{\alpha}^{2}\right)\left(s_{\alpha n}^{2}-{\alpha}^{2}n^{2}\right)\right]\right\} ,\label{eq:k20}
\end{align}
\begin{align}
k_{11} & =\frac{s^{2}}{4\alpha^{8}}(m-1)\left(s_{\alpha n}^{2}-{s}_{\alpha}^{2}n^{2}\right)\left({s_{\alpha n}c_{\alpha n}-\alpha\,n+{n}^{2}}\left({s_{\alpha}c_{\alpha}-\alpha}\right)\right)\nonumber \\
 & \times\left[m({s_{\alpha n}c_{\alpha n}-\alpha\,n})\left({\,s_{\alpha}^{2}-{\alpha}^{2}}\right)+({s_{\alpha}c_{\alpha}-\alpha})\left(s_{\alpha n}^{2}-{\alpha}^{2}n^{2}\right)\right],\label{eq:k11}
\end{align}
\begin{align}
k_{31} & =\frac{1}{16\alpha^{10}}\left({\,s_{\alpha}^{2}-{\alpha}^{2}}\right)\left(s_{\alpha n}^{2}-{\alpha}^{2}n^{2}\right)\left[({s_{\alpha}c_{\alpha}+\alpha)}\left(s_{\alpha n}^{2}-{\alpha}^{2}n^{2}\right)\right.\nonumber \\
 & \left.+m({s_{\alpha n}c_{\alpha n}+\alpha\,n})({s_{\alpha}^{2}-{\alpha}^{2}})\right]^{2},\label{eq:k31}
\end{align}
\begin{align}
k_{22} & =\frac{1}{8\alpha^{12}}\left({\,s_{\alpha}^{2}-{\alpha}^{2}}\right)\left(s_{\alpha n}^{2}-{\alpha}^{2}n^{2}\right)\left[m({s_{\alpha n}c_{\alpha n}-\alpha\,n})({s_{\alpha}^{2}-{\alpha}^{2}})\right.\nonumber \\
 & \left.+({s_{\alpha}c_{\alpha}-\alpha)}\left(s_{\alpha n}^{2}-{\alpha}^{2}n^{2}\right)\right]\left[m({s_{\alpha n}c_{\alpha n}+\alpha\,n})({s_{\alpha}^{2}-{\alpha}^{2}})\right.\nonumber \\
 & \left.+({s_{\alpha}c_{\alpha}+\alpha)}\left(s_{\alpha n}^{2}-{\alpha}^{2}n^{2}\right)\right],\label{eq:k22}
\end{align}
and 
\begin{align}
k_{13} & =\frac{1}{16\alpha^{14}}\left({\,s_{\alpha}^{2}-{\alpha}^{2}}\right)\left(s_{\alpha n}^{2}-{\alpha}^{2}n^{2}\right)\left[({s_{\alpha}c_{\alpha}-\alpha)}\left(s_{\alpha n}^{2}-{\alpha}^{2}n^{2}\right)\right.\nonumber \\
 & \left.+m({s_{\alpha n}c_{\alpha n}-\alpha\,n})({s_{\alpha}^{2}-{\alpha}^{2}})\right]^{2}.\label{eq:k13}
\end{align}
The corresponding long-wave approximations are 
\begin{equation}
k_{20}\approx k_{206}\alpha^{6}+k_{208}\alpha^{8},\label{eq:k20app}
\end{equation}
where 
\[
k_{206}=\frac{n^{4}s^{2}}{108}\varphi(n-1)(n+1)^{2}(m-n^{2})
\]
 and
\begin{eqnarray*}
k_{208} & = & \frac{s^{2}}{810}(n-1)n^{4}(n+1)^{3}\left(m^{2}(n(3n+8)+3)-4m\left(n^{2}-1\right)n^{2}\right.\\
 &  & -\left.(n(3n+8)+3)n^{4}\right),
\end{eqnarray*}
 
\begin{equation}
k_{11}\approx k_{116}\alpha^{6}+k_{118}\alpha^{8},\label{eq:k11app}
\end{equation}
where 
\[
k_{116}=\frac{n^{7}s^{2}}{81}(n-1)(n+1)^{2}(m-1)(n+m)
\]
 and
\[
k_{118}=\frac{s^{2}}{1215}(m-1)(n-1)n^{7}(n+1)^{2}(m(n(8n-3)+7)+n(n(7n-3)+8)),
\]
\begin{equation}
k_{31}\approx k_{318}\alpha^{8},\label{eq:k31app}
\end{equation}
where 
\[
k_{318}=\frac{n^{6}}{324}(n^{3}+m)^{2},
\]
\begin{equation}
k_{22}\approx k_{228}\alpha^{8},\label{eq:k22app}
\end{equation}
where 
\[
k_{228}=\frac{n^{8}}{486}(n+m)(n^{3}+m),
\]
and 
\begin{equation}
k_{13}\approx k_{138}\alpha^{8},\label{eq:k13app}
\end{equation}
where 
\[
k_{138}=\frac{n^{10}}{2916}(n+m)^{2}.
\]
The coefficient of the $\alpha^{4}$ term that appears in equation
(\ref{eq:gamSsmallAlphaApprox}) is 
\begin{align}
k_{S}= & \frac{\text{Ma}\left(n^{3}-4n^{2}+4n-1\right)}{60(m-1)}\nonumber \\
 & +\frac{\text{Ma}^{3}}{128(m-1)^{5}n^{4}(n+1)s^{2}}(n-1)\left(m^{4}(3n+1)+2m^{3}\left(-3n^{3}-2n^{2}+4n+1\right)n\right.\nonumber \\
 & +\left.4m^{2}\left(n^{3}-2n^{2}-2n+1\right)n^{3}+2m\left(n^{3}+4n^{2}-2n-3\right)n^{5}+(n+3)n^{8}\right)\nonumber \\
 & +\frac{\text{Bo}\text{Ma}^{2}}{192(m-1)^{4}n(n+1)^{2}s^{2}}\left(m^{3}\left(3n^{2}-4n-3\right)\right.+m^{2}\left(2n^{3}+13n^{2}-6n-5\right)n\nonumber \\
 & +\left.m\left(-5n^{3}-6n^{2}+13n+2\right)n^{3}+\left(-3n^{2}-4n+3\right)n^{5}\right)\nonumber \\
 & +\text{Bo}^{2}\text{Ma}\frac{n^{2}\left(-m^{2}+m(n-1)n+n^{3}\right)}{144(m-1)^{3}(n+1)^{2}s^{2}}\label{eq:ks}
\end{align}
The coefficients of the constant, quadratic and quartic terms of the
marginal wavenumber equation (\ref{eq:MAappEqn}) are 
\begin{equation}
\zeta_{0}=\frac{1}{108}s^{2}(n-1)(n+1)^{2}(m-n^{2})\varphi\text{Ma}+\frac{1}{81}n^{3}s^{2}(n-1)(n+1)^{2}(m-1)(n+m)\text{Bo},\label{eq:zeta0}
\end{equation}
\begin{align}
\zeta_{2}= & \begin{gathered}\frac{\text{Ma}}{810}(-1+n)(1+n)^{3}\left(-4mn^{2}\left(-1+n^{2}\right)+m^{2}(3+n(8+3n))-n^{4}(3+n(8+3n))\right)s^{2}\end{gathered}
\nonumber \\
+ & \text{Bo}\frac{(-1+m)(-1+n)n^{3}(1+n)^{2}(n(8+n(-3+7n))+m(7+n(-3+8n)))\ s^{2}}{1215}\nonumber \\
+ & \textrm{Bo}\frac{n^{2}}{2916}\left(3(m+n^{3})\text{Ma}+n^{2}(m+n)\text{Bo}\right)^{2}+\frac{1}{81}s^{2}n^{3}(n-1)(n+1)^{2}(m-1)(n+m)\nonumber \\
\label{eq:zeta2}
\end{align}
and 
\begin{equation}
\zeta_{4}=\frac{1}{324}n^{2}(m+n^{3})^{2}\text{Ma}^{2},\label{eq:zeta4}
\end{equation}
where only the leading order term in $s$ has been retained in $\zeta_{4}$,
so that $\zeta_{4}=\zeta_{40}$ of section \ref{subsec:Marginal-wavenumbers}.

The linear and cubic coefficients in $\text{Bo}_{c}$ of expression
(\ref{eq:MAapproxCmode}) are given by 
\begin{align}
\beta_{1}=\frac{1}{15}\left(\frac{\left(m^{2}-1\right)m}{m+n}-m^{2}+\frac{2(m-1)m}{m-n^{2}}-\frac{6(m-1)\left(3mn+m+4n^{2}\right)}{3mn+m+(n+3)n^{2}}\right.\nonumber \\
\left.+(m-7)n+4m+n^{2}-2\right)\label{eq:beta1}
\end{align}
and 
\begin{equation}
\beta_{3}=\frac{1}{36}\frac{n^{3}(n+m)\left\vert n-1\right\vert \psi^{2}}{\left[\varphi{s}(m-n^{2})(n+1)\right]^{2}\left\vert m-1\right\vert }\text{.}\label{eq:beta3}
\end{equation}
The constant, linear and cubic coefficients in $\text{Ma}_{cL}$,
$M_{0}$, $M_{1}$ and $M_{3}$, of the expression (\ref{eq:RTmaApp})
are 
\begin{equation}
M_{0}=\frac{4n^{3}(m-1)(m+n)}{3\phi(m-n^{2})},\label{eq:M0}
\end{equation}
\begin{align}
M_{1}=\frac{1}{15}\left(\frac{m(1-m^{2})}{m+n}+\frac{2(m-1)m}{n^{2}-m}+\frac{6(m-1)\left(3mn+m+4n^{2}\right)}{\phi}\right.\nonumber \\
\left.-mn+(m-4)m-n^{2}+7n+2\right)\label{eq:M1}
\end{align}
and 
\begin{equation}
M_{3}=-\frac{(n-1)\psi^{2}}{64(m-1)^{3}n^{3}(n+1)^{2}s^{2}(m+n)}.\label{eq:M3}
\end{equation}

\section{Long-wave formulas for $F_{0}$, $F_{1}$ and $F_{2}$ \label{sec:Longwave-formulas-for-F}}

The small wavenumber approximations for the case of finite thickness,
$n$, and small Marangoni number, $\Ma$ are given here. The long-wave
approximations of (\ref{eq:F2Re})-(\ref{eq:F0Im}) are first written
as polynomials in $\Ma$ and $\Bo$, then the coefficients are expanded,
so that keeping only the leading term in $\alpha$ , equations (\ref{eq:F2Re})-(\ref{eq:F0Im})
are approximately 
\begin{align}
F_{2} & =\operatorname{Re}(F_{2})\approx\frac{1}{3}\,\psi\text{,}\label{eq:F2reApprox}\\
\operatorname{Re}(F_{1}) & \approx\frac{1}{9}n^{3}(m+n){\alpha}^{4}+\frac{1}{3}n(m+n^{3}){\alpha}^{2}\text{Ma}+\frac{1}{9}\,\,\,{n}^{3}(m+n){\alpha}^{2}\text{Bo,}\label{eq:F1reApprox}\\
\operatorname{Im}(F_{1}) & \approx\frac{2}{3}n^{2}s(n+1)(1-m)\alpha\text{,}\label{eq:F1imApprox}\\
\operatorname{Re}(F_{0}) & \approx\frac{1}{36}\,{n}^{4}{\alpha}^{6}\text{Ma}+\frac{1}{36}{n}^{4}{\alpha}^{4}\text{MaBo}\,\text{,}\label{eq:F0reApprox}\\
\operatorname{Im}(F_{0}) & \approx\frac{1}{6}n^{2}s(1-n^{2}){\alpha}^{3}\text{Ma,}\label{eq:F0imApprox}
\end{align}
where $\psi$ is given by equation (\ref{eq:psi}). For $m=1$, we
find 
\begin{align}
F_{2} & =\operatorname{Re}(F_{2})\approx\frac{1}{3}\,(n+1)^{4},\\
\operatorname{Re}(F_{1}) & \approx\frac{1}{9}n^{3}(n+1){\alpha}^{4}+\frac{1}{3}n(n^{3}+1){\alpha}^{2}\text{Ma}+\frac{1}{9}\,\,\,{n}^{3}(n+1){\alpha}^{2}\text{Bo},\\
\operatorname{Im}(F_{1}) & =0\text{,}\\
\operatorname{Re}(F_{0}) & \approx\frac{1}{36}\,{n}^{4}{\alpha}^{6}\text{Ma}+\frac{1}{36}{n}^{4}{\alpha}^{4}\text{MaBo}\,\text{,}
\end{align}
and 
\begin{equation}
\operatorname{Im}(F_{0})\approx\frac{1}{6}n^{2}s(1-n^{2}){\alpha}^{3}\text{Ma.}
\end{equation}

\section{Normal modes with undisturbed surfactant \label{sec:Are-there-normal}}

Assuming that the surfactant is undisturbed, $G=0$, which implies
that $h\ne0$, it follows from the second equation of (\ref{eq:dispEqnSystem})
that $A_{21}=0$. This implies in particular that $\textrm{Im}(A_{21})=0$.
However, in expression (\ref{eq:ImA21}), each term is positive, since
each of the expressions $s_{\alpha}c_{\alpha}-\alpha$, $s_{\alpha}^{2}-\alpha^{2}$,
and $s_{\alpha n}^{2}-\alpha^{2}n^{2}$ is positive. This contradiction
shows that there are no normal modes with $G=0$ if $s\ne0$.

If, however, $s=0$, but $B$ is nonzero, then $\textrm{Im}(A_{21})=0$
identically. However, $\textrm{Re}(A_{21})=0$ yields, from equation
(\ref{eq:ReA21}), that 
\begin{equation}
m=\frac{s_{\alpha n}^{2}-\alpha^{2}n^{2}}{n^{2}(s_{\alpha}^{2}-\alpha^{2})}.\label{eq:m_AppD}
\end{equation}
This equation gives a two-dimensional manifold of normal modes (parameterized
with variables $n$ and $\alpha$). Thus, the normal modes with $G$$=0$
(and $h\neq0$) do exist, but only when $s=0$. Note that the first
equation of the system (\ref{eq:dispEqnSystem}) implies that $\gamma=-A_{11}$,
and we find, making use of (\ref{eq:m_AppD}), the growth rate for
this mode is 
\[
\gamma_{R}=-\textrm{Re}(A_{11})=\left(\frac{s_{\alpha n}c_{\alpha n}-\alpha n}{n^{2}(s_{\alpha}^{2}-\alpha^{2})}+s_{\alpha}c_{\alpha}-\alpha\right)\left(s_{\alpha n}^{2}-\alpha^{2}n^{2}\right)\frac{1}{2\alpha^{5}F_{2}}B.
\]
Thus, we have one nonzero branch of modes, which are the usual Rayleigh-Taylor
modes for the stagnant base configuration. Also, for any negative
$\textrm{Bo}$, if $B=0$, that is $\alpha^{2}=-\textrm{Bo}$, then
$A_{21}=0$ without any restrictions on $m$ and $n$. We can see
that $A_{11}=0$ in this case as well, so that $\gamma_{R}=0$, which
indicates the marginal stability mode for the Rayleigh-Taylor instability
of the stagnant base configuration.

 \bibliographystyle{plainnat}
\bibliography{couette}

\end{document}